\documentclass[twocolumn,trackchanges]{aastex63}

\newcommand{\twCO}{$^{12}$CO}   
\newcommand{\thCO}{$^{13}$CO}

\newcommand{\kms}{km~s$^{-1}$}

\begin{document}

\title{Spatially Resolved X-ray Study of Supernova Remnant G306.3$-$0.9 with Unusually High Calcium Abundance}

\correspondingauthor{Ping Zhou, Yang Chen}
\email{pingzhou@nju.edu.cn, ygchen@nju.edu.cn}

\author{Jianbin Weng}
\affiliation{School of Astronomy and Space Science, Nanjing University,
163 Xianlin Avenue,
Nanjing, 210023, China}

\author{Ping Zhou}
\affiliation{School of Astronomy and Space Science, Nanjing University,
163 Xianlin Avenue,
Nanjing, 210023, China}
\affiliation{Anton Pannekoek Institute for Astronomy, University of Amsterdam,
Science Park 904,
1098 XH Amsterdam, The Netherlands}

\author{Yang Chen}
\affiliation{School of Astronomy and Space Science, Nanjing University,
163 Xianlin Avenue,
Nanjing, 210023, China}

\author{Shing-Chi Leung}
\affiliation{TAPIR, Walter Burke Institute for Theoretical Physics,
Mailcode 350-17, Caltech,
Pasadena, CA 91125, USA}

\author{Silvia Toonen}
\affiliation{Anton Pannekoek Institute for Astronomy, University of Amsterdam,
Science Park 904,
1098 XH Amsterdam, The Netherlands}

\author{Hagai B. Perets}
\affiliation{Physics Department, Technion --- Israel Institute of Technology,
Haifa 3200003, Israel}

\author{Ken'ichi Nomoto}
\affiliation{Kavli Institute for the Physics and Mathematics of the Universe (WPI), The University of Tokyo Institutes for Advanced Study, The University of Tokyo,
Kashiwa, Chiba 277-8583, Japan}

\author{Yossef Zenati}
\affiliation{Department of Physics and Astronomy, The Johns Hopkins University,
Baltimore, MD 21218, USA}

\author{Jacco Vink}
\affiliation{Anton Pannekoek Institute for Astronomy, University of Amsterdam,
Science Park 904,
1098 XH Amsterdam, The Netherlands}
\affiliation{GRAPPA, University of Amsterdam, Science Park 904, 1098 XH Amsterdam, The Netherlands}
\affiliation{SRON, Netherlands Institute for Space Research, Sorbonnelaan 2, 3584 CA Utrecht, The Netherlands}

\begin{abstract}

G306.3$-$0.9 is an asymmetric Galactic supernova remnant (SNR), whose progenitor has been thought to be a Type Ia supernova (SN), but its high Ca abundance appears inconsistent with the Type Ia origin. Hoping to uncover the reason for its asymmetry and the origin of this SNR, we performed a spatially resolved X-ray spectroscopic analysis of XMM-{\it Newton}
and {\it Chandra} observation data. 
We divided the SNR into 13 regions and analyzed the spectra using two-temperature models (0.2 keV + 1 keV).
Compared to the southwestern regions, the northeastern regions have higher metal abundances and
a lower gas density.
This suggests that the asymmetric morphology results from the non-uniform ambient environment.
We found that neither Type Ia nor core-collapse SN models can account for the abnormally high abundance ratios of Ar/Si, Ca/Si, or the shape of the abundance curve. 
A comparison
with the Ca-rich transient models suggests that G306.3$-$0.9 is likely to be the first identified
Galactic ``Ca-rich transient'' remnant, although the theoretical production of element S is lower. We also note that the conclusion for the SNR's origin relies on the measured abundance ratios and existing nucleosynthesis models. Between two groups of Ca-rich transient explosion models, we prefer the He shell detonation for an accreting WD, rather than the merger of a white dwarf and a neutron star.

\end{abstract}

\keywords{Supernova remnants (1667) --- Interstellar medium (847) --- Explosive nucleosynthesis (503) --- Abundance ratios (11) ---  X-ray astronomy (1810)}

\section{Introduction} \label{sec:intro}

Supernovae (SNe), explosions that release a huge amount of energy, are mainly divided into two classes: thermonuclear (including Type Ia) and core-collapse (CC) SNe. Although there is no conclusion 
on the process in which SNe explode, it's widely accepted that the progenitor of a Type Ia SN is a white dwarf (WD), while that of a CC SN is a  massive star \citep[e.g.,][]{2012A&ARv..20...49V}. For several decades, Type Ia SNe have 
often been used as a synonym for the thermonuclear SNe \citep[see e.g.,][for recent reviews]{Hillebrandt2000, Nomoto2017SNIa}. However, this idea has been challenged, after recent SN searches revealing several subclasses of thermonuclear SNe, each of which has distinguished photometric and spectroscopic properties. The newly found sub-types include Type Iax SNe (faint), Super-Chandrasekhar SNe (bright; see \citealt{2019NatAs...3..706J} for a review) and so on.  

Recently, a new subtype of thermonuclear SNe called Ca-rich transient has been discovered in external galaxies \citep{2010Natur.465..322P,2010Natur.465..326K,2012ApJ...755..161K}. They were initially classified into Type Ib SNe because of the strong He features in their optical spectra but were later found to have different characteristics of low luminosity, fast temporal evolution, large Ca to O line ratios in nebular phase and show a tendency to occur in the outskirts of galaxies. Such large offsets have been explained by in-situ low-mass stars or runaway binary mergers \citep[but see][]{2021MNRAS.503.5997P}. This group of SNe was at first called ``Ca-rich gap transients'' because its luminosity lies in the ``gap'' between novae and ordinary SNe \citep{2012ApJ...755..161K,2017hsn..book..317T}. As more similar transients were reported, the gap has been populated with these objects and now they are called ``Ca-rich transients'' or ``Calcium-strong transients'' \citep{2019ApJ...887..180S}.

There have been debates on the origin of the large Ca to O line ratios. A plausible explanation is that their ejecta is rich in Ca \citep[e.g. $\sim 0.1$ M$_\sun$ Ca ejecta estimated for SN 2005E,][]{2010Natur.465..322P}. Two promising models providing large Ca yields have been proposed: i) the accreted WD Helium shell detonation \citep{2011ApJ...738...21W}; ii) the merger of a He-rich WD and a neutron star \citep[NS,][]{2012MNRAS.419..827M}. Another interpretation is that these transients are ``O-poor but Ca-normal'' SNe, including low-mass stripped CC SNe \citep[e.g. SN 2005cz,][]{2010Natur.465..326K} and ultra-stripped CC SNe. The CC origin was once disfavoured because Ca-rich transients were generally found in old stellar populations. 
However, the discovery of iPTF 15eqv and SN 2019ehk in star-forming regions 
might support the CC scenario 
\citep{2017ApJ...846...50M,2021ApJ...907L..18D}, although a WD progenitor system cannot be excluded for 2019ehk \citep{2021ApJ...908L..32J}.
The progenitor of Ca-rich transients has been increasingly controversial. One emerging opinion is that there are two subclasses (related to old or young populations) of Ca-rich transients
\citep{2017ApJ...846...50M,2021ApJ...907L..18D}.
Enlarging the small sample of Ca-rich transients or discovering  nearby remnants of this group will be
crucial to test different theoretical explanations.

So far, Ca-rich transients have only been observed in the optical band from external galaxies
while the remnants or Galactic cases of this type have not been found. However, the transient census has shown that the Ca-rich transient occurrence rate is $\sim 15\%$ of the SN Ia rate \citep{2020ApJ...905...58D}, which converts to $\sim 3$ Galactic Ca-rich transients occurring in the past 4 kyr. It is highly possible to find a Ca-rich SNR among the 300--400 known SNRs in the Galaxy, which would be the nearest lab of such an explosion.

The X-ray radiation of supernova remnants (SNRs) can provide information about the properties of the SN explosion and the surrounding environment. The metal abundances of ejecta revealed by X-ray spectroscopy can be compared with those predicted by various SN nucleosynthesis models 
\citep[see][and references therein]{2017hsn..book.2063V, Mori2018}. Similar approaches have been used in the data from the Milky Way Galaxy \citep[e.g.,][]{Kobayashi2020} and Perseus Cluster \citep{Simionescu2019} for distinguishing explosion mechanisms and progenitors.

G306.3$-$0.9 is a Galactic SNR, showing asymmetric morphology in the X-ray, infrared and radio bands (\citealt{2013ApJ...766..112R}, \citealt{2016A&A...592A.125C}, hereafter C16). C16 also suggested that it is a Type Ia SNR based on an analysis of element abundances and the centroid of the Fe-K line. Later, \citet{2017MNRAS.466.3434S} (hereafter S17) and \citet{2019PASJ...71...61S} (hereafter S19) supported its Type Ia origin based on {\it Suzaku} observation. But there exist differences in conclusions among the works above. C16 showed two-component gas with temperatures of $\sim$ 0.2 keV + 1--2 keV, while S17 suggested different temperatures of $\sim$ 0.6 keV + 3 keV. Both studies estimated the distance of the SNR at $\sim$ 8 kpc, but S19 suggests that it is located at $\sim$ 20 kpc and has a stratified ejecta structure, where the hot Fe-group elements with short ionization timescales are the third component separated from cool intermediate-mass elements (IMEs) with longer ionization timescales.

The proposed Type Ia origin of G306.3$-$0.9, together with its asymmetric morphology, appears at odds with the idea that Type Ia SNRs tend to be symmetric. 
Statistically, Type Ia SNRs are more symmetric than CC SNRs in X-ray morphology \citep{2011ApJ...732..114L}, either due to a more symmetric explosion process or a more uniform environment.
Nevertheless, a few SNRs with asymmetric morphologies have also been proposed to have a Type Ia origin, e.g. SNR 3C397 \citep{1999ApJ...520..737C, 2015ApJ...801L..31Y, 2018ApJ...861..143L,2020arXiv200608681M,2021ApJ...913L..34O} and SNR W49B \citep{2018A&A...615A.150Z, 2020ApJ...904..175S}. It would be of interest to explore if the aspherical morphology resulted from an intrinsically asymmetric Type Ia explosion or a non-uniform ambient medium. 

SNR G306.3$-$0.9 also shows strong Ca lines and unusually large Ca abundance in its X-ray spectra, which has not been well discussed in previous work. The Ca-rich ejecta property makes it an intriguing target to search for a connection to the newly found SN group of Ca-rich transients.

To find out the progenitor, we revisited the XMM-{\it Newton} and {\it Chandra} observations of SNR G306.3$-$0.9. To unveil the origin of the asymmetric morphology, we performed a spatially resolved spectroscopy using an adaptive spatial binning method. The data and binning methods used in this study are described in \S \space \ref{sec:data}. We present in \S \space \ref{sec:results} the details of spectral analysis and results. In \S \space \ref{sec:discussion}, we discussed the possibility of the SNR being the first Ca-rich transient remnant and Galactic case found, after comparing the fitting abundances with models. The paper is closed in \S \space \ref{sec:conclusions} with concluding remarks.

\section{Data and Methods} \label{sec:data}

\subsection{X-ray Data} \label{subsec:data}

XMM-{\it Newton} performed an observation towards G306.3$-$0.9 on March 2, 2013 (obs. ID: 0691550101, PI: Miller, J.), the total exposure time of the MOS1, MOS2 and pn data is 56.8 ks, 56.8 ks and 54.9 ks, respectively, after removing the periods with proton flares. The Science Analysis System (SAS) software (vers. 16.1.0)\footnote{\url{https://www.cosmos.esa.int/web/xmm-newton/sas}}, was used to reproduce data and extract spectra
. We also retrieved and reprocessed two archival {\it Chandra} ACIS-S observation data (obs. ID: 13419, exposure time: 5.04 ks, PI: Miller, J.; obs. ID: 14812, exposure time: 47.7ks, PI: Miller, J.) using Chandra Interactive Analysis of Observations (CIAO) software (vers. 4.12)\footnote{\url{https://cxc.harvard.edu/ciao/}}. Xspec (vers. 12.10.1f)\footnote{\url{https://heasarc.gsfc.nasa.gov/xanadu/xspec}} was used for spectral analysis based on atomic data from ATOMDB 3.0.9\footnote{\url{http://www.atomdb.org}} and ionization balance calculation from ATOMDB 3.0.7.

\subsection{Spectra Extraction and Spatial Binning Method} \label{subsec:binning}

First, the global X-ray spectra for the SNR were extracted. The selected source and background regions are shown in Appendix \ref{sec:region}. 
The XMM-{\it Newton} image does not reveal any clear point-like sources, but three soft sources were shown in the {\it Chandra} X-ray image (C16). We removed these point-like sources in our spectral analysis using three circular regions with a radius of $8''$. 
The global spectra extracted 
were binned with the optimal binning scheme based on the response matrices \citep{2016A&A...587A.151K}.
We combined the spectra of two {\it Chandra} observations using {\tt\string addascaspec} command in the ASCA FTOOL\footnote{\url{https://heasarc.gsfc.nasa.gov/docs/software/ftools}} package, considering the inadequate photons from the shorter observations.

Before we perform spatially resolved spectroscopy, we used the weighted Voronoi tessellations (WVT) method \citep{2006MNRAS.368..497D} to separate the SNR into multiple regions. WVT is an adaptive spatial binning method, which allows us to divide X-ray image into regions with the required S/N. By using S/N $\sim$ 70 ($\sim$ 4900 photons), we obtained 13 regions from the 0.3 -- 8.0 keV pn image, which has the most photons among the five sets of data. Besides, WVT tends to produce circular regions so we adjusted the rim regions manually so that they could coincide with the SNR. The final region division is shown in Figure \ref{fig:region}.

We extracted spectra for all 13 regions, with a similar method as used for the global spectra. Because the photon counts in small regions are much lower than that in the global spectra, MOS1 and MOS2 spectra were also combined using {\tt\string addascaspec} command.

\subsection{APEX \twCO\ (2-1) observation} \label{subsec:apex}

We performed a molecular observation toward G306.3$-0.9$ using the Atacama Pathfinder Experiment (APEX) 12-m radio telescope. 
We observed \twCO~and \thCO~(2-1) simultaneously in a $9'\times 9'$ region using the nFLASH receiver. At the frequency of 230~GHz, the telescope provides a half-power beam width of $29''$, a main-beam efficiency of 68\%,  and a velocity resolution of 0.079~\kms. After resampling the data to a velocity resolution of 0.5~\kms, we obtained an average RMS of 0.16~K/0.14~K for the \twCO/\thCO\ data cube.

The observation aimed to explore whether the SNR is associated with molecular gas. 
We have not found clear evidence to support the association. Therefore, this paper focuses on the X-ray analysis, and 
only briefly discusses the the molecular gas along the line-of-sight in Appendix~\ref{sec:molecular}.

\section{Results} \label{sec:results}

\subsection{Spectral Fitting} \label{subsec:fitting}

S19 proposed that there exist separated Fe-rich ejecta based on the study of Fe-K lines. Therefore we applied a similar triple thermal component model {\it apec}+{\it vnei}+{\it vnei} to the global spectra, plus the {\it tbabs} absorption model \citep{2000ApJ...542..914W}. Here the {\it apec} model is a collisional ionization equilibrium model to describe the shock-heated interstellar medium (ISM) with the solar abundance. Two {\it vnei} models are used to characterize the shocked ejecta and the hot Fe-rich ejecta in non-equilibrium ionization. The abundances of Mg, Si, S, Ar, Ca, and Fe in $vnei_{\rm h}$ (subscript ``h'' indicates the hot gas) were set as free parameters, while we fixed all abundances of elements but that of Fe to solar value  in $vnei_{\rm Fe}$ (subscript ``Fe'' indicates the Fe-rich gas). The abundance of Ni was tied to that of Fe in both {\it vnei} components. 
In this paper, solar abundances from \citet{2009ARA&A..47..481A} were used. A cross-normalization constant was added to all models to account for the instrument-dependent flux discrepancy. It turns out that the fitted flux of {\it Chandra} observation is $\sim$ 10\% higher than that of XMM-{\it Newton}, which is consistent with previous cross-calibration study (see \citealt{2015A&A...575A..30S,2017AJ....153....2M,2017A&A...597A..35P} for details). 

\begin{figure}[t!]
\includegraphics[height=\columnwidth, angle=270]{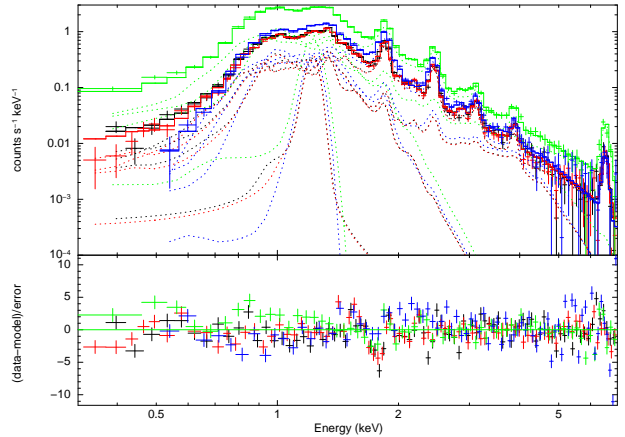}
\caption{G306.3$-$0.9 global spectra and fitting residuals based on the triple-temperature model (black: XMM-{\it Newton} MOS1; red: XMM-{\it Newton} MOS2; green: XMM-{\it Newton} pn; blue: {\it Chandra}). The dashed lines depict different components of the models and Gaussian lines added at 1.22 keV.
\label{fig:globalspectra}}
\end{figure}

The first best-fit results showed large residuals around 1.2 keV (with $\chi^{2}$/d.o.f. $\sim$ 4.40), which is likely caused by Fe L-shell line flux deficits in the atomic data of Xspec code. By adding a Gaussian line at $\sim$ 1.2 keV, we obtained a line centroid $\mu$ = 1.22 keV with a line width $\sigma$ = 21 eV, which are consistent with the fitting from S19. These deficits were also found in a few earlier studies of other sources, e.g. SNR W28 \citep{2012PASJ...64...81S}, SNR Kes 79 \citep{2016ApJ...831..192Z} and Capella binary \citep{2000ApJ...530..387B}. Hence, in all subsequent spectral fits, a Gaussian line at $\mu$ = 1.22 keV with $\sigma$ = 21 eV was added. 
Considering that there were not many hard X-ray photons received and there are not enough bins on the high energy end of the spectra, especially after region division, all subsequent fittings including that for global spectra were based on C-statistic \citep{1979ApJ...228..939C}. Global spectra are shown in Figure \ref{fig:globalspectra} and the best-fit results are shown in Table \ref{tab:globalspectra}. 

\begin{figure}[t!]
	\plotone{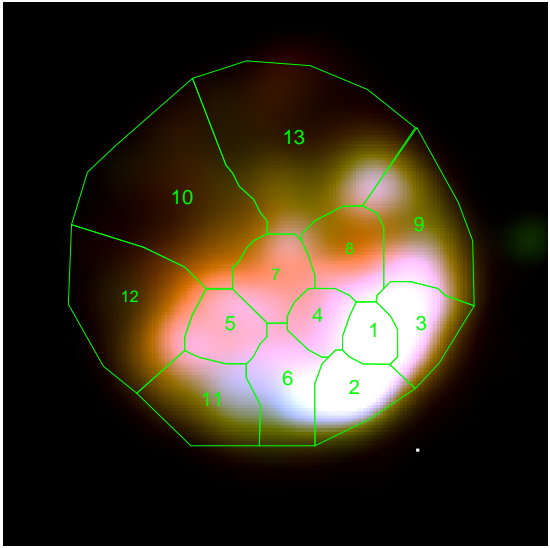}
	\caption{Composite tricolor pn image (red: 0.5--1.5 keV; green: 1.5--3.5 keV; blue: 3.5--7.0 keV) and the 13 regions divided based on the WVT method. The number represent the regions for spectral analysis. The best-fit parameters are shown in Table \ref{tab:apecvnei}.
		\label{fig:region}}
\end{figure}

\begin{deluxetable}{cc|cc}
	\tablenum{1}
	\tabletypesize{\footnotesize}
	\tablecaption{Global spectral fitting results with 90\% confidence (model: {\it apec} + {\it vnei} + {\it vnei}).\label{tab:globalspectra}}
	\tablewidth{0pt}
\startdata
\\[0.3pt]
$\chi_{\nu}^{2}$/d.o.f.
&      2.42 / 310
&        Ar
&        4.73$_{      -0.70}^{+       0.69}$
\\ [8pt]
C-stat
&        724.64
&        Ca
&        6.69$_{      -1.46}^{+       1.64}$
\\ [8pt]
\parbox{0.2\columnwidth}{\centering $N_{\rm H}$ \\ $(10^{22} {\rm cm}^{-2})$}
&		 1.86$_{      -0.04}^{+       0.04}$
&        Fe
&        0.77$_{      -0.08}^{+       0.30}$
\\ [8pt]
\parbox{0.2\columnwidth}{\centering $kT_{\rm c}$ \\ (keV)}
&        0.21$_{      -0.01}^{+       0.01}$
&        \parbox{0.2\columnwidth}{\centering$ \tau_{\rm h}$ \\ $(10^{11} {\rm cm}^{-3}{\rm s})$}
&        6.16$_{      -2.13}^{+       6.77}$
\\ [8pt]
\parbox{0.2\columnwidth}{\centering ${\rm norm}_{\rm c}$ \\ $(10^{-1} {\rm cm}^{-5})$}
&	1.64$_{  -0.41}^{+   0.51}$
&	\parbox{0.2\columnwidth}{\centering ${\rm norm}_{\rm h}$ \\ $(10^{-2} {\rm cm}^{-5})$}
&	1.71$_{      -0.20}^{+       0.25}$
\\ [8pt]
\parbox{0.2\columnwidth}{\centering $kT_{\rm h}$ \\ (keV)}
&	0.72$_{      -0.06}^{+       0.05}$
&	\parbox{0.2\columnwidth}{\centering $kT_{\rm Fe}$ \\ (keV)}
&	3.06$_{      -0.24}^{+       0.47}$
\\ [8pt]
Mg
&	0.73$_{      -0.08}^{+       0.08}$
&	Fe
&	11.42$_{  -1.88}^{+   2.51}$
\\ [8pt]
Si
&	1.14$_{      -0.08}^{+       0.09}$
&	\parbox{0.2\columnwidth}{\centering$ \tau_{\rm Fe}$ \\ $(10^{10} {\rm cm}^{-3}{\rm s})$}
&	3.29$_{  -0.24}^{+   0.24}$
\\ [8pt]
S
&	2.91$_{      -0.24}^{+       0.24}$
&	\parbox{0.2\columnwidth}{\centering ${\rm norm}_{\rm Fe}$ \\ $(10^{-4} {\rm cm}^{-5})$}
&	8.25$_{  -2.20}^{+   2.13}$
\\
\enddata
\tablecomments{The abundances of element X in this table and all subsequent spectral fits are defined as $(n({\rm X})/n({\rm H}))/(n({\rm X})/n({\rm H}))_\sun$, where $n$ is the atom density. Subscripts ``c'', ``h'' and ``Fe'' indicate the cool, hot and Fe-rich gas, respectively}
\end{deluxetable}


We also tried to apply the double thermal component model {\it apec}+{\it vnei} similar to C16 and S17, but the triple component model always returned the better fit and passed the F-test (with probability $< 10^{-13}$).

To reveal the spatial variation of the parameters across the SNR, we analyzed the spectra in 13 regions (see Figure \ref{fig:region}). 
We first tried to fit regional spectra with 3-component models as we did for the global fits. 
However, because of the small number of hard X-ray photons,  we could not constrain the ejecta components, especially the hottest Fe-rich gas that dominates the photons above 5 keV (see Figure \ref{fig:globalspectra} and S19). 
Therefore, we ignored the Fe-rich gas and set the photon energy upper limit at 5.0 keV before fitting the spectra in individual small regions. 
We used the {\it vnei} model to describe the ejecta and added the {\it apec} to account for the relatively cool shocked ISM, because the remnant is 
interacting with the ISM (C16). 
The Gaussian lines at 1.22 keV and the cross-calibration constants were also added and fitted. F-test also showed that it is reasonable to add a cool component to the single-component {\it vnei} model (probabilities $< 0.1\%$), 
except region 3 and 10 (probabilities $\sim 5\%$ and $\sim 1\%$, respectively). The best-fit results of the two-component model are shown in Table \ref{tab:apecvnei} and the spatial distributions of parameters are shown in Figure \ref{fig:apecvnei}. The spectra and residuals based on the {\it apec}+{\it vnei} model are shown in Appendix \ref{sec:regionalspectra}.

\begin{figure}[t!]
\includegraphics[width=\columnwidth]{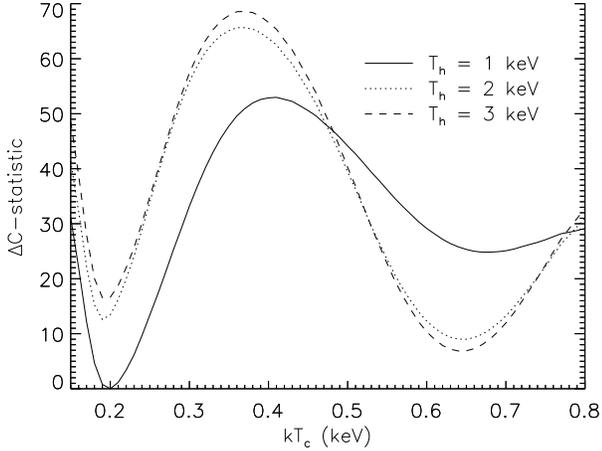}
\caption{C-statistic curves of region 4 spectral fitting. The x-axis corresponds to the temperature of the component {\it apec} and the y-axis corresponds to the C-statistic minus the minimum. The solid line, dotted line and dashed line depict various {\it vnei} temperatures.
\label{fig:cstat}}
\end{figure}

\begin{splitdeluxetable*}{cccccccBcccccccc}
\tabletypesize{\footnotesize}
\tablenum{2}
\tablecaption{Double-component model {\it apec}+{\it vnei} spectral fitting results with 90\% confidence\label{tab:apecvnei}}
\tablehead{\colhead{Region} & \colhead{C-stat} & \colhead{$N_{\rm H}$ $(10^{22} {\rm cm}^{-2})$} & \colhead{$kT_{\rm c}$ (keV)} & \colhead{${\rm norm_c}$ $(10^{-14} {\rm cm}^{-5})$} & \colhead{$kT_{\rm h}$ (keV)} & \colhead{Mg}  & \colhead{Region} & \colhead{Si} & \colhead{S} & \colhead{Ar} & \colhead{Ca} & \colhead{Fe} &\colhead{$\tau$ $(10^{11} {\rm cm}^{-3}{\rm s})$} & \colhead{${\rm norm_h}$ $(10^{-14} {\rm cm}^{-5})$}
}
\startdata
reg1
&      149.89
&        1.78$_{      -0.15}^{+       0.14}$
&        0.25$_{      -0.04}^{+       0.10}$
&    4.22e-03$_{  -3.67e-03}^{+   8.30e-03}$
&        0.98$_{      -0.06}^{+       0.09}$
&        1.78$_{      -0.37}^{+       0.57}$
&reg1
 &        1.51$_{      -0.37}^{+       0.62}$
 &        3.35$_{      -0.75}^{+       1.17}$
 &        5.84$_{      -1.94}^{+       2.82}$
 &        6.88$_{      -2.84}^{+       3.84}$
 &        3.13$_{      -0.93}^{+       1.43}$
 &        2.38$_{      -0.65}^{+       1.09}$
 &    5.20e-04$_{  -1.19e-04}^{+   1.42e-04}$
\\
reg2
 &      148.70
 &        2.01$_{      -0.12}^{+       0.10}$
 &        0.21$_{      -0.02}^{+       0.02}$
 &    2.82e-02$_{  -1.36e-02}^{+   2.23e-02}$
 &        1.06$_{      -0.10}^{+       0.09}$
 &        0.61$_{      -0.16}^{+       0.15}$
&reg2
 &        0.97$_{      -0.15}^{+       0.19}$
 &        1.52$_{      -0.23}^{+       0.28}$
 &        1.79$_{      -0.71}^{+       0.81}$
 &        3.28$_{      -1.42}^{+       1.75}$
 &        1.15$_{      -0.25}^{+       0.28}$
 &        1.38$_{      -0.32}^{+       0.46}$
 &    1.54e-03$_{  -2.68e-04}^{+   3.60e-04}$
\\
reg3
 &      216.46
 &        1.71$_{      -0.22}^{+       0.15}$
 &        0.24$_{      -0.06}^{+       0.07}$
 &    5.35e-03$_{  -4.41e-03}^{+   9.56e-03}$
 &        0.95$_{      -0.03}^{+       0.08}$
 &        1.04$_{      -0.16}^{+       0.18}$
&reg3
 &        0.89$_{      -0.19}^{+       0.23}$
 &        2.12$_{      -0.39}^{+       0.44}$
 &        3.20$_{      -1.07}^{+       1.18}$
 &        5.65$_{      -2.11}^{+       2.65}$
 &        1.44$_{      -0.64}^{+       0.50}$
 &        1.87$_{      -0.44}^{+       0.32}$
 &    1.31e-03$_{  -2.48e-04}^{+   2.44e-04}$
\\
reg4
 &      161.71
 &        1.82$_{      -0.09}^{+       0.08}$
 &        0.20$_{      -0.02}^{+       0.02}$
 &    1.46e-02$_{  -6.84e-03}^{+   1.29e-02}$
 &        0.91$_{      -0.05}^{+       0.07}$
 &        1.24$_{      -0.41}^{+       0.57}$
&reg4
 &        1.56$_{      -0.41}^{+       0.51}$
 &        3.08$_{      -0.79}^{+       0.92}$
 &        4.37$_{      -1.97}^{+       2.36}$
 &       10.22$_{      -3.98}^{+       4.80}$
 &        3.03$_{      -0.76}^{+       0.83}$
 &        4.05$_{      -1.48}^{+       3.06}$
 &    4.91e-04$_{  -9.80e-05}^{+   1.31e-04}$
\\
reg5
 &      157.93
 &        1.90$_{      -0.10}^{+       0.08}$
 &        0.20$_{      -0.02}^{+       0.02}$
 &    2.14e-02$_{  -1.06e-02}^{+   1.72e-02}$
 &        0.93$_{      -0.06}^{+       0.10}$
 &        0.93$_{      -0.15}^{+       0.35}$
&reg5
 &        1.36$_{      -0.28}^{+       0.38}$
 &        2.93$_{      -0.58}^{+       0.72}$
 &        4.35$_{      -1.54}^{+       2.14}$
 &        5.63$_{      -2.65}^{+       3.80}$
 &        2.70$_{      -0.54}^{+       0.78}$
 &        2.09$_{      -0.75}^{+       1.10}$
 &    6.97e-04$_{  -1.45e-04}^{+   1.27e-04}$
\\
reg6
 &      175.43
 &        1.97$_{      -0.11}^{+       0.09}$
 &        0.20$_{      -0.02}^{+       0.02}$
 &    2.81e-02$_{  -1.33e-02}^{+   2.05e-02}$
 &        1.24$_{      -0.14}^{+       0.19}$
 &        0.58$_{      -0.17}^{+       0.18}$
&reg6
 &        1.12$_{      -0.20}^{+       0.26}$
 &        1.85$_{      -0.29}^{+       0.35}$
 &        1.60$_{      -0.78}^{+       0.91}$
 &        1.42$_{      -1.40}^{+       1.77}$
 &        1.05$_{      -0.24}^{+       0.30}$
 &        0.95$_{      -0.25}^{+       0.37}$
 &    9.34e-04$_{  -2.23e-04}^{+   2.55e-04}$
\\
reg7
 &      198.77
 &        1.95$_{      -0.10}^{+       0.09}$
 &        0.24$_{      -0.02}^{+       0.04}$
 &    1.11e-02$_{  -6.10e-03}^{+   8.03e-03}$
 &        0.91$_{      -0.10}^{+       0.13}$
 &        0.71$_{      -0.32}^{+       0.45}$
&reg7
 &        1.22$_{      -0.29}^{+       0.39}$
 &        3.22$_{      -0.70}^{+       0.85}$
 &        3.87$_{      -1.66}^{+       2.35}$
 &        8.23$_{      -3.52}^{+       5.32}$
 &        3.27$_{      -0.76}^{+       1.19}$
 &        2.07$_{      -0.81}^{+       1.45}$
 &    5.89e-04$_{  -1.34e-04}^{+   2.27e-04}$
\\
reg8
 &      186.61
 &        1.87$_{      -0.11}^{+       0.08}$
 &        0.24$_{      -0.03}^{+       0.05}$
 &    8.55e-03$_{  -4.80e-03}^{+   6.27e-03}$
 &        0.90$_{      -0.06}^{+       0.06}$
 &        1.28$_{      -0.34}^{+       0.54}$
&reg8
 &        1.48$_{      -0.39}^{+       0.70}$
 &        3.59$_{      -0.82}^{+       1.13}$
 &        5.44$_{      -1.99}^{+       2.58}$
 &        9.73$_{      -3.61}^{+       4.88}$
 &        3.19$_{      -0.78}^{+       1.03}$
 &        3.75$_{      -1.09}^{+       2.13}$
 &    5.82e-04$_{  -1.40e-04}^{+   1.80e-04}$
\\
reg9
 &      218.09
 &        1.95$_{      -0.11}^{+       0.09}$
 &        0.21$_{      -0.02}^{+       0.03}$
 &    2.04e-02$_{  -1.07e-02}^{+   1.74e-02}$
 &        1.07$_{      -0.07}^{+       0.09}$
 &        1.08$_{      -0.18}^{+       0.21}$
&reg9
 &        1.01$_{      -0.19}^{+       0.25}$
 &        2.05$_{      -0.33}^{+       0.42}$
 &        1.94$_{      -0.82}^{+       0.97}$
 &        2.30$_{      -1.22}^{+       1.52}$
 &        1.82$_{      -0.37}^{+       0.44}$
 &        1.90$_{      -0.38}^{+       0.59}$
 &    1.09e-03$_{  -2.00e-04}^{+   2.98e-04}$
\\
reg10
 &      200.62
 &        2.00$_{      -0.10}^{+       0.09}$
 &        0.21$_{      -0.04}^{+       0.03}$
 &    1.19e-02$_{  -7.40e-03}^{+   1.83e-02}$
 &        0.77$_{      -0.03}^{+       0.02}$
 &        1.20$_{      -0.38}^{+       0.42}$
&reg10
 &        1.90$_{      -0.39}^{+       0.58}$
 &        4.39$_{      -0.85}^{+       1.25}$
 &        8.07$_{      -2.30}^{+       3.12}$
 &       13.40$_{      -4.99}^{+       6.32}$
 &        3.63$_{      -0.92}^{+       1.21}$
 &        3.05$_{      -0.75}^{+       1.35}$
 &    8.90e-04$_{  -1.88e-04}^{+   2.32e-04}$
\\
reg11
 &      169.88
 &        1.96$_{      -0.14}^{+       0.10}$
 &        0.20$_{      -0.02}^{+       0.02}$
 &    2.66e-02$_{  -1.37e-02}^{+   1.88e-02}$
 &        1.15$_{      -0.11}^{+       0.16}$
 &        0.32$_{      -0.17}^{+       0.19}$
&reg11
 &        0.86$_{      -0.16}^{+       0.22}$
 &        1.25$_{      -0.21}^{+       0.27}$
 &        1.17$_{      -0.72}^{+       0.83}$
 &        1.48$_{      -1.46}^{+       1.88}$
 &        0.63$_{      -0.20}^{+       0.24}$
 &        1.00$_{      -0.31}^{+       0.44}$
 &    9.91e-04$_{  -2.31e-04}^{+   2.32e-04}$
\\
reg12
 &      186.84
 &        1.90$_{      -0.11}^{+       0.08}$
 &        0.19$_{      -0.03}^{+       0.02}$
 &    1.97e-02$_{  -1.06e-02}^{+   2.10e-02}$
 &        0.84$_{      -0.04}^{+       0.04}$
 &        1.43$_{      -0.28}^{+       0.36}$
&reg12
 &        1.37$_{      -0.25}^{+       0.31}$
 &        3.15$_{      -0.53}^{+       0.72}$
 &        5.41$_{      -1.63}^{+       1.95}$
 &        9.03$_{      -3.35}^{+       4.15}$
 &        2.72$_{      -0.58}^{+       0.77}$
 &        2.93$_{      -0.34}^{+       0.87}$
 &    9.73e-04$_{  -1.77e-04}^{+   1.93e-04}$
\\
reg13
 &      211.39
 &        1.99$_{      -0.09}^{+       0.08}$
 &        0.23$_{      -0.02}^{+       0.03}$
 &    2.07e-02$_{  -9.76e-03}^{+   1.50e-02}$
 &        0.93$_{      -0.07}^{+       0.12}$
 &        0.85$_{      -0.19}^{+       0.21}$
&reg13
 &        1.39$_{      -0.24}^{+       0.31}$
 &        3.00$_{      -0.46}^{+       0.57}$
 &        3.81$_{      -1.19}^{+       1.47}$
 &        4.57$_{      -2.00}^{+       2.49}$
 &        1.97$_{      -0.39}^{+       0.51}$
 &        2.23$_{      -0.62}^{+       0.82}$
 &    1.34e-03$_{  -3.12e-04}^{+   3.44e-04}$
\\
\enddata
\end{splitdeluxetable*}

\begin{figure*}[ht!]
\gridline{\fig{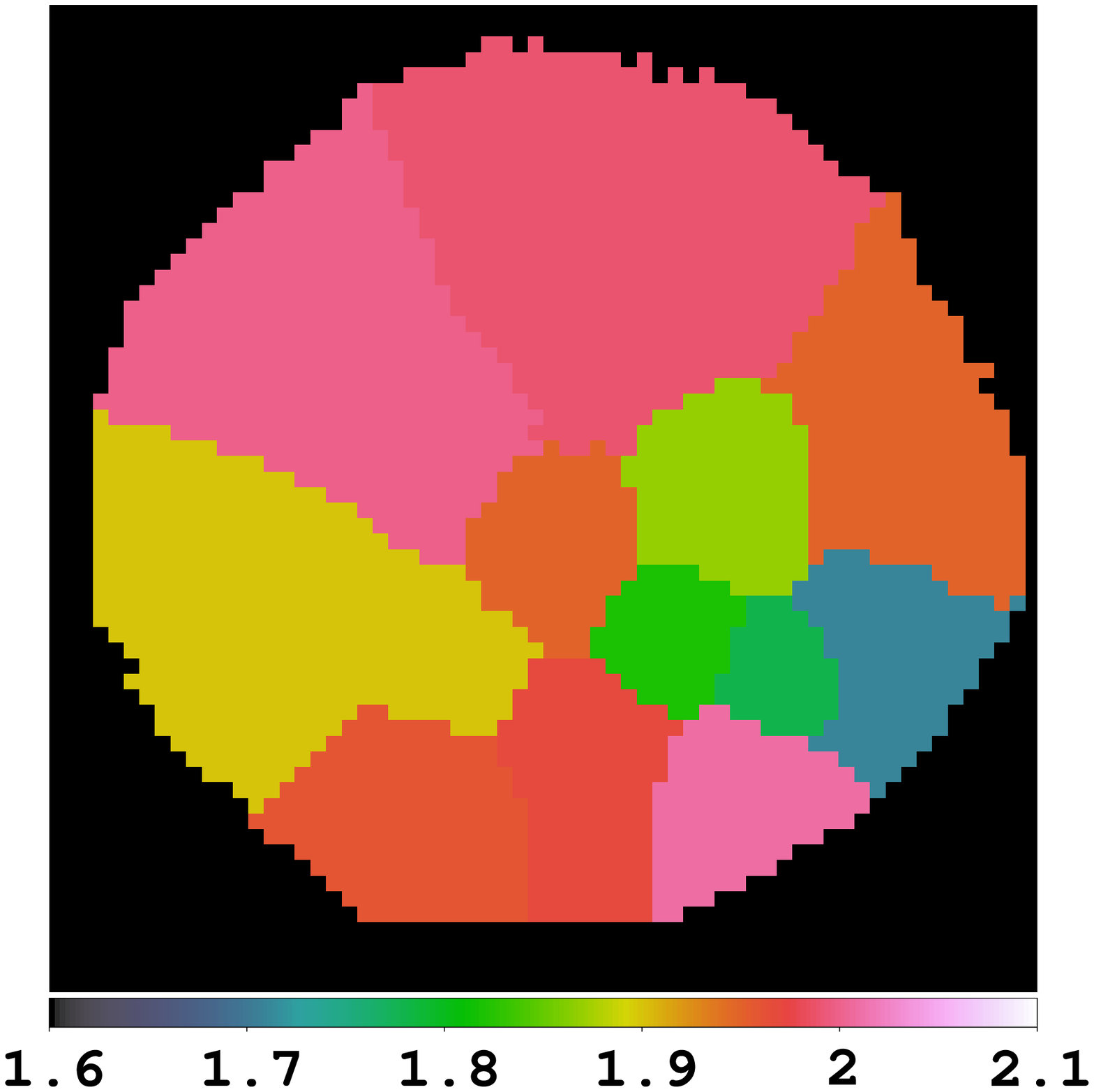}{0.2\textwidth}{$N_{\rm H}$ $ (10^{22} {\rm cm}^{-2})$}
          \fig{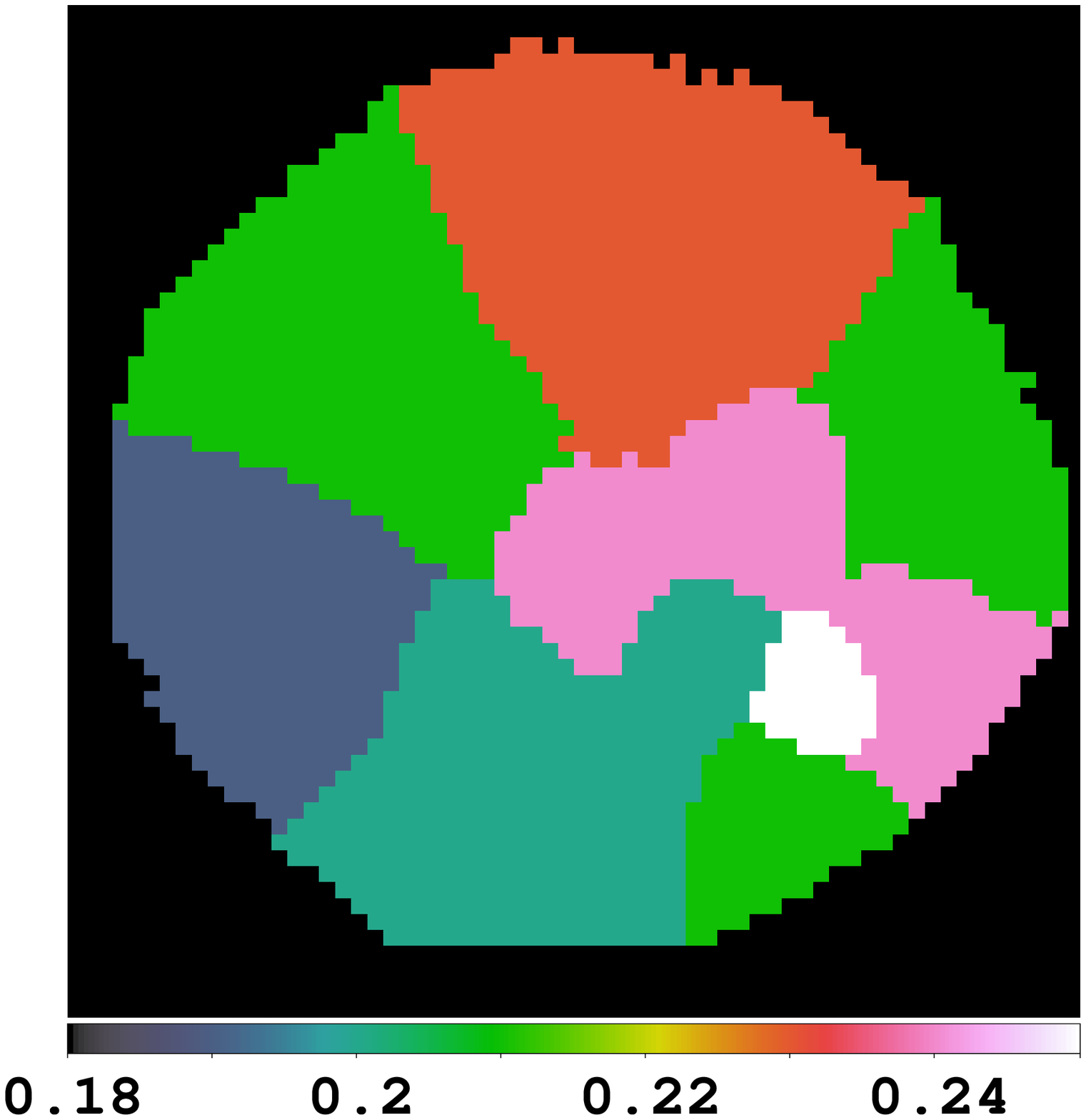}{0.2\textwidth}{$kT_{\rm c}$ (keV)}
          \fig{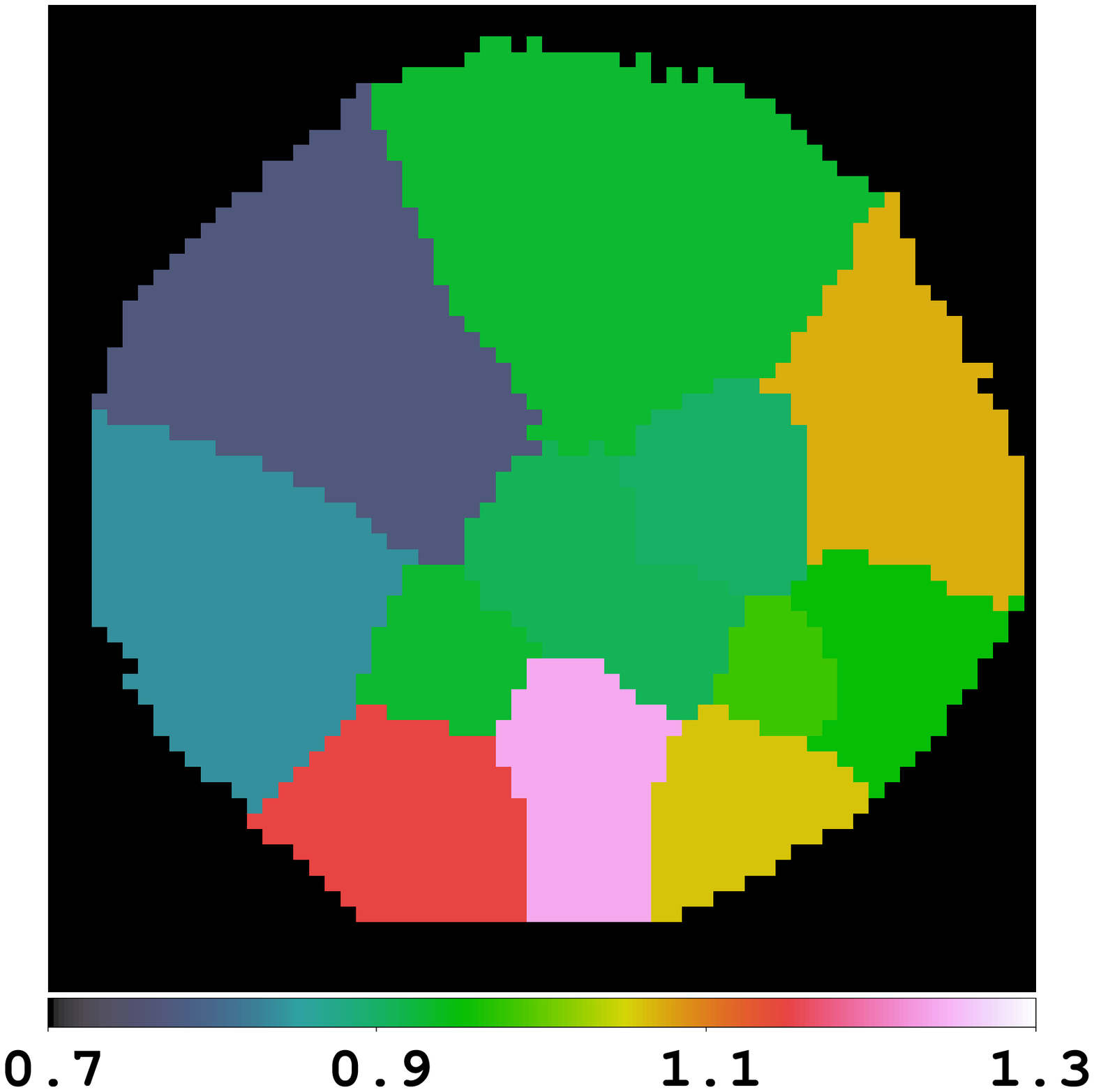}{0.2\textwidth}{$kT_{\rm h}$ (keV)}
          \fig{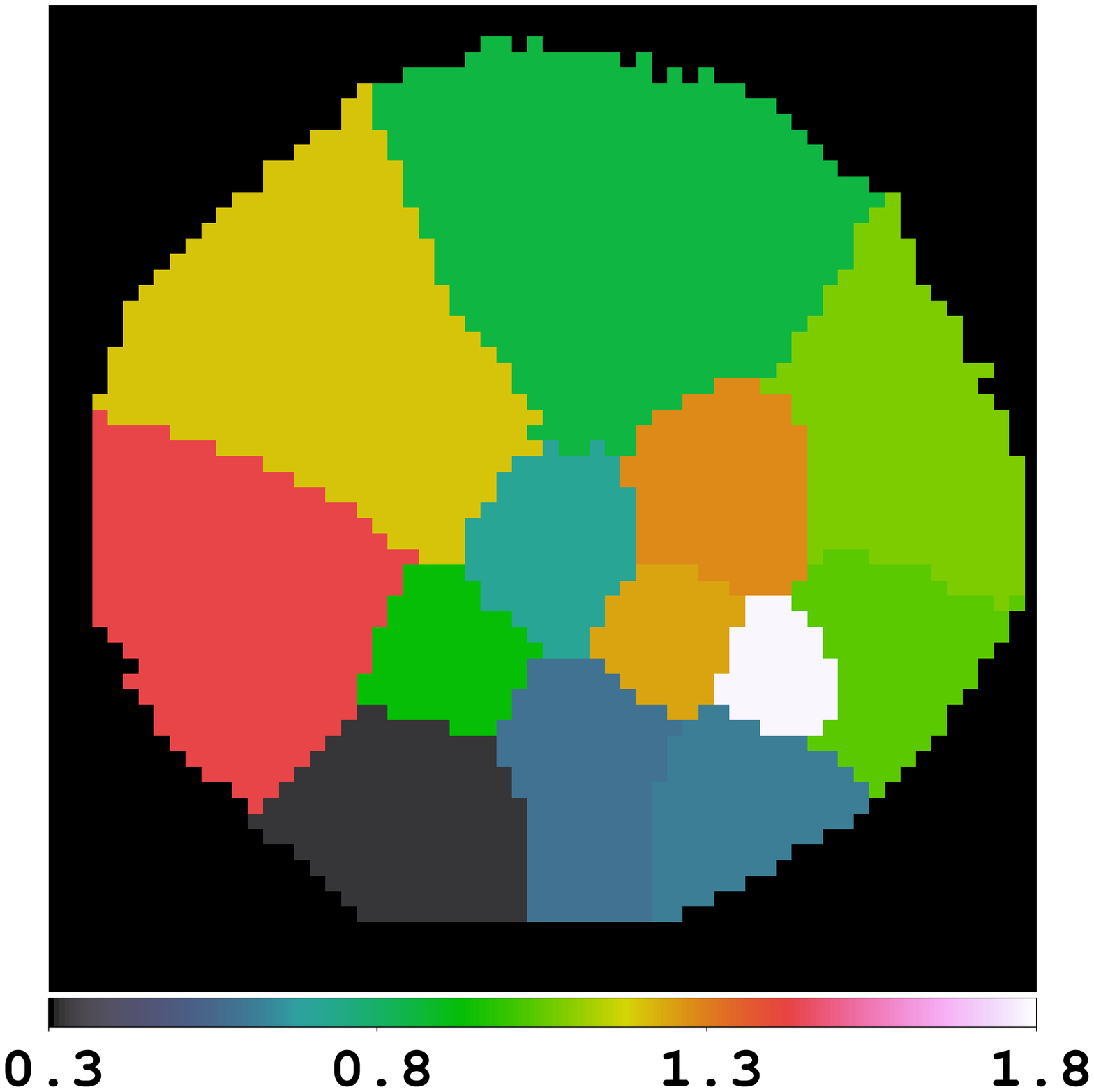}{0.2\textwidth}{Mg}
          \fig{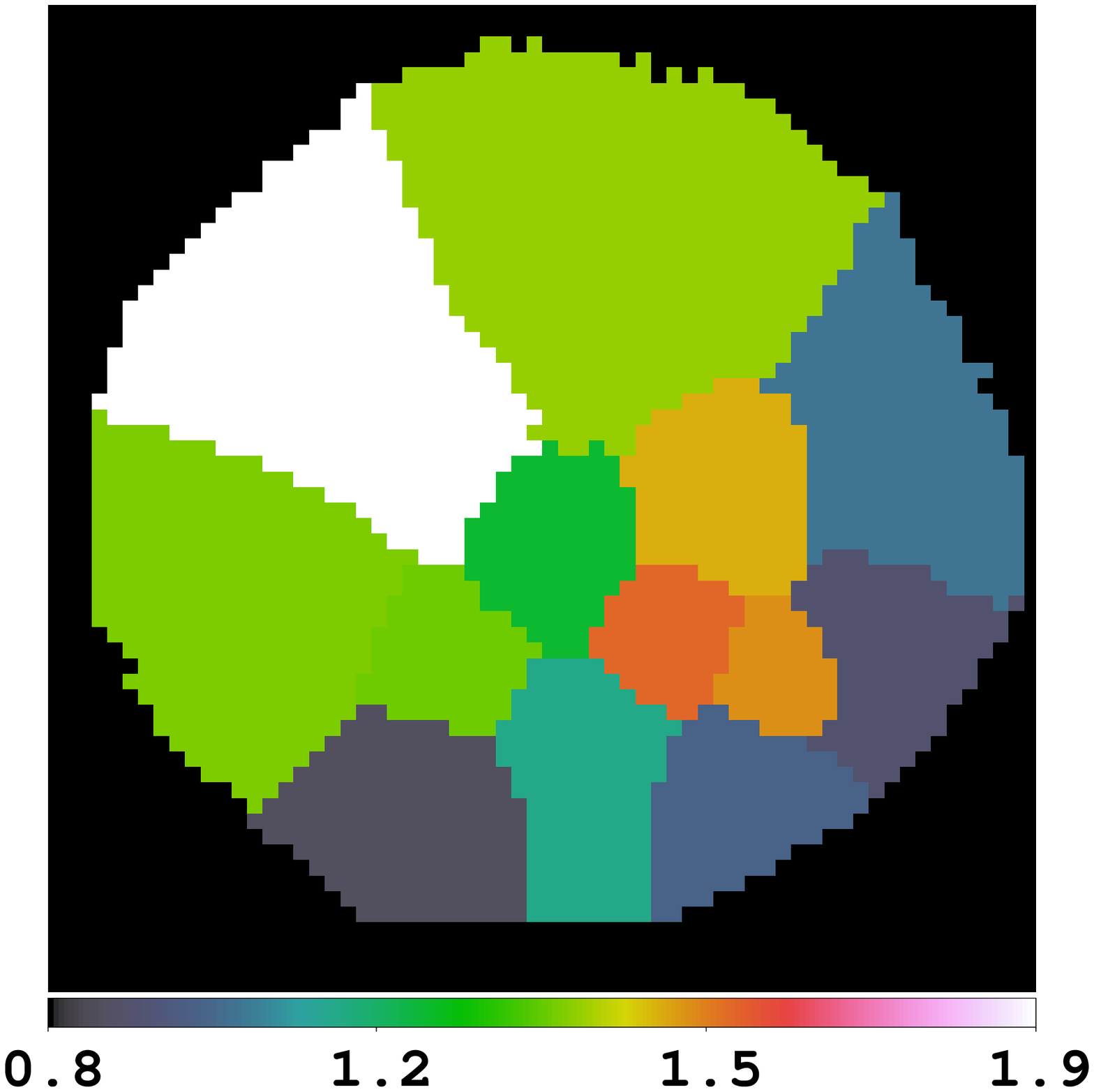}{0.2\textwidth}{Si}
          }
\gridline{\fig{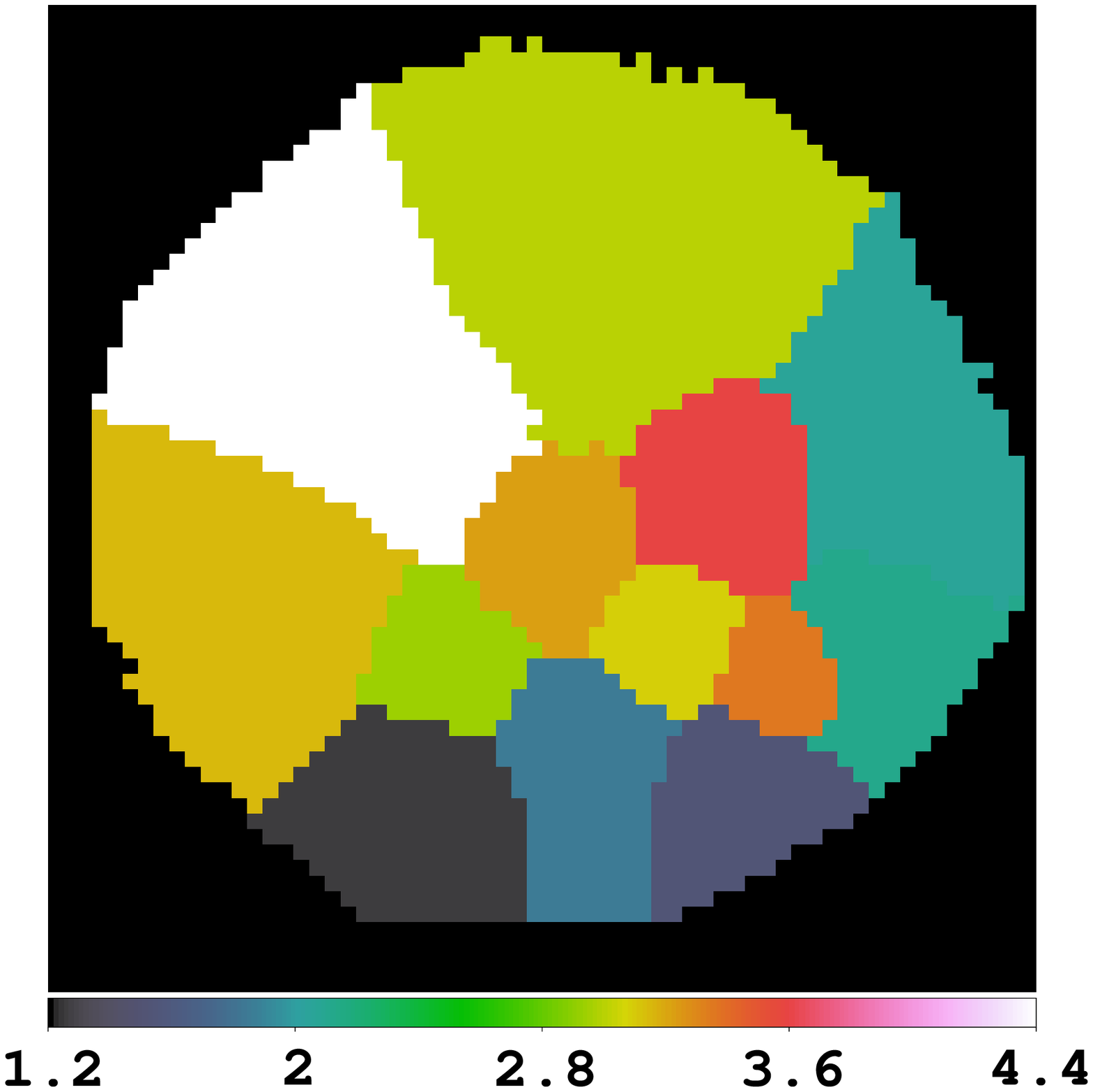}{0.2\textwidth}{S}
          \fig{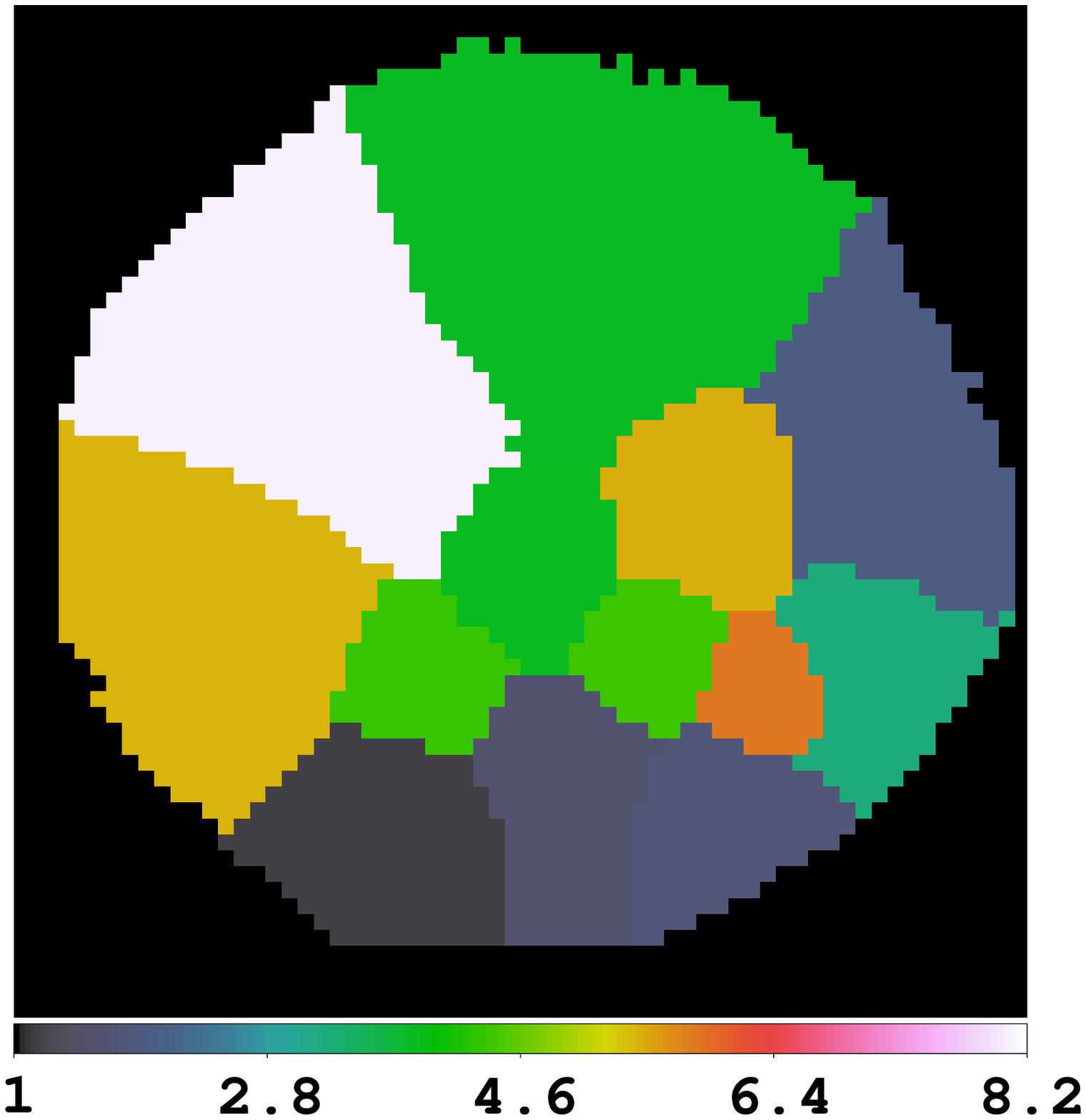}{0.2\textwidth}{Ar}
          \fig{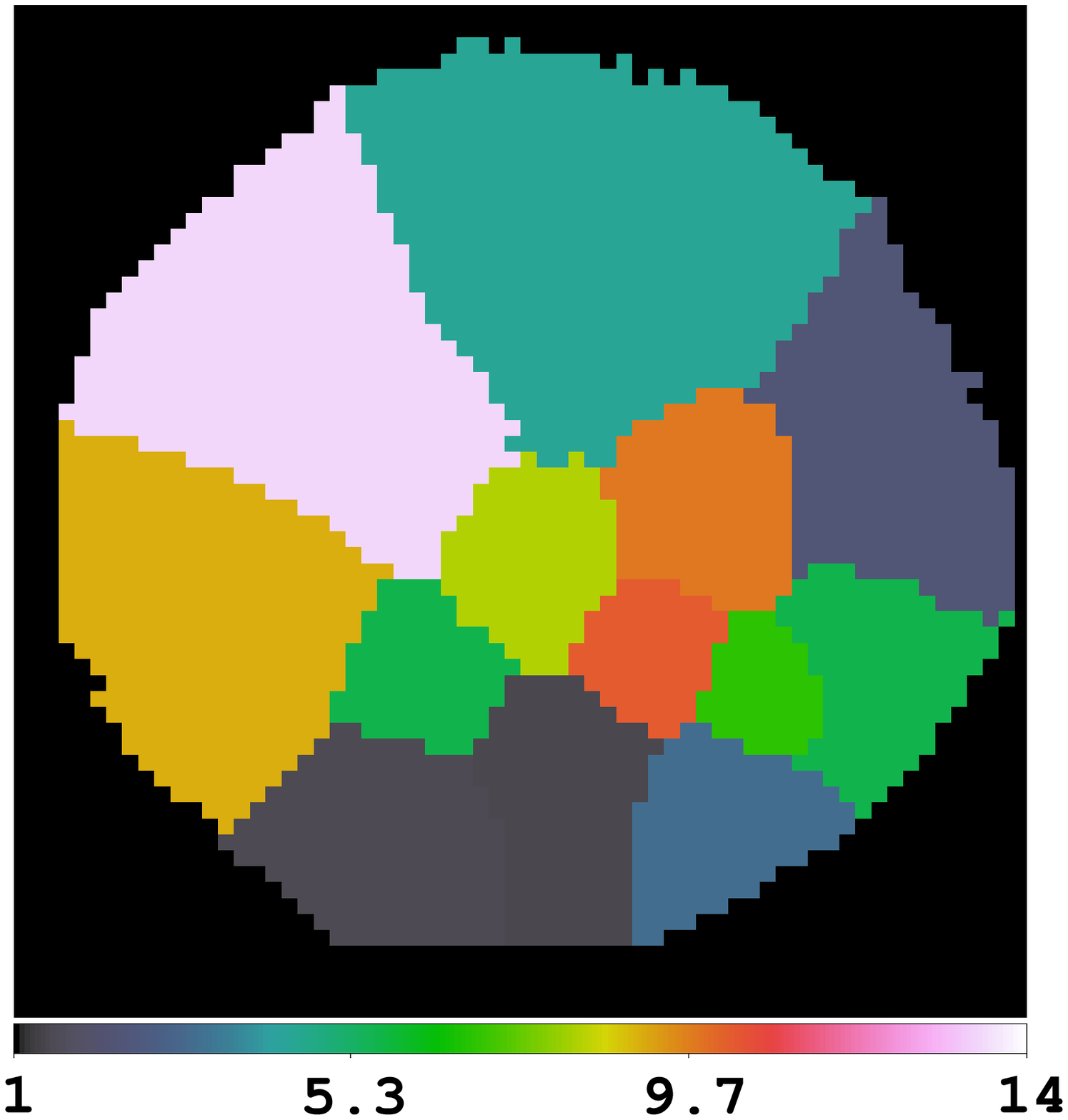}{0.2\textwidth}{Ca}
          \fig{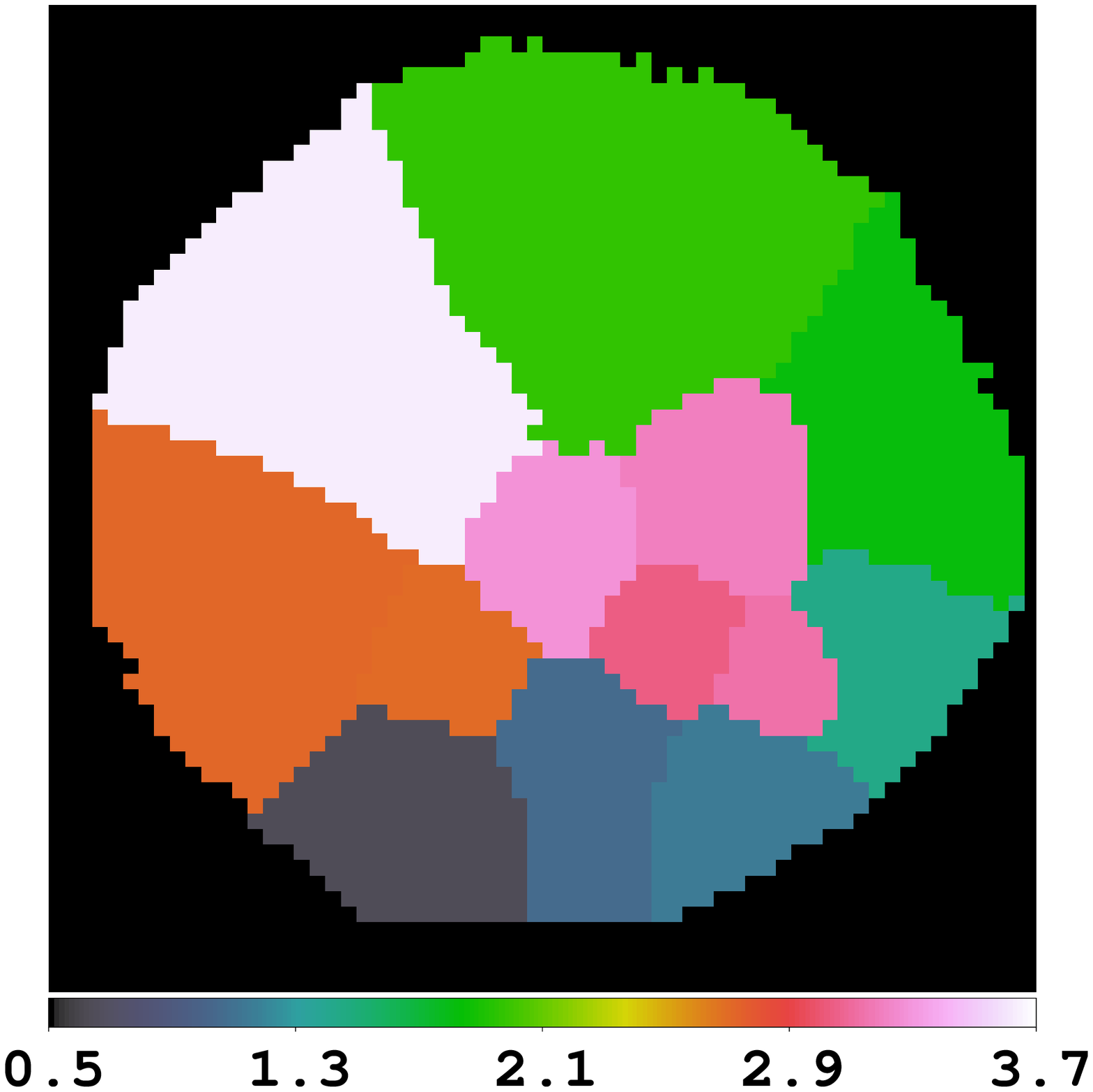}{0.2\textwidth}{Fe}
          \fig{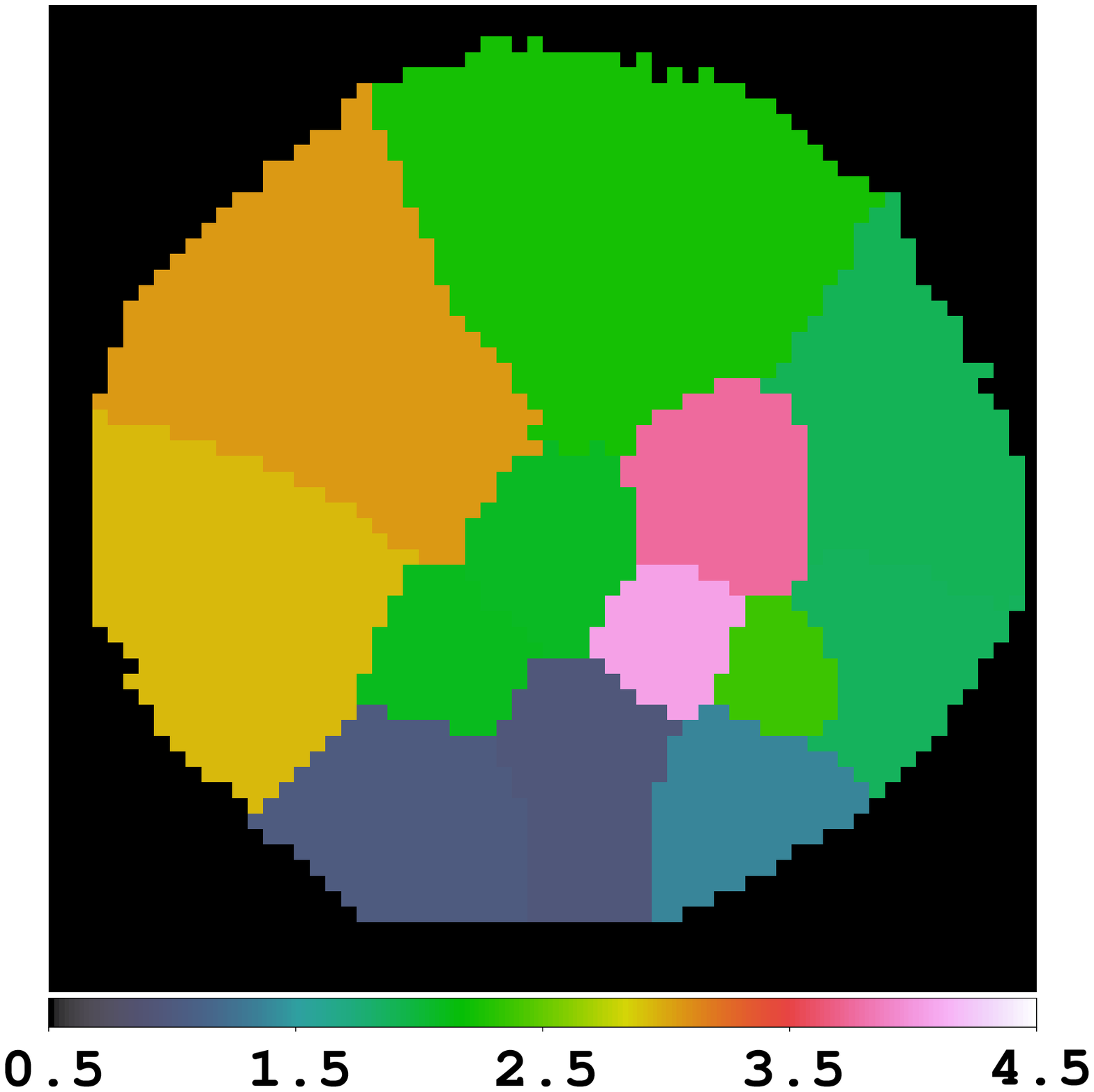}{0.2\textwidth}{$\tau$ $(10^{11} {\rm cm}^{-3}{\rm s})$}
          }
\gridline{\rightfig{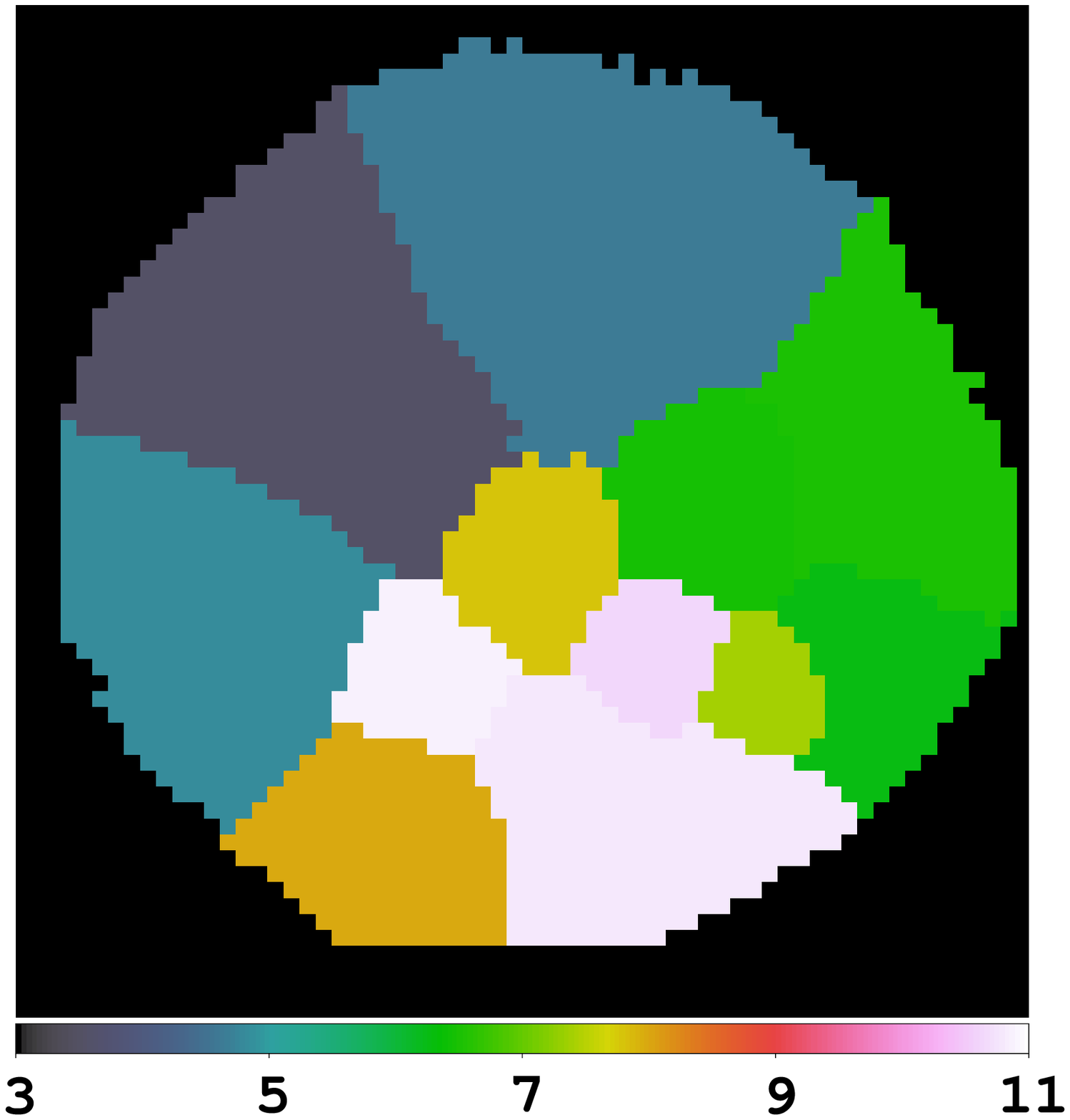}{0.2\textwidth}{${\rm H_c}$ denisty $({\rm cm}^{-3})$}
          \fig{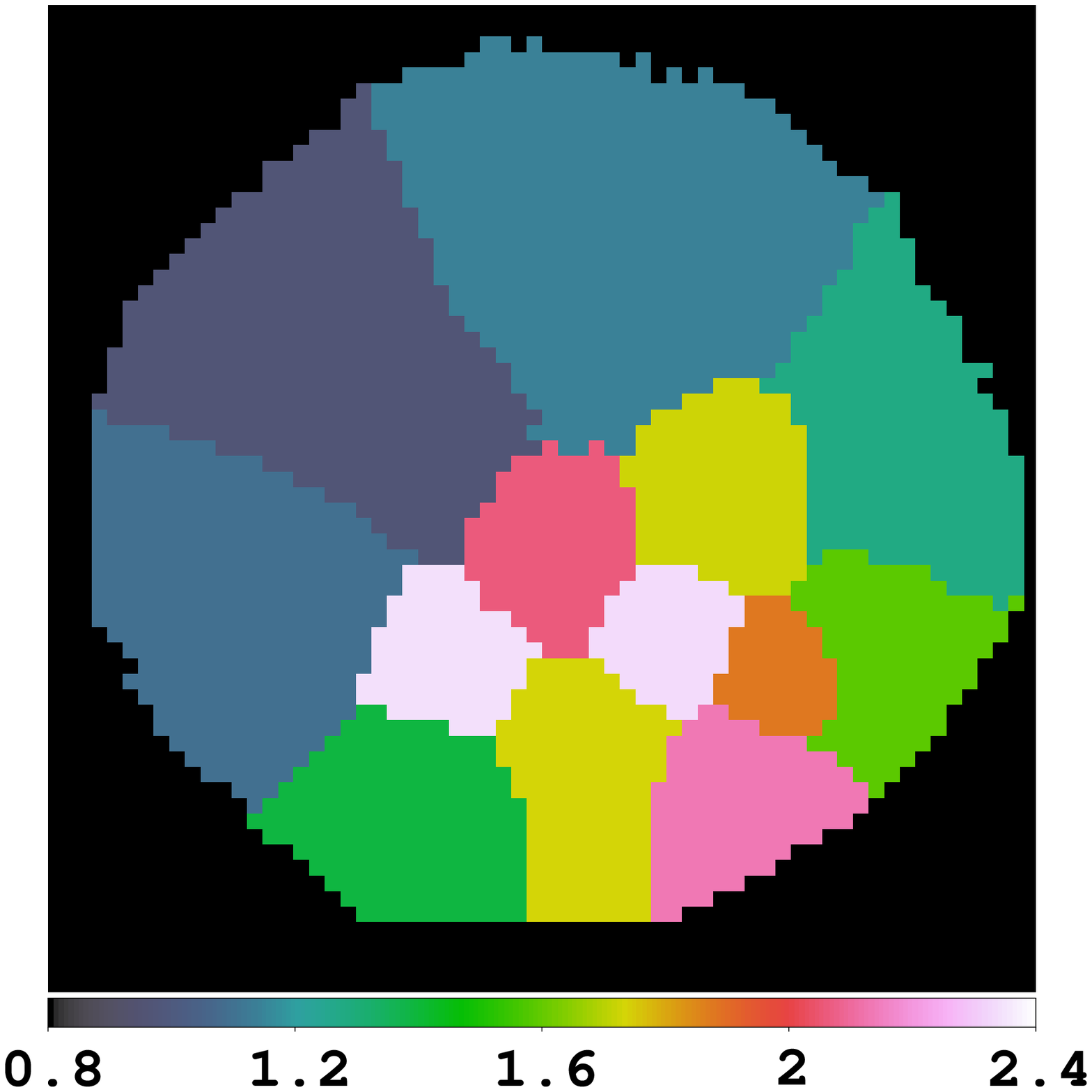}{0.2\textwidth}{${\rm H_h}$ denisty $({\rm cm}^{-3})$}
          \leftfig{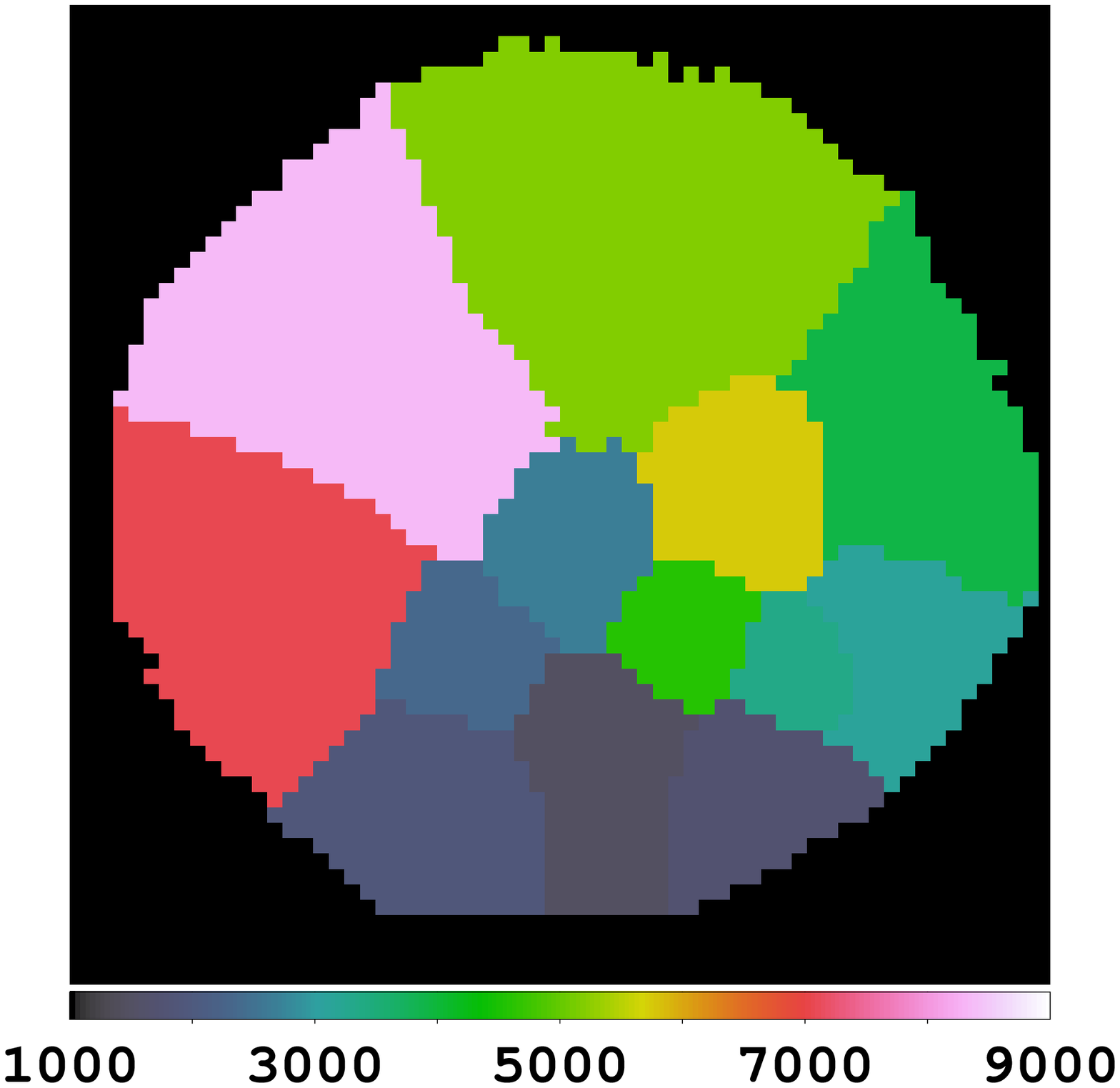}{0.2\textwidth}{Ionization Age (yr)}
          }
\caption{Spatial distributions of fitting parameters of the double-component model {\it apec}+{\it vnei}.
\label{fig:apecvnei}}
\end{figure*}

\subsection{Gas Temperature} \label{subsec:temperature}

The initial best-fit temperatures of the gas in regions 5, 7, 10 and 13 are $\sim$ 0.6 keV + $\sim$ 3 keV, which are 
consistent with the results of S17, obtained from {\it Suzaku} data, while other regions show best-fit temperature of $\sim$ 0.2 keV + $\sim$ 1 keV, which are consistent with the results of C16. This is likely due to degeneracy between the two groups of models in fitting the X-ray spectra, where we find two local minimum points of C-statistic at both 0.2 keV + 1 keV and 0.6 keV + 3 keV. Figure \ref{fig:cstat} shows the C-statistic curves of the region 4's spectra. The curves of other regional spectra have a similar trend but in some cases, the 0.6 keV + 3 keV is the minimum point. Because it is mainly the soft X-ray photons that suffer from absorption, there is a degeneracy between the cool component temperature $kT_{\rm c}$ and $N_{\rm H}$. Therefore, the fitted $N_{\rm H}$ also showed a bimodal distribution. In some regions, when the {\it apec} component was fitted at $\sim$ 0.6 keV, the fitted temperature of the {\it vnei} component would become unreasonably high.

S19 has also discussed about this issue, and proposed the existence of three components in the {\it Suzaku} spectra: $\sim$ 0.2 keV {\it apec}, $\sim$ 1 keV {\it vnei} and a 5 keV (fixed) hot Fe-rich component. 
Our joint fits of XMM-{\it Newton} and {\it Chandra} global spectra showed similar results.
However, note that after dividing the SNR into small regions, the hard X-ray photons in regional spectra 
become too few to constrain or support the $\sim$ 3 keV hot Fe component, which dominate the $> 5$ keV spectra. 
In a circular region of the SNR radius $\sim 2 \arcmin$, the net count rate in 6--7 keV (Fe-K lines dominate) of the XMM-{\it Newton} observation is $6.535 \times 10^{-3}$ cnt s$^{-1}$, while that of background is $1.838 \times 10^{-2}$ cnt s$^{-1}$. Thus the SNR can be divided into 5 regions at most for regional S/N ratio $>$ 4, which is too coarse for spatial-resolved analysis. 

Considering the facts above, it is reasonable that we ignored the $> 5$ keV channels and constrained all regions' temperature to be around 0.2 + 1 keV, which agree with the previous global fits.

\subsection{Comparisons with Spectral Fit Results of Previous Studies} \label{subsec:spectralcomparison}
The triple-component model fits of global spectra generally agree with those of S19, with some minor differences. 
S19 fixed the temperature of the Fe-rich component to 5~keV, while our spectral fit gives a temperature of $\sim 3$~keV. Moreover, S19 used the earlier solar abundance table by \citet{1989GeCoA..53..197A}.
These might cause discrepancies in the best-fit temperature of the less hot ejecta component, foreground absorption, and abundances.
In this work, the Mg abundance is lower by $38\%$, while the S, Ar and Ca abundances are higher by $21\%$, $33\%$ and $40\%$, respectively. 
The Fe abundances in the $vnei_{\rm h}$ and $vnei_{\rm Fe}$ are lower by $37\%$ and $57\%$ than S19, respectively\footnote{The comparisons above were made based on the solar abundance table of \citet{2009ARA&A..47..481A}.}.

Our regional spectral fit and C16 show significant spatial variations of properties across the SNR, demonstrating the importance of performing spatial-resolved analysis. 
As we divided the SNR into more regions, it is not easy to make a detailed comparison between the results of the two studies.
Our fits showed ISM component temperatures consistent with C16, but theirs gave much more drastic variations in the hot-component temperatures and the abundances across the SNR.
Moreover, we obtained much lower metal abundances and also find the highest abundances are in the northeast parts of the SNR instead of the centrally peaked distribution in C16 (see Figure \ref{fig:apecvnei}).  

\subsection{Gas Density, Shock Age and Explosion Energy} \label{subsec:gascalculation}

The fitting parameter {\it norm} ($10^{-14}$/(4$\pi d^{2}$)$\int n_{e}n_{\rm H}{\rm d}V$) was used for calculations of the mean gas density of each region, where $d$ is the distance to the SNR, $n_{e}$ and $n_{\rm H}$ are electron density and H density in each region volume $V$, respectively. The distance $d$ was adopted as 20 kpc according to S19. $n_{e}$ = 1.2$n_{\rm H}$ (for fully ionized plasma with solar abundances) was adopted. The X-ray emission was assumed to come from shell-like volumes. For a uniform density case and a shock compression ratio of 4, mass conservation suggests that the thickness of the shell $\Delta R = 1/12 R$ ($R$ is the radius of the SNR, which has an angular radius of $2\farcm{2}$). Every region was treated approximately as a prism with a depth of $2(\sqrt{R^{2}-r^{2}}-\sqrt{(11/12 R)^{2}-r^{2}})$, where $r$ is the projection distance from the region to the SNR centre. Based on all the assumptions mentioned above, the volume of small regions and the gas density can be calculated. As for the double thermal component model, the two-temperature gas was presumed to fill the whole column volume ($f_{\rm c}$ + $f_{\rm h}$ = 1) and to be in pressure balance ($n_{\rm c}T_{\rm c}$ = $n_{\rm h}T_{\rm h}$), where $f$ is the filling factor and $n$ is the H density
.
The total mass of cool gas $M_{\rm c}$ is 130$_{-21}^{+31}$ M$_\sun$, while that of hot gas $M_{\rm h}$ is 34$_{-6}^{+4}$ M$_\sun$.

With gas density acquired, the ionization age $t_{\rm i}$ was also calculated from ionization timescales $\tau = n_e t_{\rm i}$, as shown in Figure \ref{fig:apecvnei}.

It is reasonable to assume that the SNR is in Sedov-Taylor phase \citep{1950RSPSA.201..159T,1959sdmm.book.....S,1988RvMP...60....1O} according to the calculated ionization age ($\sim$ 5000 yr). The forward shock velocity is estimated as $v_{\rm s} = [16kT_{\rm c}/(3\mu m_{\rm p})]^{1/2} = 410(kT_{\rm c}/0.2$ ${\rm keV})^{1/2}$ km $\rm s^{-1}$, where $\mu$ = 0.61 is the average particle mass for a fully ionized solar-abundant plasma, in units of the proton mass $m_{\rm p}$. Hence we obtained a Sedov age $t_{\rm Sedov} = 2 R_{\rm s}/5v_{\rm s} = 12d_{20}(kT_{\rm c}/0.2$ ${\rm keV})^{-1/2}$ kyr, where $R_{\rm s}$ is the shock radius and $d_{20}$ = $d$/(20 kpc) is the distance scaled to 20 kpc. From the total mass of the shocked gas $M = M_{\rm c} + M_{\rm h} = 140$ M$_\sun$, we estimated the pre-shock medium density $\rho_0 = M_{\rm c}/(4/3\pi R_{\rm s}^3)$, and thus the explosion energy $E = 25\rho_0R_{\rm s}^3v_{\rm s}^2/(4 \xi) = 4.5 \times 10^{50}d_{20}^{2.5}(kT_{\rm c}/0.2$ ${\rm keV})$ erg, where $\xi = 2.026$ for monoatomic gas ($\gamma = 5/3$). These estimations of Sedov evolution are different from those of S19, primarily because S19 adopted an explosion energy of $10^{51}$ erg for a typical Type Ia SN.

\section{Discussion} \label{sec:discussion}

\subsection{Distance} \label{subsec:distance} 

\citet{2013ApJ...766..112R}, C16 and S17 assumed the distance to be $\sim$ 8 kpc (the fiducial distance to the Galactic centre). However, S19 proposed the distance being $\sim$ 20 kpc based on the column density derived from H$_{\rm I}$ 21cm emission and X-ray analysis.

Another method to determine distance is to use the relation between the radio surface brightness of SNR $\Sigma$ and its diameter $D$, although this method has a large uncertainty and it is generally accepted that there are different $\Sigma-D$ relations for SNRs that evolve in different environments. The H density for G306.3$-$0.9 calculated is not considerably high and there is no evidence for interaction with a dense molecular cloud as yet, also, no gamma-rays originating from the hadronic process has been detected (S17). Furthermore, we performed a molecular observation toward the SNR and have not found an interaction between the SNR and the molecular gas (see Appendix \ref{sec:molecular}). Therefore we applied the observed surface brightness $\Sigma_{\rm 1GHz}=1.7 \times 10^{-21}{\rm Wm^{-2}Hz^{-1}sr^{-1}}$ \citep{2013ApJ...766..112R} to a $\Sigma-D$ relation for SNRs in low-density environment $\Sigma_{\rm 1GHz}=1.89 \times 10^{-16}D^{-3.5}{\rm Wm^{-2}Hz^{-1}sr^{-1}}$ \citep{2013ApJS..204....4P}. The result turns out to be 26 kpc, in favour of the distance of 20 kpc from S19. If the SNR is in a much denser environment the calculated distance from a different $\Sigma-D$ relation will be even longer. \citet{2014SerAJ.189...25P} also calculated the distance to G306.3$-$0.9 as 26 kpc, using a new $\Sigma-D$ relation from an updated Galactic SNR sample.

\subsection{Progenitor} \label{subsec:progenitor}

Global spectral fitting results show overabundances of various elements, which can be used to reveal the composition of the ejecta. On the assumptions of the gas geometry mentioned in Section \ref{subsec:gascalculation}, the gas and metal masses in small regions can be calculated. 

We obtain mass-weighted abundances ${\rm X} = \sum{m_i{\rm X}_i}/\sum{m_i}$ ($m_i$ and ${\rm X}_i$ are the gas mass and the abundance of element X for the $i^{\rm th}$ region, respectively), which take into account the asymmetric gas and ejecta distributions. By comparing abundance ratios X/Si with various SN models, the progenitor can be identified. However, the Fe abundances in individual regions are mainly determined from the Fe-L lines because the hottest Fe-rich component which has bad spectral quality above 5 keV has been ignored in regional spectral fit. 
The triple-component fits for global spectra (see Figure \ref{fig:globalspectra}) also showed that part of the Fe-L lines emission might come from a hotter ejecta component. These mean that the mass-weighted Fe/Si ratio might have larger uncertainty than that we presented in this paper.
Therefore, the Fe/Si ratio from the global spectra, which still shows an obvious Fe-K$\alpha$ line, is also used for comparison.
We also calculated the Fe/Si ratio based on the analysis of {\it Suzaku} observation data in S19, because of the better spectral quality at around 6.5 keV\footnote{The fitting results of model(c) in S19 which consists of two components of Fe with different ionization timescales are adopted. The new calculations are based on the same solar abundance \citep{2009ARA&A..47..481A}, the same shell-like volume assumptions with filling factors taken into account and pressure balance assumption as discussed in section \ref{subsec:gascalculation}, instead of the original solar abundance and assumptions in S19.}. However, the Fe abundance of the 3rd (hottest) component in S19 still cannot be well-constrained because the temperature was not fitted but assumed to be 5 keV. Moreover, the SNR was treated as uniform due to low spatial resolution of {\it Suzaku}. So this calculated Fe/Si ratio might still be less reliable than presented.

It should be mentioned that C16, S17 and S19 all compare abundance ratios with various SN models but could not find a perfect match, 
and then concluded that the progenitor of G306.3$-$0.9 is a Type Ia SN, mainly based on the study of the Fe-K$\alpha$ line.
Here we compared the results with a larger collection of SN models, as shown in Table \ref{tab:model}.

\begin{deluxetable}{cc}
	\tablenum{3}
	\tabletypesize{\footnotesize}
	\tablecaption{SN models compared with SNR G306.3$-$0.9 \label{tab:model}}
	\tablewidth{0pt}
	\tablehead{
		\colhead{Model\tablenotemark{\scriptsize a}} & \colhead{Number of models} 
	}
	\startdata
	Core-collapse
	& 5\tablenotemark{\scriptsize c}
	\\[8pt]
	\parbox{0.5\columnwidth}{\centering Normal Type Ia (near-Chandrasekhar)}
	& 6\tablenotemark{\scriptsize d}
	\\[8pt]
	\parbox[c]{0.6\columnwidth}{\centering White dwarf double-detonation (sub-Chandrasekhar)}
	& 3\tablenotemark{\scriptsize e}
	\\[8pt]
	Ca-rich transient\tablenotemark{\scriptsize b}
	& 8\tablenotemark{\scriptsize f}
	\\
	\enddata
	\tablenotetext{a}{Super-Chandrasekhar Type Ia SN from double white dwarf merger model of \citealt{2010Natur.463...61P} was also considered but was not categorized in this table.}
	\tablenotetext{b}{Some models in this category does not necessarily produce Ca-rich ejecta, but have similar SN mechanisms to He shell detonations or NS-WD mergers.}
	\tablenotetext{c}{\citealt{1995ApJS..101..181W,1996ApJ...460..408T,2003ApJ...598.1163M,2006NuPhA.777..424N,2016ApJ...821...38S}}
	\tablenotetext{d}{\citealt{1997NuPhA.621..467N,1999ApJS..125..439I,2003ApJ...593..358B,2010ApJ...712..624M,2013MNRAS.429.1156S,2018ApJ...861..143L}}
	\tablenotetext{e}{\citealt{2011ApJ...734...38W,2019ApJ...878L..38T,2020ApJ...888...80L}}
	\tablenotetext{f}{He-shell detonation: \citealt{2011ApJ...738...21W,2012MNRAS.420.3003S}. NS-WD merger: \citealt{2016MNRAS.461.1154M,2019MNRAS.488..259F,2019MNRAS.486.1805Z,2021arXiv210403415B}. Ultra-stripped CC: \citealt{2017MNRAS.466.2085M,2018Sci...362..201D}}
\end{deluxetable}

\begin{figure}[ht!]
	\gridline{\fig{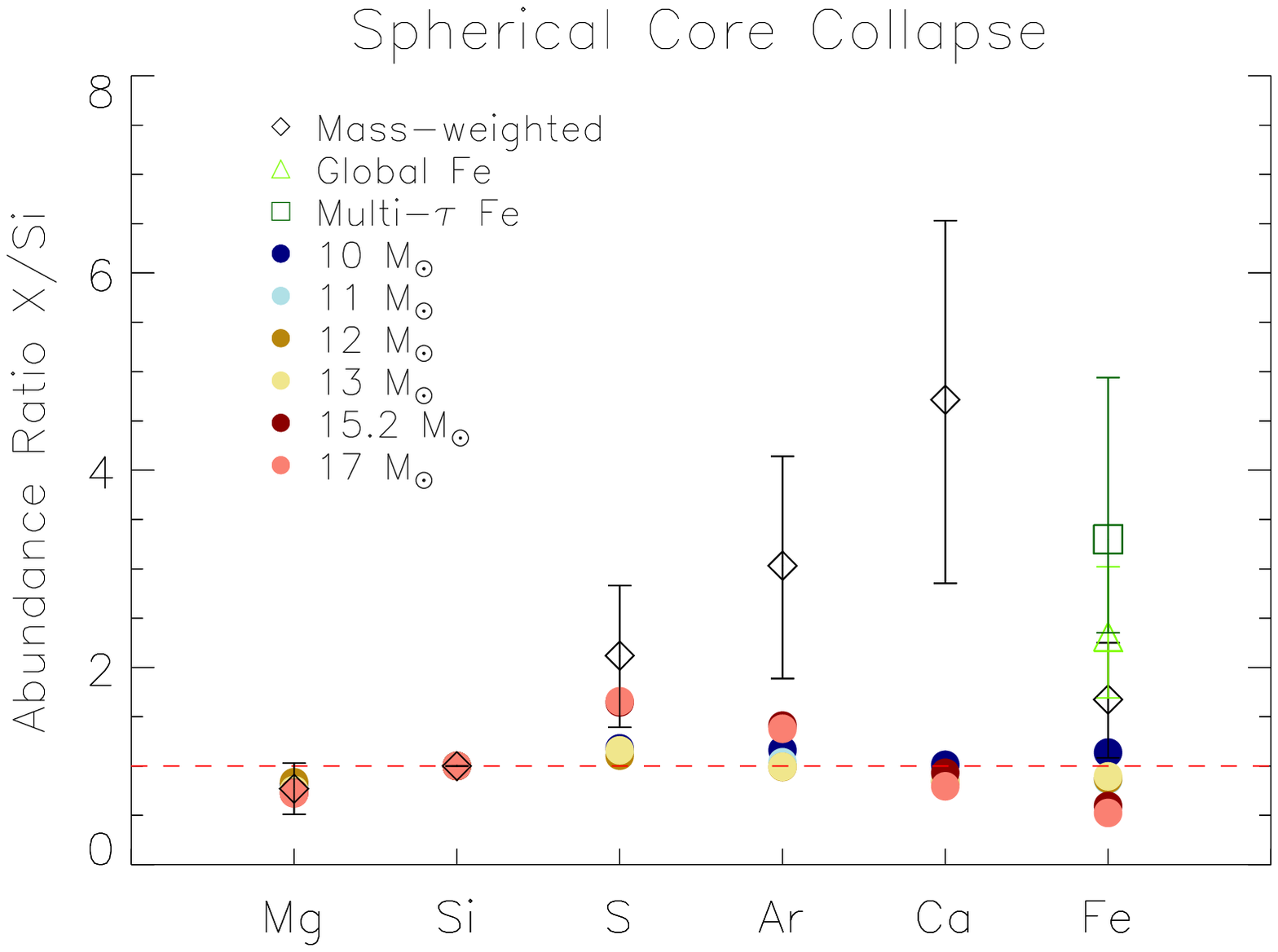}{0.89\columnwidth}{(a)}
	}
	\gridline{\fig{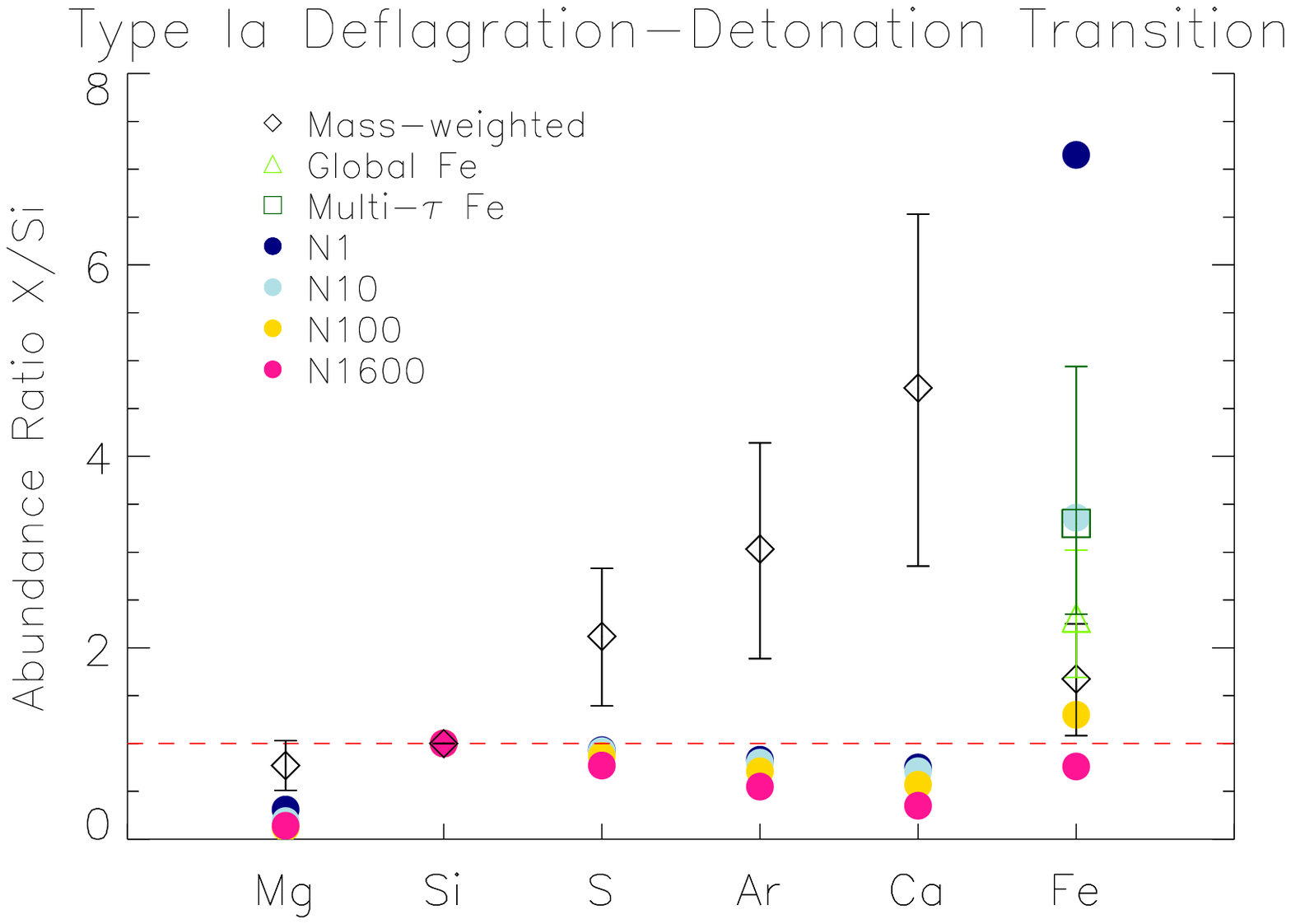}{0.89\columnwidth}{(b)}
	}
	\gridline{\fig{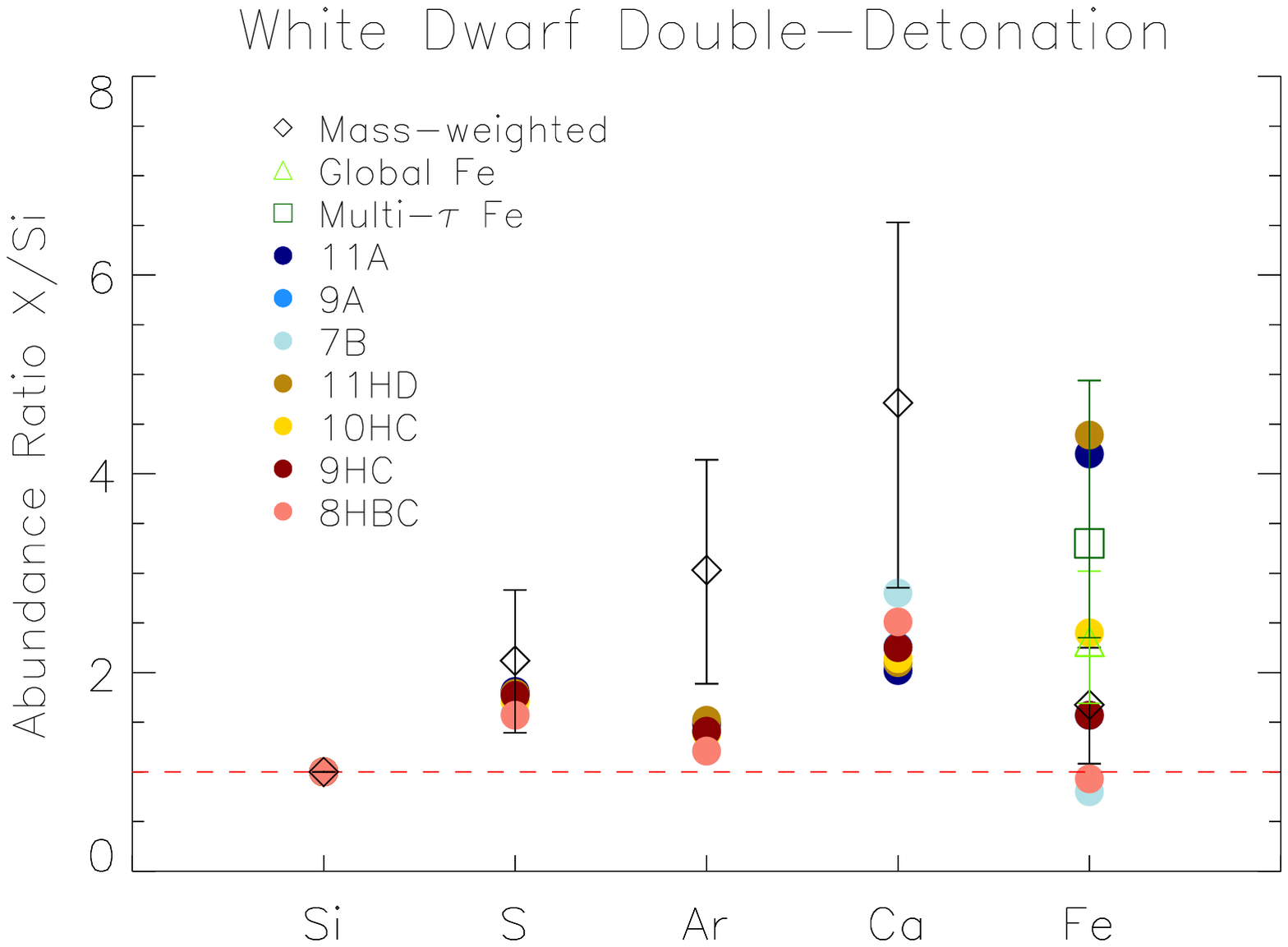}{0.89\columnwidth}{(c)}
	}
	\caption{Comparisons between abundance ratios and SN models (a: \citealt{2016ApJ...821...38S} W18 model; b: \citealt{2013MNRAS.429.1156S}; c: \citealt{2011ApJ...734...38W}). All models are mixed with 35 M$_\sun$ ISM. Diamond, triangle and square icons with error bars indicate the values from mass-weighted results, global spectra fits and \citealt{2019PASJ...71...61S}. The abundance ratio X/Si of element X is defined as $(n({\rm X})/n({\rm Si}))/(n({\rm X})/n({\rm Si}))_\sun$, where $n$ is the atom density.
		\label{fig:iaccsub}}
\end{figure}

\subsubsection{Core-collapse SNe} \label{subsubsec:cc}

We compared the abundance ratios with 
typical CC SN models \citep{2016ApJ...821...38S}, as shown in Figure \ref{fig:iaccsub}. 
Before the comparisons, the metals predicted in all SN models were mixed with 35 M$_\sun$ ISM, the amount of hot gas calculated in section \ref{subsec:gascalculation} (other masses are tested for Ca-rich transient models, as stated below). The mixing with ISM is sometimes omitted in studies, which consider the observed abundances represent the ejecta composition. However, if the mass of ejecta is small and the ejecta abundance ratios deviate greatly from solar abundance, the ISM mixing can cause a great difference in the ratios, as is the case for Ca-rich transient discussed below.
One should note that the ejecta-ISM mixing tend to make the abundance ratios closer to 1, i.e. the solar abundance, which is depicted by the red dashed line in Figure \ref{fig:iaccsub} and \ref{fig:carich}. However, the mixing will not change the trend of
metal ratios, 
which means the trend of ratios will be smoothed but not reversed. 
For example, the CC SN models in Figure \ref{fig:iaccsub} predict abundance ratios of S/Si $>$ Ca/Si (S/Ca $>$ 1), the mixing of ISM will make S/Si and Ca/Si closer to 1, but cannot change the relation to S/Si $<$ Ca/Si (S/Ca $<$ 1). 

Figure~\ref{fig:iaccsub} shows that CC SN models 
fail to explain the observed high Ca/Si and Ar/Si ratios,
although they explain the S/Si ratios better than Type Ia models (discussed below).

\subsubsection{Thermonuclear SNe:\\Normal Type Ia; WD double-detonation} \label{subsubsec:ia}

We subsequently compared the observed metal ratios in G306.3$-0.9$ with thermonuclear SN models in Table \ref{tab:model}, which include ``normal Type Ia'' models involving Chandrasekhar-mass WDs and double-detonation models for sub-Chandrasekhar-mass WDs.
It is noted that these models cover a large range of parameters and thus are not necessarily limited to Type Ia SNe.

Figure \ref{fig:iaccsub}~(b) shows the comparison between abundance ratios and three-dimensional deflagration-detonation transition (DDT) models for Chandrasekhar-mass WDs \citep{2013MNRAS.429.1156S}. The observed abundance ratios in G306.3$-$0.9 do not match any of the theoretical results for Type Ia DDT SNe, which predict too-small S/Si, Ar/Si, and Ca/Ar ratios. Moreover, the Fe mass calculated from the global spectral fits is $0.10^{+0.05}_{-0.04}d_{20}^{2.5}$ M$_\sun$, which is too small for DDT models.

We also compared the observed metal pattern with those predicted by sub-Chandrasekhar models (Figure \ref{fig:iaccsub} and Table \ref{tab:model}), since recent observations and theoretical work are challenging the $M_{\rm Ch}$ scenario. 
In these models, He detonations occur on the accreted He shell before the mass of WD approaches the Chandrasekhar limit. Such detonation leads to a second detonation in the CO core, due to either the strong oblique shock into the inner core or the converging shock at the other half of the WD. Hence these sub-$M_{\rm Ch}$ models are also called ``double-detonation'' models. 
Sub-Chandrasekhar models produce a wide mass range of Fe, thus some Fe/Si ratios can lie within the error bars. 
Although the Ar/Si and Ca/Si are over-solar, they are still insufficient, and even after we remove the dilution of ISM, the Ar/Si ratios still lie far away from the fitting results. The Ca/Si of ``7B'' model is slightly larger than the lower limit of error without ISM mixing, but this model's Fe production is deficient and leads to a sub-solar Fe/Si ratio. Moreover, the ratio trend of S/Si $>$ Ar/Si does not match the fitting results.

Among the SN models in Table \ref{tab:model}, CCSN, normal Type Ia (near-$M_{\rm Ch}$) or WD double-detonation (sub-$M_{\rm Ch}$) models are not able to account for the high Ca/Si ratio of this SNR. Furthermore, CCSN and near-$M_{\rm Ch}$ models all show abundance patterns of Ca/Si $<$ 1 and mostly S/Si $>$ Ar/Si $>$ Ca/Si (except for a few cases, e.g. the 12 M$_\sun$ and 15 M$_\sun$ models by \citet{1995ApJS..101..181W} show a trend of S/Si $<$ Ar/Si and Ar/Si $>$ Ca/Si, the 25 M$_\sun$ hypernova model by \citet{2006NuPhA.777..424N} shows S/Si $>$ Ar/Si and Ar/Si $<$ Ca/Si). By comparison, this SNR shows just an opposite trend of S/Si $<$ Ar/Si $<$ Ca/Si. Although there are limited bins around Ca lines $\sim$ 3.8keV in regional spectra, the global spectral fitting results also show high Ca/Si raito (= 5.50) and a trend of S/Si $<$ Ar/Si $<$ Ca/Si, consistent with  mass-weight abundance pattern. The peculiar ratios of Si, S, Ar and Ca cannot be explained by O burning and incomplete Si burning nucleosynthesis in common supernovae, but shows more resemblance to that of He burning, where $\alpha$-particles react and form elements along the $\alpha$-chain, up to $^{40}$Ca. In short, the models above are not able to explain either the Ca/Si or the growth trend. We need other models that have processes to produce Ca-rich ejecta.

\subsubsection{Peculiar Thermonuclear SNe:\\Ca-rich Transients} \label{subsubsec:carich}

\begin{figure*}[ht!]
	\gridline{\fig{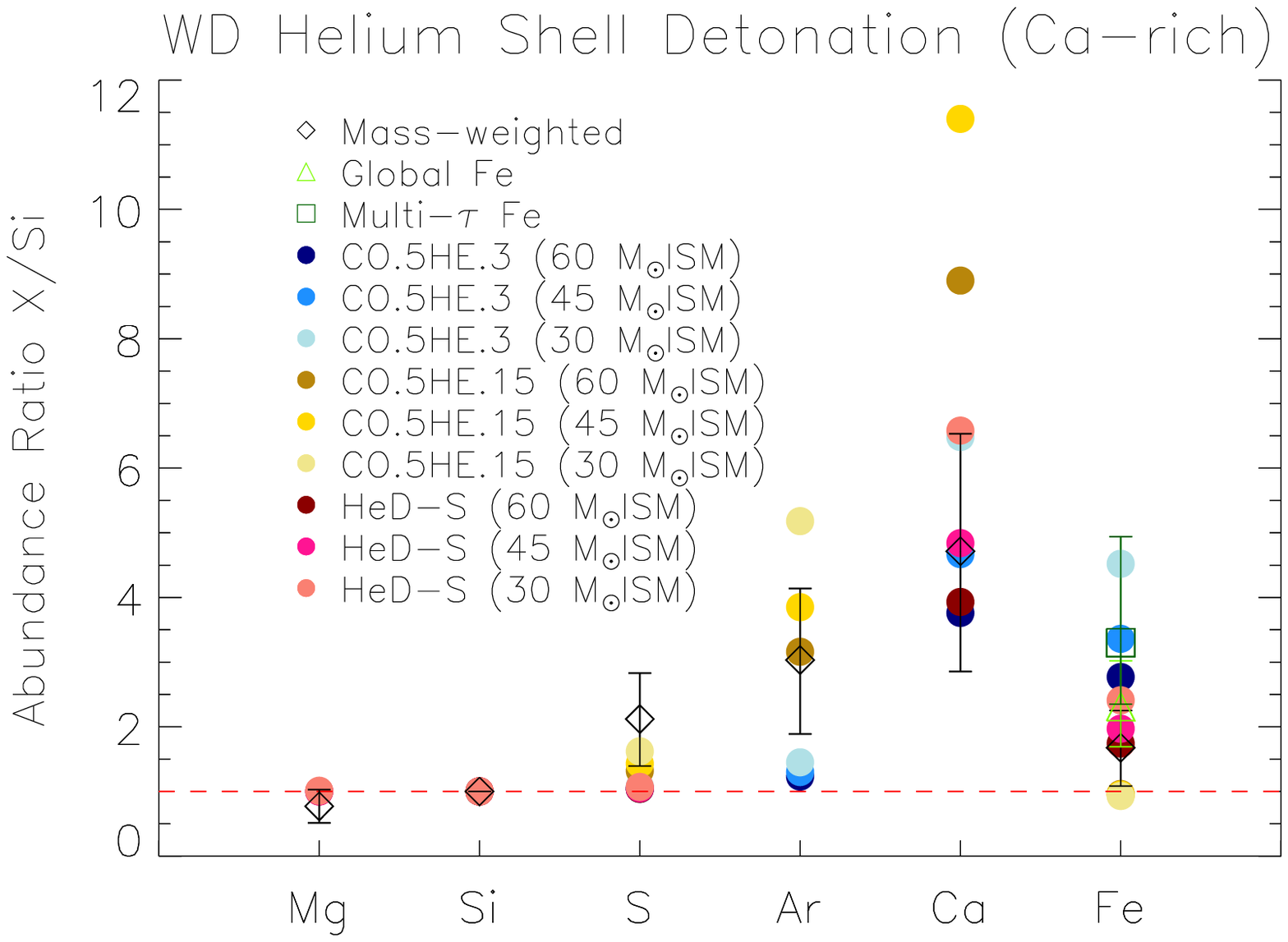}{0.5\textwidth}{(a)}
		\fig{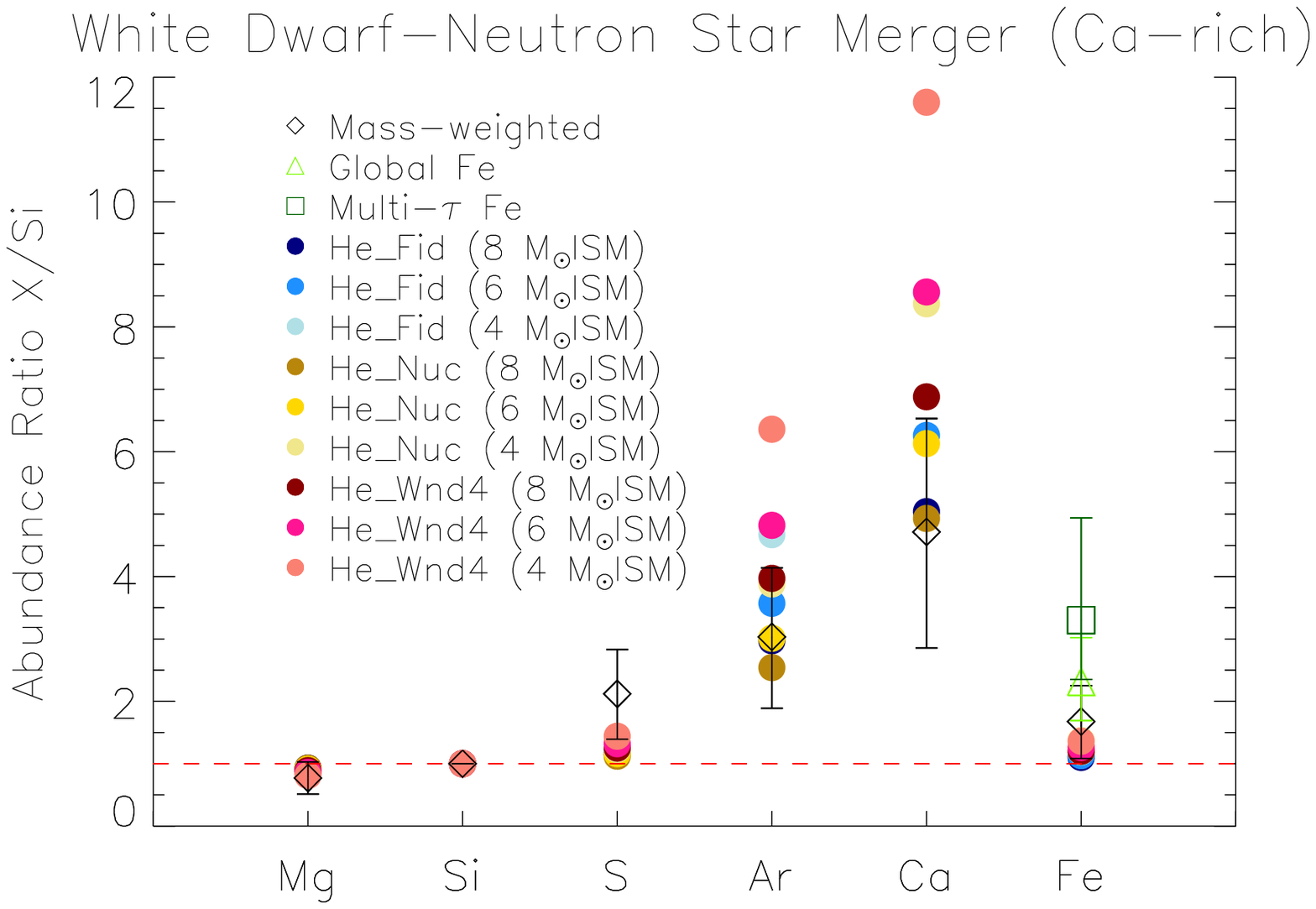}{0.5\textwidth}{(b)}
	}
	\caption{Comparisons between abundance ratios relative to Si and Ca-rich transient models (a: \citealt{2011ApJ...738...21W,2012MNRAS.420.3003S}; b: \citealt{2016MNRAS.461.1154M}) mixed with different mass of ISM. Diamond, triangle and square icons with error bars indicate the values from mass-weighted abundance, global spectra fitting and \citealt{2019PASJ...71...61S}. The abundance ratio X/Si of element X is defined as $(n({\rm X})/n({\rm Si}))/(n({\rm X})/n({\rm Si}))_\sun$, where $n$ is the atom density. The red dashed line indicates the solar abundance. The Ca/Si ratio of CO.5HE.15 mixed with 30 M$_\sun$ ISM is missing because the theoretical value is beyond the y-axis range, i.e. $>$ 12.
		\label{fig:carich}}
\end{figure*}

Recently, a new type of thermonuclear SN called ``Ca-rich transient'' has been found in extragalactic systems. One of its significant characteristics is the Ca-rich ejecta and the mass of the Ca element may reach $\sim 10^{-1}$ M$_\sun$. A natural explanation for Ca-rich ejecta is the burning of He-rich matter and there are two leading nucleosynthesis models so far: i) the accreted WD He shell detonation \citep{2011ApJ...738...21W}; ii) the merger of a He-rich WD and a NS \citep{2016MNRAS.461.1154M, 2020MNRAS.493.3956Z}. 
Both models predict very large Ar/Si and Ca/Si ratios in the ejecta. 
Nevertheless, because both models involve low-mass ejecta (typically $<\sim$ 0.3 M$_\sun$), the ISM dilution can change the values of abundance ratios greatly and how we approach the dilution will be critical. Figure \ref{fig:carich} shows the comparisons with the two models mixed with various ISM masses.

It had been recognized that He detonation in the WD accreted shell would always lead to a second detonation in the CO core (i.e., double-detonation models, e.g., \citealt{2019ApJ...873...84P, 2019arXiv191007532P}). However, \citet{2011ApJ...738...21W} assumed that the second detonation does not happen if the WD mass is low and the He detonation occurs in the low-density regime, or the actual axial symmetry is not perfectly aligned. Figure \ref{fig:carich}~(a) shows comparisons with the He shell detonation models (one-dimensional models by \citealt{2011ApJ...738...21W} and two-dimensional models by \citealt{2012MNRAS.420.3003S}). Three best-fit models among the total of 14 are shown, ``CO.5HE.3''/``CO.5HE.15'' stands for a WD with a 0.5 M$_\sun$ CO core and a 0.3/0.15 M$_\sun$ He shell, and ``HeD-S'' corresponds to a 0.45 M$_\sun$ core and a 0.21 M$_\sun$ shell. The Mg/Si ratio in CO.5HE.3/CO.5HE.15 and Ar/Si ratio in HeD-S are missing because these models did not provide the yields of Mg and Ar. Figure \ref{fig:carich} shows that, with $\sim$ 30--60 M$_\sun$ of ISM mixed, the theoretical predictions can generally match the observed metal pattern. Moreover, models all show a trend of S/Si $<$ Ar/Si $<$ Ca/Si, which is consistent with that in the SNR. 

We note that one or two abundance ratios in the SNR still cannot perfectly match existing Ca-rich transient models. Firstly, the S productions of CO.5HE.3 and CO.5HE.15 models are slightly low. As for HeD-S, the S production is even lower and the S/Si ratio is already close to the solar abundance. Secondly, because the SNR spectra shows strong Fe-K$\alpha$ lines, we should also pay attention to the Fe/Si ratio. CO.5HE.3 and HeD-S models have Fe ratios that match the spectral results but much lower S/Si and Ar/Si ratios. Alternatively, CO.5HE.15 matches the IME/Si ratios best, but it produces little iron-group elements and thus has a Fe/Si ratio close to 1. However, there are many further complications of the models. A higher density of He layer would lead to a more complete burning, i.e. a larger mass of $^{56}$Ni (decays to $^{56}$Fe) and a lower yield of IMEs and $^{44}$Ti (decays to $^{44}$Ca). There can also be mixing of C-rich material into the He-envelope, and a larger mass fraction of C would lower the mean atomic weight of the $\alpha$-capture products, which means more IMEs and less Fe are produced \citep[][and Leung et al.\ in prep.]{2011ApJ...738...21W, 2012MNRAS.420.3003S}. Among the highly diverse outcomes expected from the He shell detonation, it might be possible that at certain conditions it can produce the amount of IMEs and Fe that well explain each abundance ratios of SNR G306.3$-$0.9.


Next, we consider the possibility of the merger of a NS and a WD as the progenitor. In this scenario, a He WD is tidally disrupted as it approaches the NS companion. The WD debris forms an accretion disc and the mid-plane temperature would become high enough for nuclear fusion as the gravitational energy converts to internal energy \citep{2012MNRAS.419..827M}. The nucleosynthesis products will be ejected as disc wind, along with a large fraction of unburnt He. Hence, the production of metal elements of NS-WD merger is generally lower by an order of magnitude than that of He shell detonation model (see references in Table \ref{tab:model}). Therefore, when we compare this model with the observed metal abundances, the mixed ISM mass which allows the theoretical ratios to lie within the observed error bars is much smaller. The nucleosynthesis model of \citet{2016MNRAS.461.1154M} is shown in Figure \ref{fig:carich}~(b), where ``He\_Fid'' corresponds to a fiducial model of a 0.3 M$_\sun$ WD -- 1.2 M$_\sun$ NS merger, ``He\_Nuc'' has the same configurations as ``He\_Fid'' but with nuclear heating suppressed, and ``He\_Wnd4'' has a larger wind cooling efficiency $\eta_{\rm w}$, which corresponds to the ratio of the launched wind velocities to the local escape speed. When the mixed ISM mass is small, the theoretical values can match the fit results, except that they underpredict the S/Si and Fe/Si ratios. It seems that the small innermost region of the disc where $^{56}$Ni is synthesized cannot produce sufficient Fe to account for the Fe yields of the SNR. Although the 3D simulations of \citet{2021arXiv210403415B} show that a larger mass of the WD donor can increase the yield of Fe (up to $\sim 0.1$ M$_\sun$), a lack of He-rich material in the massive CO/ONe WDs cannot give the Ca-rich abundance pattern of this SNR.

\begin{deluxetable}{cccccc}
	\tablenum{4}
	\tabletypesize{\footnotesize}
	\tablecaption{The ISM mass needed for the theoretical values of Ca-rich transient models falling within error bars of fitted abundance ratios relative to Si (X/Si), calculated based on the mass-weighted fitting results of the whole SNR in units of the solar mass. The results are rounded to the  nearest integers. ``$\times$'' is labelled if abundance ratios cannot lie within the error bars by ISM mixing and ``-'' for elements of which nucleosynthesis results were not published originally. \label{tab:mass}}
	\tablewidth{0pt}
	\tablehead{
		\colhead{Model} & \colhead{Mg/Si} & \colhead{S/Si} & \colhead{Ar/Si} & \colhead{Ca/Si} & \colhead{Fe/Si}
	}
	\startdata
	CO.5HE.3
	&	-
	&	1--5
	&	4--15
	&	30--89
	&	87--1339
	\\
	CO.5HE.15
	&	-
	&	9--49
	&	41--151
	&	87--263
	&	$\times$
	\\
	HeD-S
	&	$>$ 3
	&	$<$ 2
	&	-
	&	31--96
	&	35--579
	\\
	He\_Fid
	&	$>$ 1
	&	$<$ 2
	&	5--18
	&	6--18
	&	1--10
	\\
	He\_Nuc
	&	$>$ 1
	&	$<$ 2
	&	4--14
	&	6--17
	&	1--20
	\\
	He\_Wnd4
	&	$>$ 1
	&	1--4
	&	8--29
	&	9--27
	&	1--21
	\\
	\enddata
\end{deluxetable}

The mixed ISM mass is still a parameter adjusted manually so far. Table \ref{tab:mass} shows the ISM mass needed for mixing, to account for the fitted abundance ratios. 
The hot gas component {\it vnei} mass is calculated as 34$_{-6}^{+4}d_{20}^{2.5}$ M$_\sun$, from the fitted parameter {\it norm}, allowing us to examine whether enough ISM exists to account for the dilution.
At 20 kpc \citep{2019PASJ...71...61S}, the hot gas mass roughly meets the requirement of He shell detonation model. As for 8 kpc \citep[][C16, S17]{2013ApJ...766..112R}, the hot gas mass turns out to be 3.4$_{-0.5}^{+0.5}$ M$_\sun$, roughly meeting the requirement of the WD+NS model. It should also be noted that the amount of ISM mixed is not very certain, it depends on the calculation of the hot gas mass, which is based on the assumptions of volume and pressure balance. Moreover, the ejecta might not be fully mixed with the ISM, and the existence of pure ejecta might also make the calculations deviate from the fact. However, based on what we could obtain from the X-ray analysis,
Ca-rich transient models can explain the abundance ratios of G306.3$-$0.9 better than any other models.

The Ca-rich transient origin of the SNR can be further confirmed if any compact remnant is revealed by follow-up searches. In the He shell detonation scenario, a hot WD can be left if the second detonation does not occur, while a spin-up NS remains after the NS-WD merger.

There have been other models proposed for Ca-rich transients, e.g. ultrastripped-envelope CC SNe and tidal disruption of a WD by an intermediate-mass black hole (IMBH). Two ultrastripped-envelope CC SNe models are included in Table \ref{tab:model}. However, because they are still CC SNe in nature and do not have Ca-rich ejecta, their abundance ratios cannot provide a good fit and thus are not shown here. Besides, the SNR is more likely associated with old stellar populations. The closest prominent infrared object is the G305 star-forming complex  \citep{2012MNRAS.426..402F}, which is $\sim 1\arcdeg$ away. But the complex is located at 4 kpc, which is far below the current distance estimation of 20 kpc. Such a distance also makes the SNR about 350 pc above the Galactic plane, where is almost outside the thin disc.

Since there is no evidence of an IMBH or a global cluster (GC) habiting near the SNR and the known Ca-rich transients are not found near GCs or dwarf galaxies \citep{2019ApJ...887..180S}, the IMBH-WD model is not considered. 

\subsubsection{IMEs and Fe Yields of SNe} \label{subsubsec:cas}

\begin{figure*}[t!]
	\centering
	\includegraphics[width=0.8\textwidth]{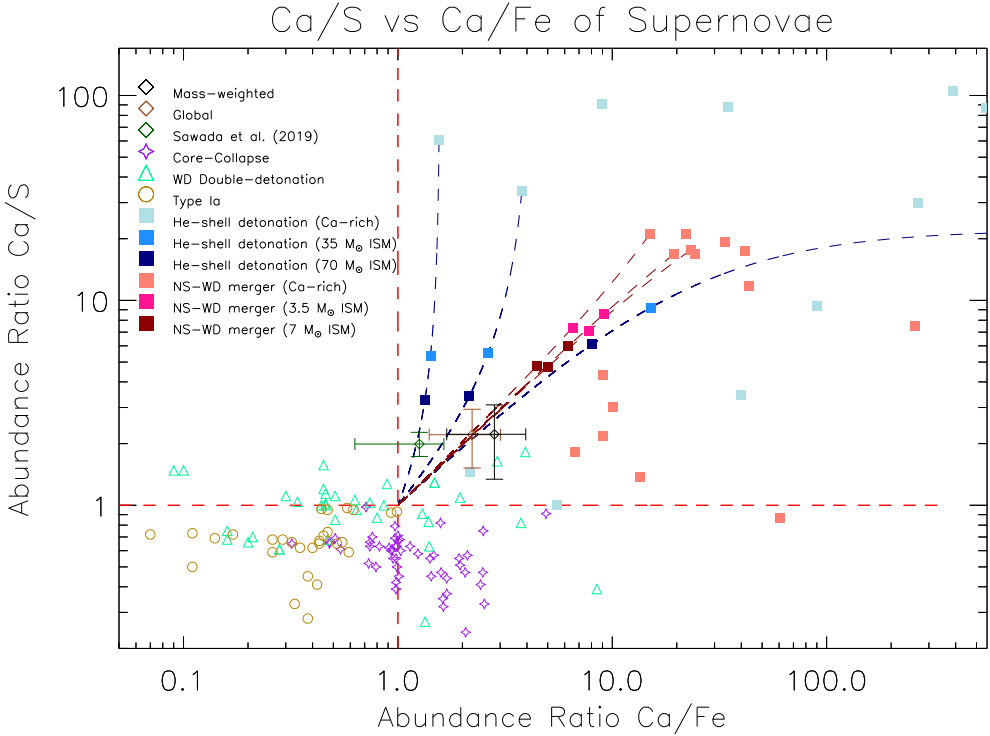}
	\caption{
	Ca/S and Ca/Fe abundance ratios of G306.3$-0.9$ compared with those from different SN models. Black, brown and dark green diamond icons with error bars indicate the values from mass-weighted abundance, global spectra fitting and \citealt{2019PASJ...71...61S}. Navy blue and deep red dashed lines correspond to the abundance ratio variation curves of Ca-rich models in Figure \ref{fig:carich} mixed with an increasing amount of ISM. The red dashed line indicates the solar abundance. All abundance ratios taken from SN models are calculated considering only pure ejecta without ISM mixed unless otherwise stated. 
		\label{fig:cascafe}}
\end{figure*}

We stress that the unusually high Ca abundance and the 1 $<$ S/Si $<$ Ar/Si $<$ Ca/Si and Ca/Si $>$ Fe/Si $>$ 1 patterns of G306.3$-$0.9 are crucial to inferring its progenitor. These characteristics are explained better by He burning. In normal SNe, Si, S, Ar and Ca are mainly products of explosive O burning and incomplete Si burning, and iron-group elements are produced by incomplete Si burning and nuclear statistical equilibrium \citep[e.g.][]{1996ApJ...460..408T,1999ApJS..125..439I}. As for Ca-rich transient, these elements come from He burning.

Figure \ref{fig:cascafe} shows that Ca/S and Ca/Fe abundance ratios of pure ejecta from the SN models in Table \ref{tab:model}. Each CC, normal Type Ia and sub-Chandrasekhar model is included in the figure except those which did not explore extended parameter space or were already included and refined in other literature. He shell detonation and NS-WD merger for Ca-rich transients were also depicted in the plot. Because S is over-abundant in mass-weighted results, its lines lie near the Ca lines, which makes the relative ratio less dependent on temperature and it is also usually included in SN nucleosynthesis studies, the ratio Ca/S is selected to illustrate the pattern. The mass-weighted and global ratios are compared with the models and the results from \citet{2019PASJ...71...61S} are also included after the same calculations discussed above. But again, it should be noted that the mass-weighted Fe abundance is 
obtained by fitting the Fe-L lines with a single shocked ejecta model, while the global fit and S19 treat the asymmetric SNR as uniform. It can be seen that while Ca/S of the three different fitting results agree with each other
, the Ca/Fe ratios show some differences.

As shown in Figure \ref{fig:cascafe}, near-$M_{\rm Ch}$ Type Ia and CC SNe occupy the Ca/S $<$ 1 area. 
Type Ia SNe are Fe factories and thus gather around the Ca/Fe $<$ 1 corner. The range for CC SNe is large, from sub-solar to over-solar, and most points lie around the solar abundance.

As for sub-Chandrasekhar models, the initial conditions such as the mass of the WD core, the accretion rates, the He detonation configurations and the metallicity, could influence the products. Therefore, the points of sub-Chandrasekhar models are quite dispersed and can account for various peculiar transients \citep{2011ApJ...734...38W,2020ApJ...888...80L}.
Two of the double-detonation model points, ``7B'' and ``8HBC'' from \citet{2011ApJ...734...38W}, are near the fitting results of the SNR. However, the large Ca/Fe ratios of these two models are mainly due to the low productions and sub-solar abundances of Fe. 
The ISM-ejecta mixed cases can be seen in Figure \ref{fig:iaccsub}, which shows that these two models could not match the observed abundance ratios.

Ca-rich transient models provide Ca/Fe$>1$ and Ca/S $>1$. 
Note that the pure ejecta points of Ca-rich transients lie far away from (1,1) because of the unusually high Ca production. 
Yet with ISM mixing taken into account, the data points would approach the solar abundance point rapidly due to low ejecta mass.
In Figure~\ref{fig:carich}, we use dashed lines 
to illustrate how the Ca/Fe and Ca/S ratios change as a function of the mixed ISM mass (only for a few exemplified Ca-rich transient models).
The filled boxes with darker colors mark the Ca-rich transient models mixed with more ISM masses (e.g., 35, 70 $M_\odot$).
The dashed lines can pass through the observed metal ratios in G306.3$-0.9$.
This means that with a proper amount of ISM mixed, some Ca-rich transient models could explain observed ratios. And the filled boxes fall within the error bars of fitted Ca/Fe ratios but have higher Ca/S ratios, revealing the problem that current Ca-rich transient models produce less element S than that is shown in the SNR.

\subsubsection{Disadvantages of NS+WD Merger} \label{subsubsec:nswd}

One key parameter for Ca-rich transients successfully explaining the abundances ratios of G306$-$0.9 is the amount of ISM mixed with ejecta. The amount of ISM needed for dilution varies by one order of magnitude between the He shell detonation and NS-WD merger models (see Table \ref{tab:mass}). Hence, the actual distance can help us determine the ISM mass and thus the Ca-rich transient subclass of the SNR. Although there still exist uncertainties about the distance, S19 and $\Sigma-D$ relation have suggested the distance to be $\gtrsim 20$ kpc (see Section \ref{subsec:distance}). At such a distance, the $\sim 35$ M$_\sun$ ISM will dilute the NS+WD merger ejecta so strongly that the theoretical abundances become too low to account for the observed values (e.g. the abundances of ``He\_Fid'' model become Mg = 1.00, Si = 1.02, S = 1.06, Ar = 1.51, Ca = 2.02, Fe = 1.04). In this respect, the He shell detonation model is preferred over the NS-WD merger.

Using the Sedov-Taylor self-similar solution,
the SN explosion energy is
calculated (see Section \ref{subsec:gascalculation}): $E = 4.5 \times 10^{50}d_{20}^{2.5}$ erg.
One can see that 
the SNR distance highly influences the SN explosion energy.
At 8 kpc, where the ISM mixing requirement of NS-WD merger models can be satisfied, the SN explosion energy would be only $\sim 4.5 \times 10^{49}$ erg, which is too low for a typical SN explosion or the 1D NS-WD merger model of \citet{2016MNRAS.461.1154M}. Although the more sophisticated 2D simulations of \citet{2019MNRAS.486.1805Z,2020MNRAS.493.3956Z} obtained similarly low explosion energy of $10^{48}$--$10^{49}$ erg for NS-CO WD mergers, they showed that the He WD cases cannot lead to any thermonuclear explosion or mass ejection. On the contrary, the explosion energy at 20 kpc is consistent with Ca-rich transient SN 2005E and He shell detonation models \citep{2010Natur.465..322P,2011ApJ...738...21W}.
 
Noteworthily, so far Ca-rich transients have been found at the outskirts of the galaxies. Therefore, a distance of $\sim 20$~kpc is reasonable if 
G306.3$-$0.9 is indeed a Ca-rich transient SNR.

Another consideration about the NS-WD merger scenario is the evolution channel and population.
If the SNR's progenitor was a NS-WD binary,
it was more likely to be a main-sequence star binary that finally evolved into a WD-NS system. 
As there is no GC found at around SNR G306.3$-0.9$, it is difficult to form a NS-WD binary via binary exchange or NS capture.
However, the binary population synthesis simulations of a main-sequence star binary evolving into a NS-WD system have shown that the rate of NS-WD merger is only 2--6\% of the observed Type Ia SNe, while those with a He WD account for a very small fraction (0.3--1.4\%) of NS-WD mergers \citep{2018A&A...619A..53T}.

\subsubsection{Comparisons with other Type Ia SNRs} \label{subsubsec:iasnr}
The above discussions on the progenitor are mainly based on the measured abundance ratios. However, it is noted that some young Type Ia SNRs also manifest enhancements of Ca/Si or Ca/Fe abundance ratios, e.g. SN 1006 \citep{2013ApJ...771...56U}, \emph{Kepler} \citep{2015ApJ...808...49K, 2019ApJ...872...45S} and N103B \citep{2021ApJ...910L..24Y}. The discrepancy between the actual abundances and theoretical results still requires more studies and convincing explanations. Nevertheless, G306.3-0.9 shows a higher Ca/Si ratio ($\sim 4.7$) than these young Type Ia remnants mixed with fewer ISM given the young ages.
With the ISM excluded, G306.3$-$0.9 would show a dramatically higher Ca abundance. Besides, one should be aware that only the outer layers of the ejecta of those young Ia remnants have been heated to emit X-ray emission, while the interior ejecta is still of low temperature.
It means that the observed metal abundance ratios cannot be directly used to compare with SN nucleosynthesis models unless different metal layers are well mixed.
Besides, according to Type Ia models (e.g. pure deflagration models by \citealt{1999ApJS..125..439I} and DDT models by \citealt{2013MNRAS.429.1156S}), Fe metals have smaller initial velocities than the lighter elements such as Ca and S and thus tend to be found in the interior.
Therefore, one would expect to see lower Ca/Fe ratios in these Type Ia SNRs after the reverse shocks sweep and heat the innermost Fe ejecta and the free expansion phase ends.

The abundance ratios of mature Type Ia SNRs with similar ages are also different from those of G306.3$-$0.9. Type Ia SNR 3C 397 \citep{2015ApJ...801L..31Y, 2018ApJ...861..143L} does not show enhanced abundance ratios of Ca to other elements Ca/X but has a large Fe/Ca ratio \citep{2005ApJ...618..321S, 2021ApJ...913L..34O}.
W49B has been proposed as a Type Ia SNR (still under debate) and has a smooth abundance ratio trend of S/Si $\sim$ Ar/Si $\sim$ Ca/Si and Ca/Fe $\lesssim 1$ \citep{2018A&A...615A.150Z, 2020ApJ...904..175S}. 

The Fe-rich hot ejecta with a shorter ionization timescale revealed by the global fits and S19 still cannot be constrained by our spatial resolved study, and its nature is still unknown. S19 attributed it to the stratified ejecta structure of Type Ia SNe. However, it might also come from the innermost slow expanding Fe ejecta in the He shell detonation \citep{2012MNRAS.420.3003S}. Another possible origin is the potential second detonation in the WD core, but the IMEs from the outer He burning still dominate the emitting ejecta.

We stress that our conclusions for the SNR's progenitor rely on spectral abundance ratios and existing SN nucleosynthesis models. The abundance pattern given by our study might be of interest for testing future SN models (e.g., Leung et al.\ in preparation; Zenati et al.\ in preparation).

\subsection{Origin of Asymmetry} \label{subsec:aymmetry}

The X-ray morphology of G306.3$-$0.9 is asymmetric. It can be roughly divided into a northeastern dim part and a southwestern bright part. Accordingly to our spatially resolved study, the northeastern part of the SNR has higher Mg, Si, S, Ar, Ca, Fe abundances, but lower H density (see Figures \ref{fig:apecvnei}). A reasonable explanation is the existence of an asymmetric ISM distribution, with denser ISM in the southwest. The ejecta can be strongly diluted due to a mixing with a dense medium, which results in the observed lower metal abundances in the southwest regions. The shocked dense gas can also greatly enhance the X-ray flux. 
Although it does not necessarily rule out the factor of intrinsic explosion asymmetry \citep[see e.g.,][for application of using explosion asymmetry in constraining explosion mechanism]{Ferrand2019, Leung2021SN2014J}, at least the ISM environment plays a significant role in the formation of the asymmetric morphology of SNR G306.3$-$0.9.

\section{Conclusions} \label{sec:conclusions}

We performed spatially resolved X-ray spectroscopy on SNR G306.3$-$0.9 using an adaptive spatial binning method. According to the fitted abundance ratios, G306.3$-$0.9 is likely to be the first found Galactic SNR of a Ca-rich transient. The fitted results reveal that the existence of a southwest denser ISM environment is the reason for its asymmetry.

1. We divided the SNR to 13 regions to examine the spatial variation of plasma properties.
The X-ray plasma in small regions can be well characterized by two-temperature gas ($\sim 0.2$ keV + $\sim 1$ keV). The cool-component plasma with solar abundances has reached collisional ionization equilibrium, while the hot component has revealed clearly enhanced metal abundances, with an ionization timescale in the range of 1--$4\times10^{11}$ cm$^{-3}$ s.

2. The total mass of cool and hot gas is 130$_{-21}^{+31}d^{2.5}_{20}$ M$_\sun$ and 34$_{-6}^{+4}d^{2.5}_{20}$ M$_\sun$, respectively. The density of the bright southwest regions is higher than that of the dim northeast regions but metal abundance has an opposite gradient. 
It explains the asymmetry of G306.3$-$0.9 by the existence of the denser ISM environment in the southwest.

3. Mass-weighted abundances relative to the solar values of the whole SNR are obtained as 
Mg = 0.97$_{-0.23}^{+0.23}$, Si = 1.26$_{-0.30}^{+0.30}$, S = 2.67$_{-0.66}^{+0.63}$, Ar = 3.82$_{-1.12}^{+1.06}$, Ca = 5.94$_{-1.87}^{+1.80}$, Fe = 2.11$_{-0.55}^{+0.52}$; and abundance ratios are 
Mg/Si = 0.77$_{-0.26}^{+0.26}$, S/Si = 2.12$_{-0.73}^{+0.71}$, Ar/Si = 3.03$_{-1.15}^{+1.11}$, Ca/Si = 4.71$_{-1.86}^{+1.82}$, Fe/Si = 1.67$_{-0.59}^{+0.57}$. 

4. Type Ia or CC SN models are unable to explain the abundance ratios or patterns in G306.3$-0.9$. Some double-detonation models marginally fit S/Si and Ca/Si ratios, but fail to explain the Fe/Si ratio. The best matched models are Ca-rich transient models, despite that the theoretical element S production is slightly low. After considering the mixing of ejecta and ISM, these models can explain the unusually high Ar/Si, Ca/Si ratios and the trend of S/Si $<$ Ar/Si $<$ Ca/Si. He shell detonation is preferred over the NS-WD model as the progenitor. 

\acknowledgments
J.Weng acknowledges the help and advice on spectral analysis from Lei Sun.

This study is partially based on observations collected at the European Organisation for Astronomical Research in the Southern Hemisphere under ESO programme 0103.D-0387.
J. Weng, P.Z., and Y.C. thank the supports by NSFC grants 11773014, 11633007, 11851305, 11503008 and 11590781.
P.Z. also acknowledges support from the Nederlandse Onderzoekschool Voor Astronomie (NOVA) and the NWO Veni Fellowship grant No. 639.041.647.
Shing-Chi Leung acknowledges support from NASA grants HST-AR-15021.001-A and 80NSSC18K1017.
HBP acknowledges support for this project from the European Union's Horizon 2020 research and innovation program under grant agreement No 865932-ERC-SNeX.
S.T. acknowledges support from the Netherlands Research Council NWO (VENI 639.041.645 grants). K. Nomoto has been supported by the World Premier International Research Center Initiative (WPI Initiative), MEXT, Japan, and JSPS KAKENHI Grant Numbers JP17K05382, JP20K04024, and JP21H04499.

\software{
ATOMDB \citep{2001ApJ...556L..91S,2012ApJ...756..128F},
CIAO \citep[vers. 4.12,][]{2006SPIE.6270E..1VF}, 
DS9 \citep{2003ASPC..295..489J},
SAS \citep[vers. 16.1.0,][]{2004ASPC..314..759G},
XSPEC \citep[vers.\ 12.10.1f,][]{1996ASPC..101...17A}.}

\bibliography{sample63}{}
\bibliographystyle{aasjournal}

\appendix

\section{Source and Background Regions} \label{sec:region}

\begin{figure*}[ht!]
\gridline{\fig{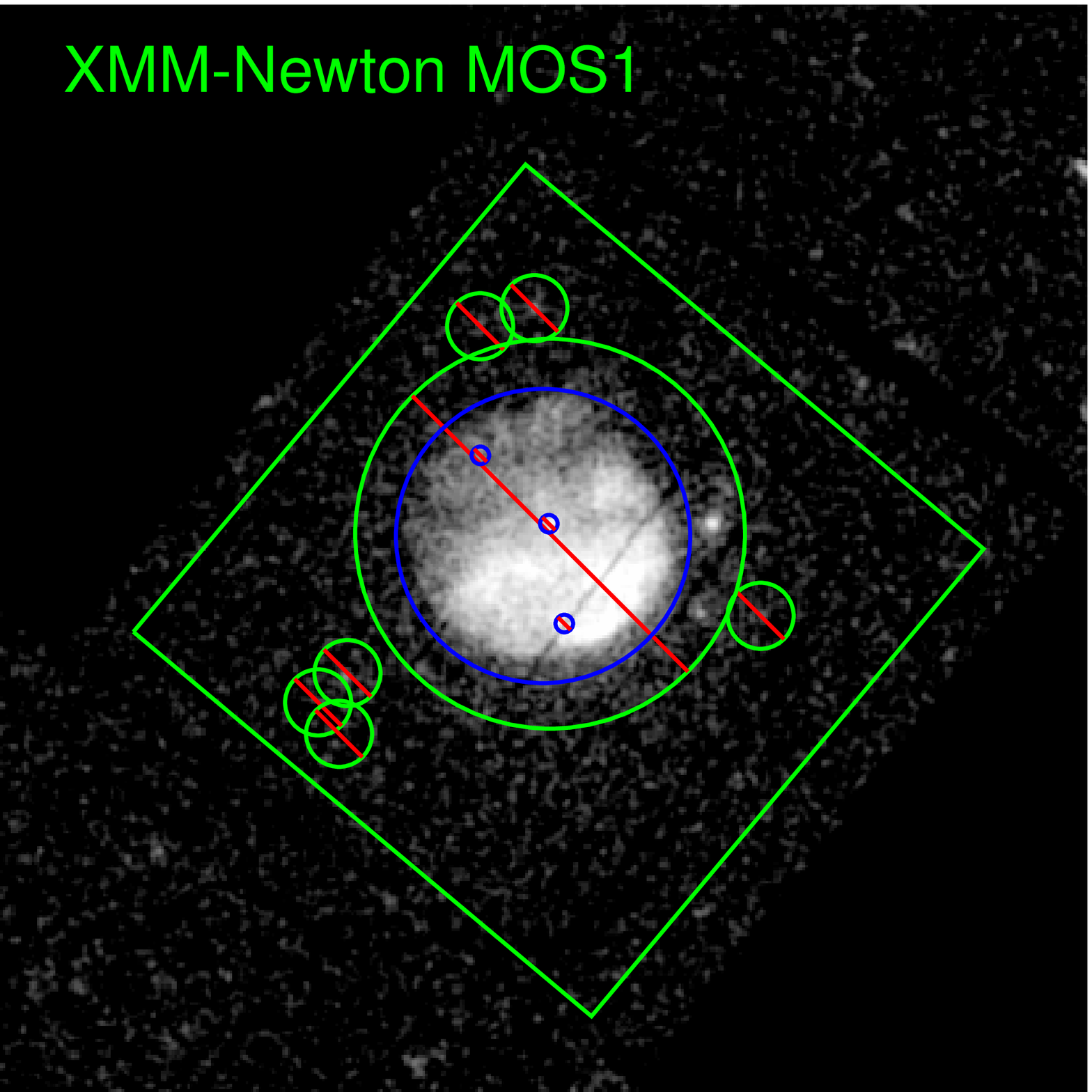}{0.3\textwidth}{(a)}
          \fig{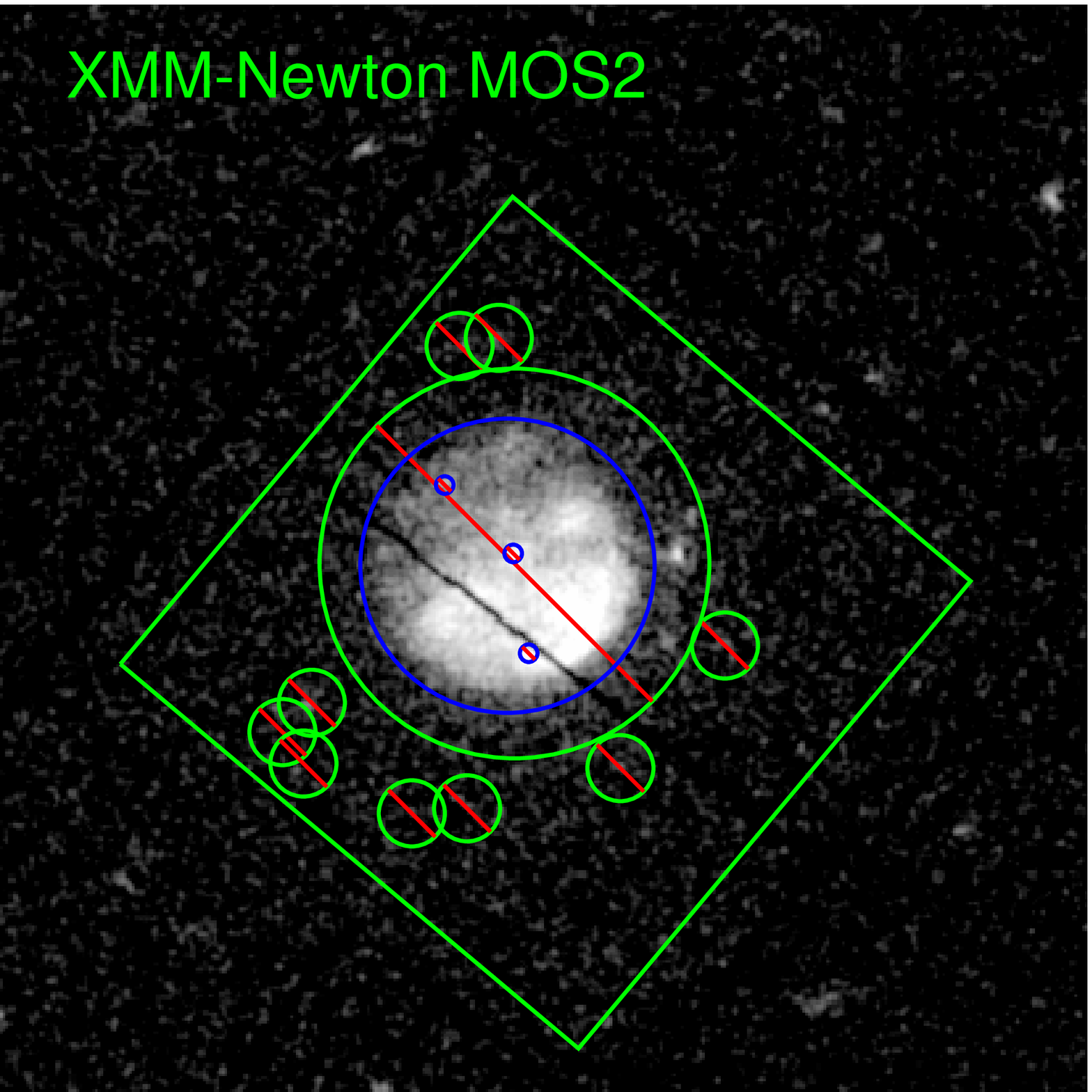}{0.3\textwidth}{(b)}
          \fig{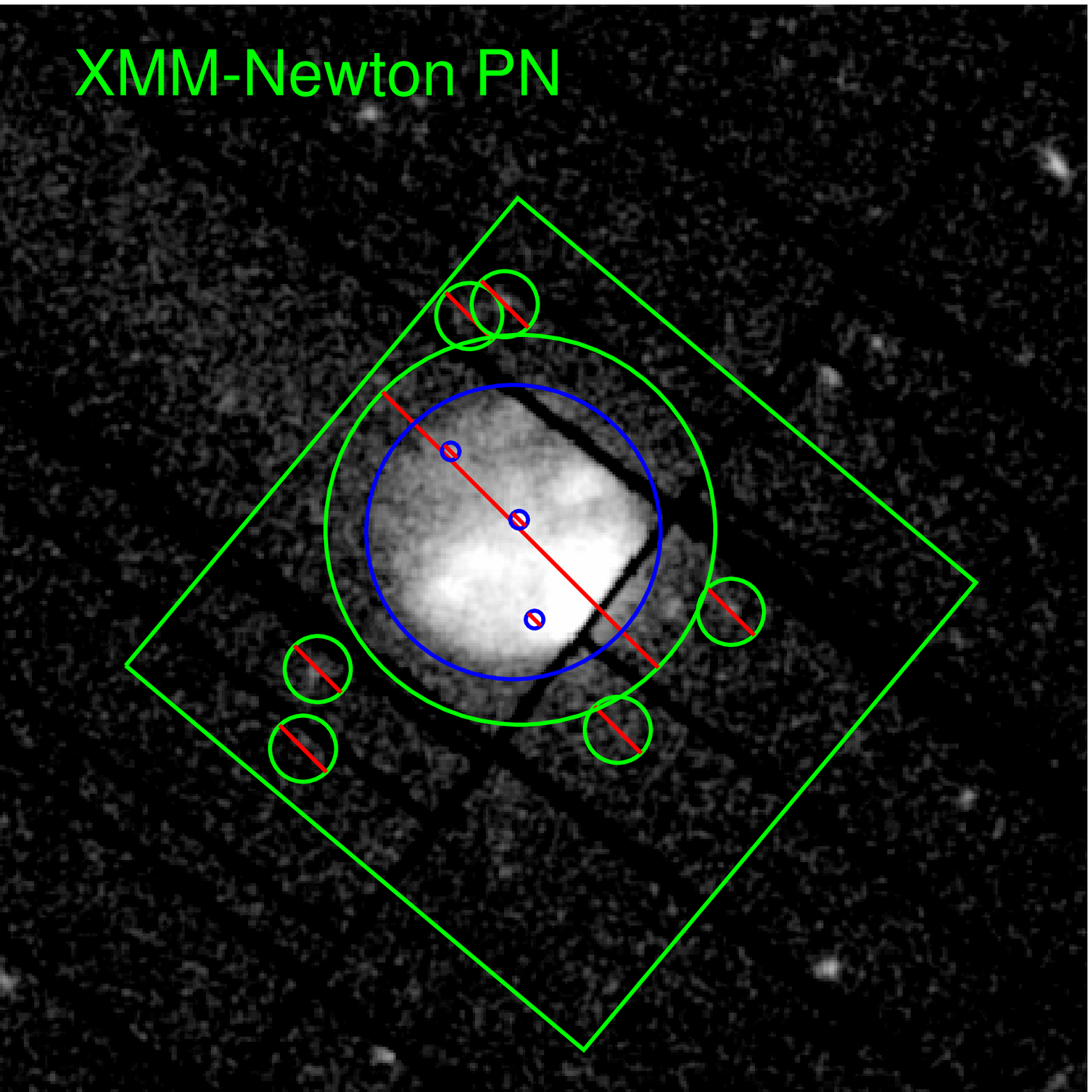}{0.3\textwidth}{(c)}
          }
\gridline{\fig{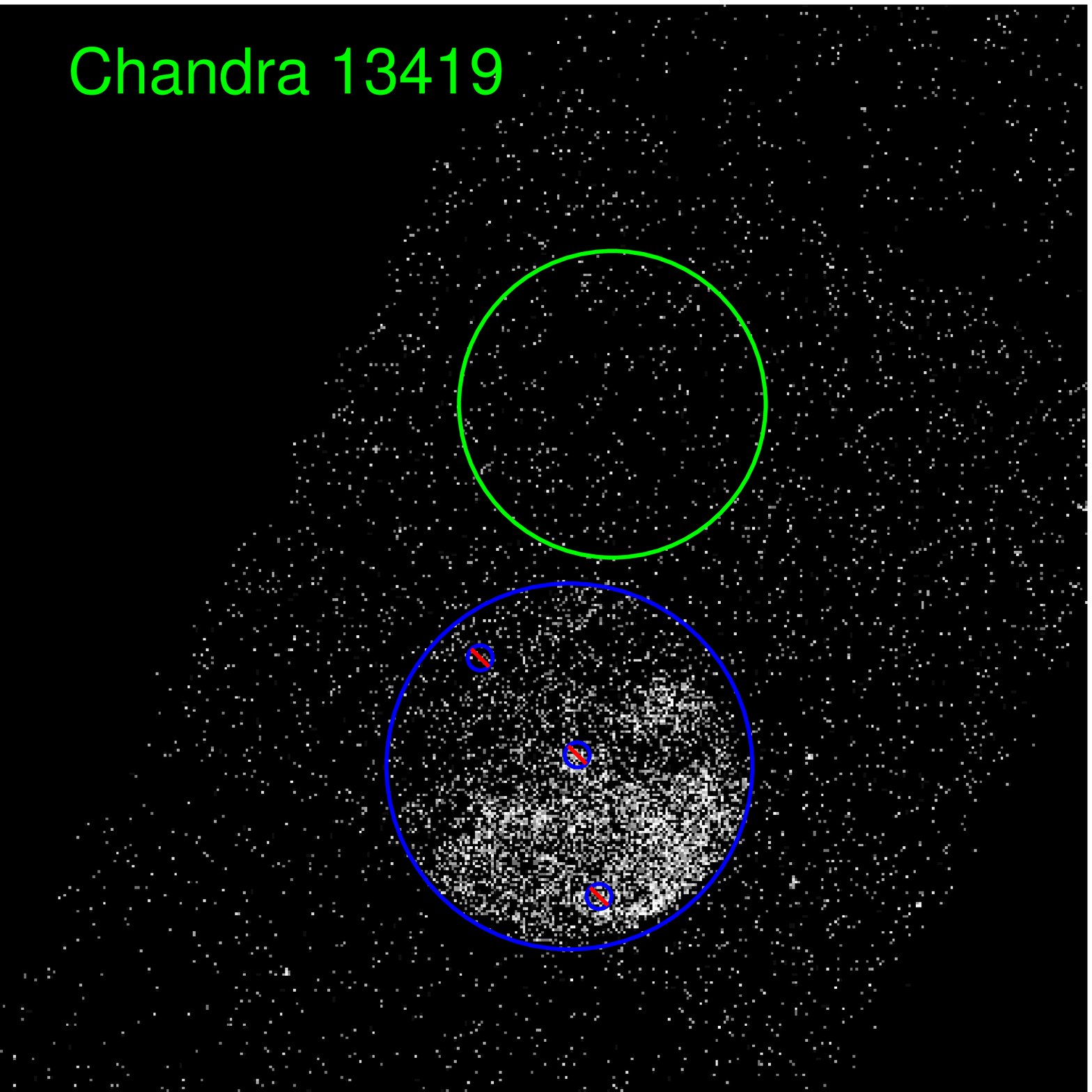}{0.3\textwidth}{(d)}
          \fig{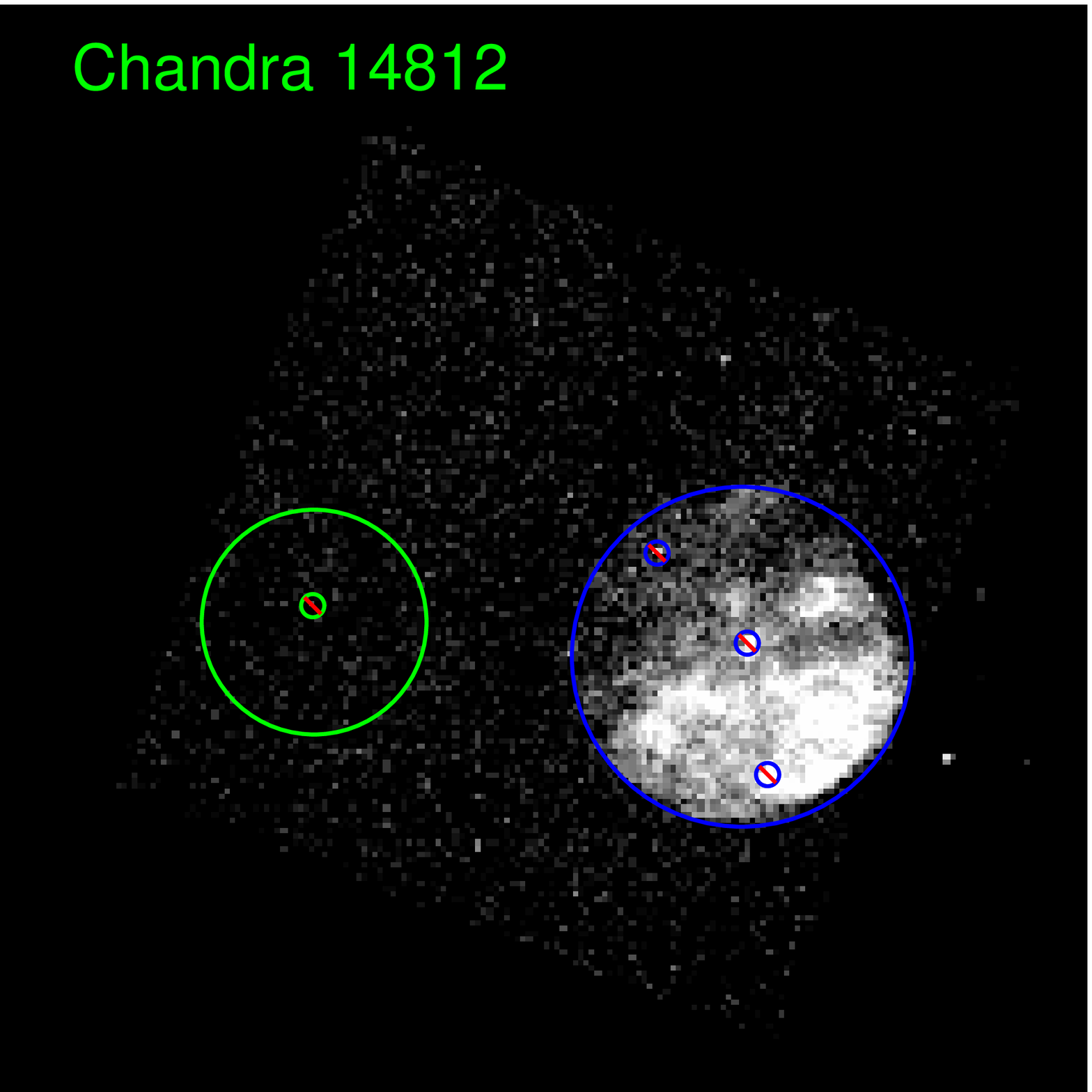}{0.3\textwidth}{(e)}
          }
\caption{Source regions (blue region with a few point sources removed according to C16) and background regions (green region with point sources excluded) for XMM-{\it Newton} MOS1 (a), MOS2 (b), pn (c) and {\it Chandra} 13419 (d), 14812 (e) data.
\label{fig:globalimg}}
\end{figure*}

Figure \ref{fig:globalimg} shows the source regions and background regions used for spectra extraction. Three point-like sources within the SNR were detected in C16 (from north to south indicated as PS\_N, PS\_C and PS\_S). Hence three circular regions with radii of 8 arcsec at the same coordinates (PS\_N: $\alpha_{\rm J2000.0} = 13^{\rm h}21^{\rm m}59^{\rm s}.2$, $\delta_{\rm J2000.0} = 63^\circ32'35''.1 $; PS\_C: $\alpha_{\rm J2000.0} = 13^{\rm h}21^{\rm m}49^{\rm s}.9$, $\delta_{\rm J2000.0} = 63^\circ33'37''.2 $; PS\_S: $\alpha_{\rm J2000.0} = 13^{\rm h}21^{\rm m}47^{\rm s}.8$, $\delta_{\rm J2000.0} = 63^\circ35'07''.8 $) were excluded. Nearby point-sources in the background region were also excluded. 

\section{Molecular Observation} \label{sec:molecular}

\begin{figure*}[ht!]
\includegraphics[width=\linewidth]{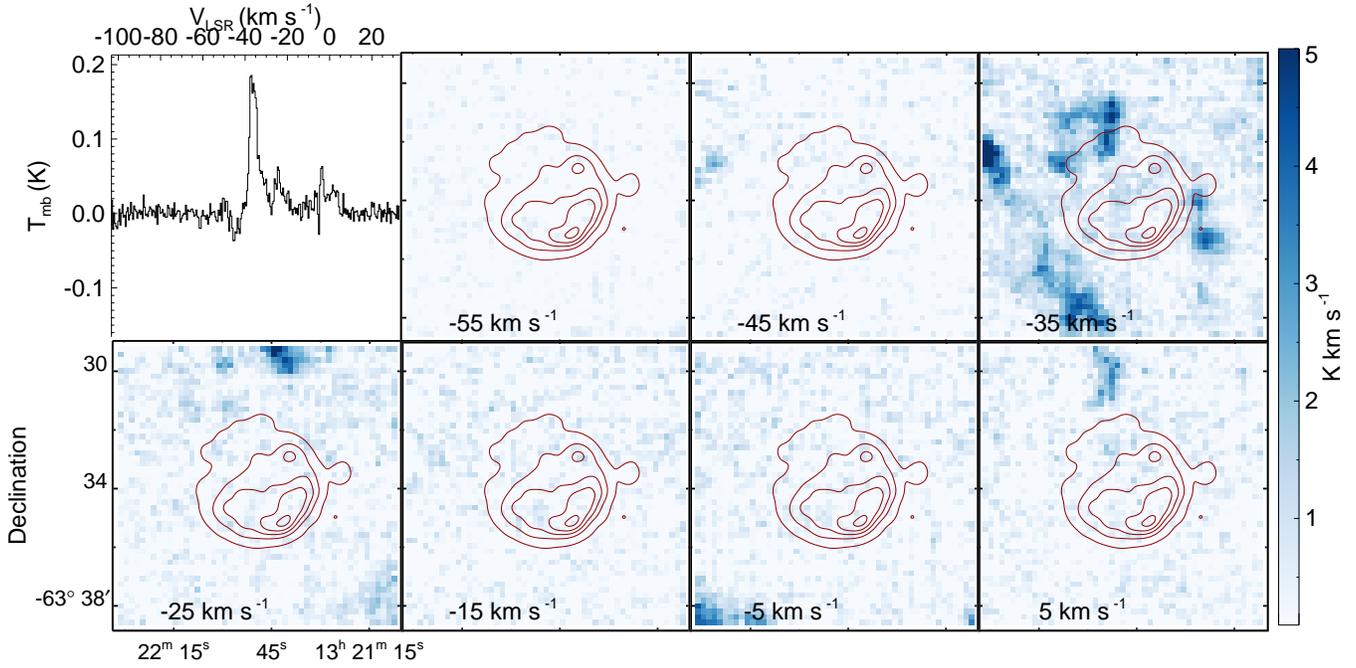}
	\caption{$^{12}$CO(2-1) velocity-integrated temperature maps ($\int T_{\rm MB} dv$) in the velocity range V$_{\rm LSR}=-60$ to 10~km~s$^{1}$ with a step of 10~km~s$^{-1}$. The red contours are taken from the XMM-{\it Newton} 1.5--6~keV X-ray data. The upper-left panel gives the CO spectrum averaged across the field-of-view.
		\label{fig:mc}}
\end{figure*}

Figure \ref{fig:mc} displays the distribution of \twCO\ (2-1) emission at the local standard of rest (LSR) velocity $V_{\rm LSR}=-60$--10~$\rm km s^{-1}$.
Molecular clouds are only found at $V_{\rm LSR}\sim  -35$ and $\sim 0~\rm km s^{-1}$.
Assuming a flat Galactic rotation curve \citep{reid14},
the former velocity corresponds to a distance of 2.8~kpc or 7.1~kpc, while the latter velocity correspond to 9.9~kpc.

We have not found clear evidence to support an association between either of these clouds and the SNR. There is no clear morphology correspondence between them. Moreover, the \twCO\ lines are narrow (less than a few km~s$^{-1}$) and fainter than 4~K, implying that these clouds are quiescent, cold, and thus unshocked by the SNR.

\section{Regional Spectra} \label{sec:regionalspectra}

The regional spectra and their fitting residuals of {\it apec}+{\it vnei} model are shown below.

\begin{figure*}[h!]
\centering
\gridline{\fig{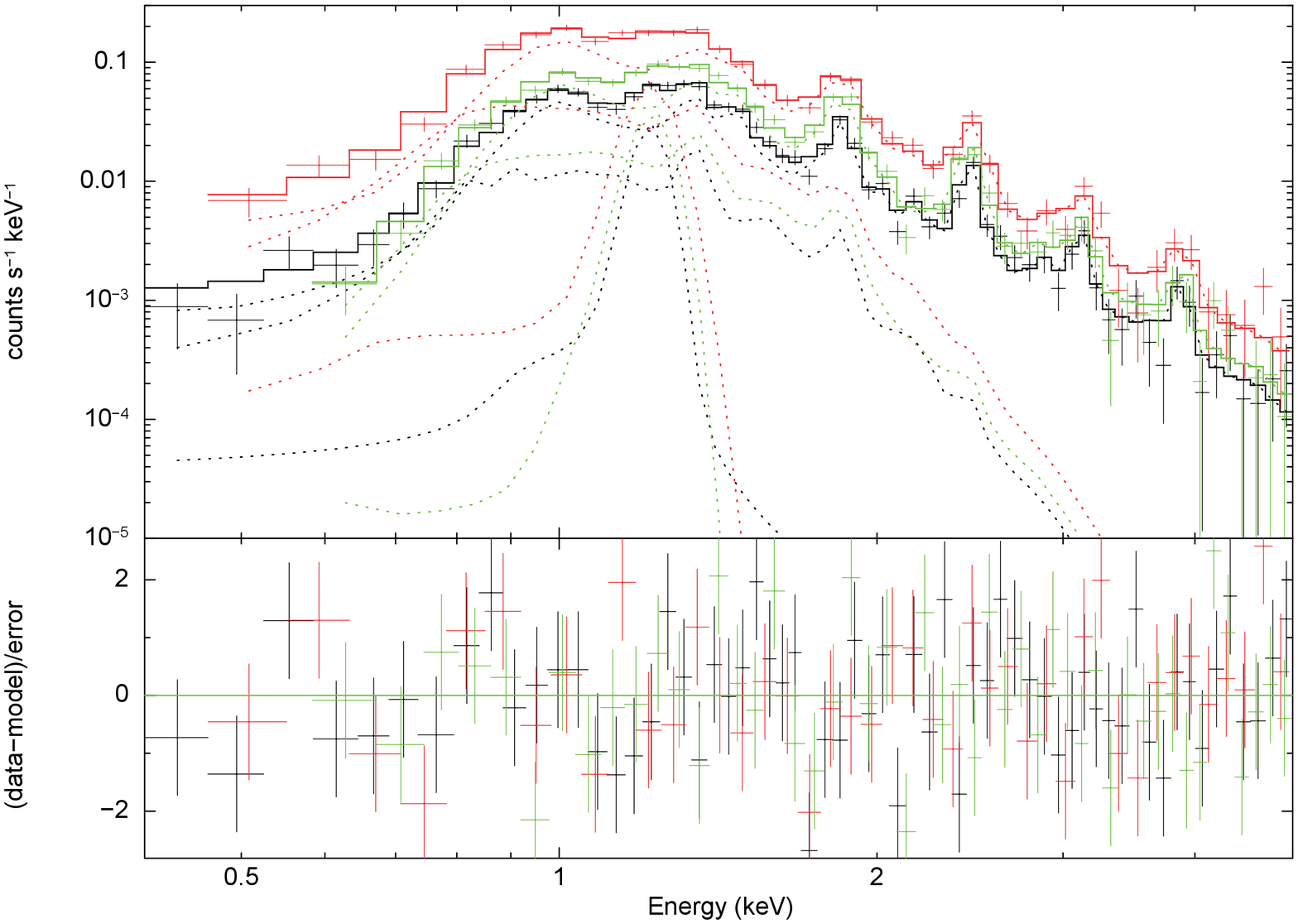}{0.45\textwidth}{reg1}
          \fig{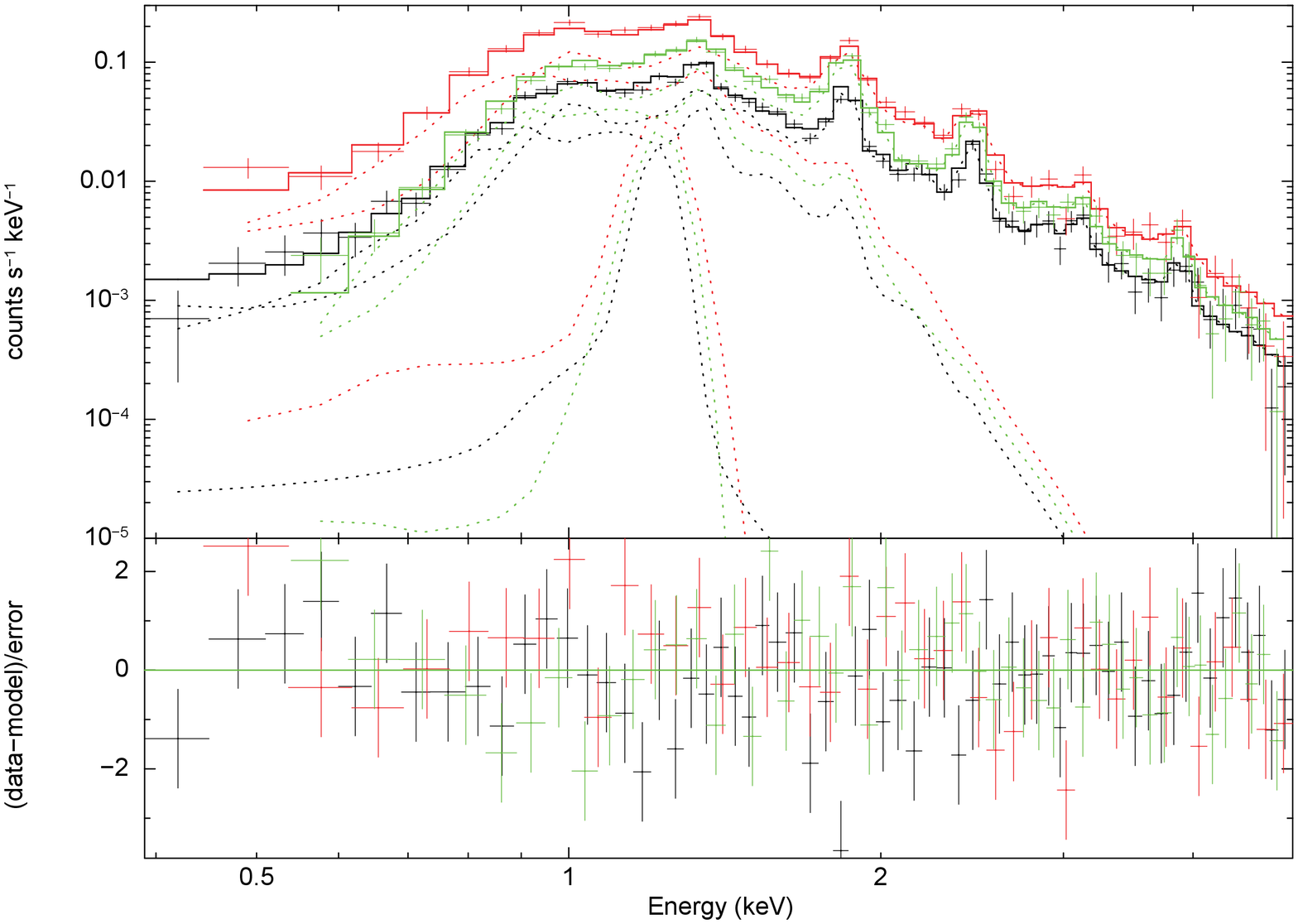}{0.45\textwidth}{reg2}
          }
\caption{The regional spectra and the fitting models. The dashed lines indicate the {\it apec}, {\it vnei} and {\it Gaussian} model components. The red, black and green bins indicate the pn spectra, the combined MOS1/2 spectra, and the combined {\it Chandra} spectra, respectively.}
\end{figure*}

\begin{figure*}[ht!]
\centering
\gridline{\fig{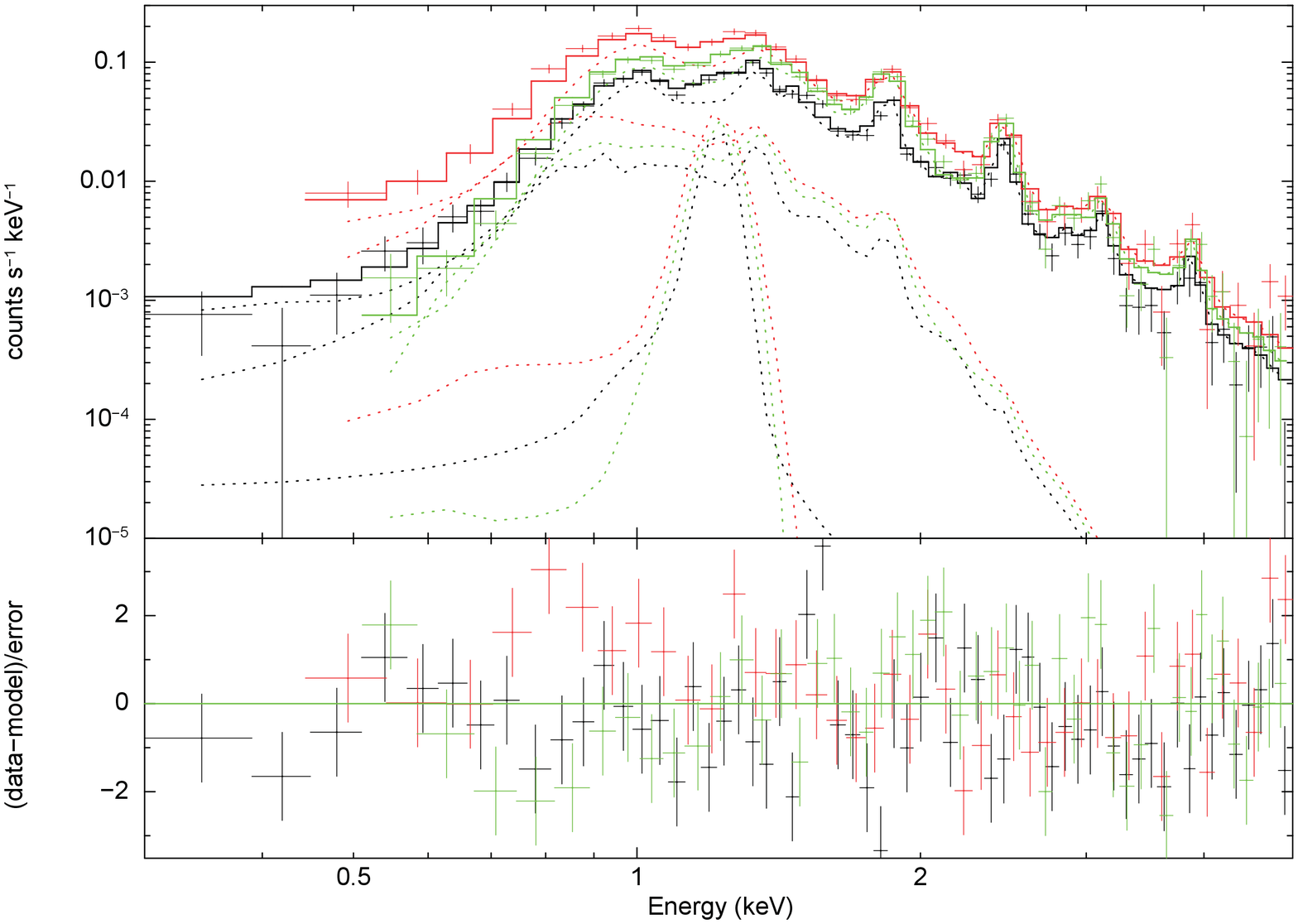}{0.45\textwidth}{reg3}
          \fig{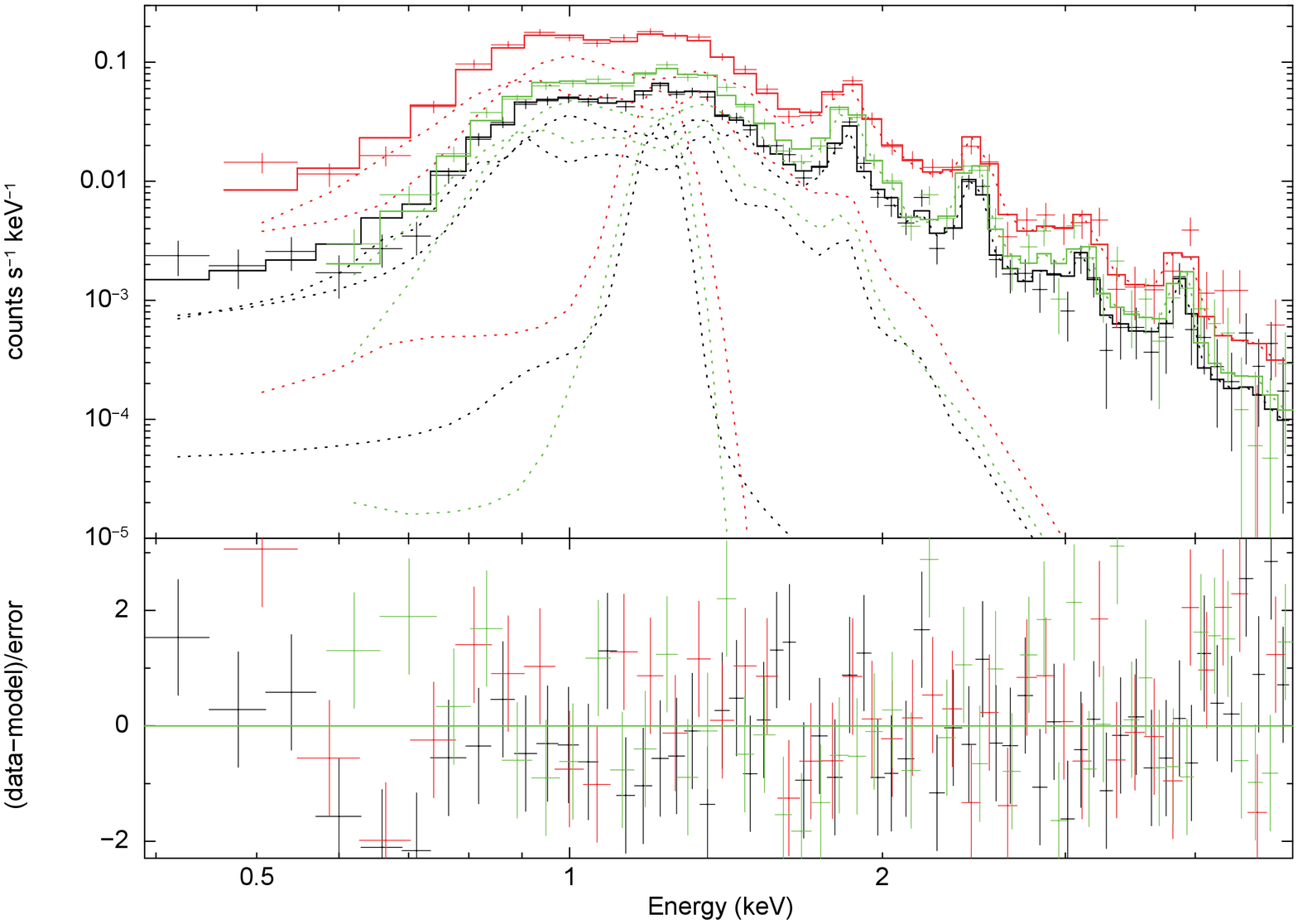}{0.45\textwidth}{reg4}
          }
\gridline{\fig{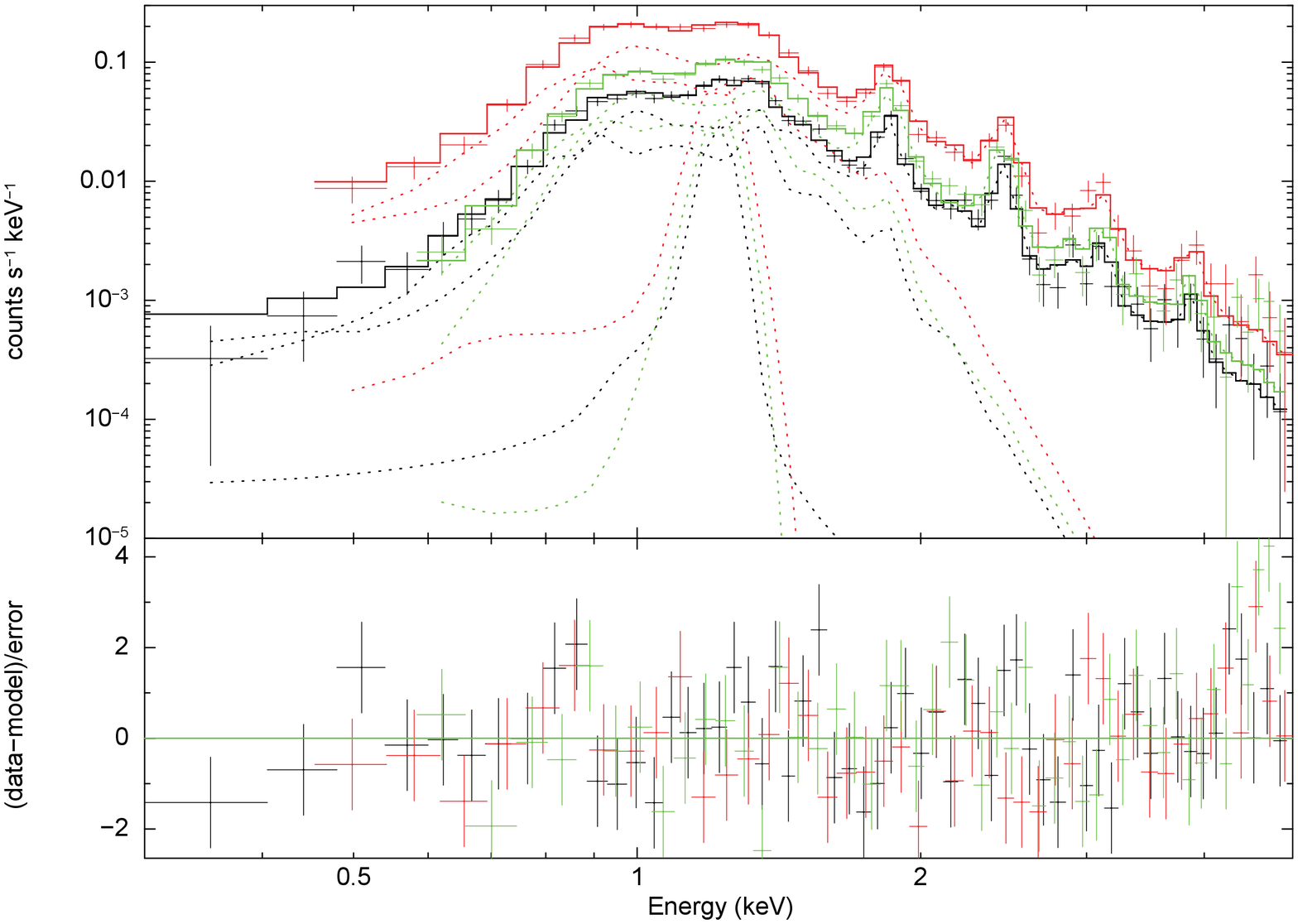}{0.45\textwidth}{reg5}
          \fig{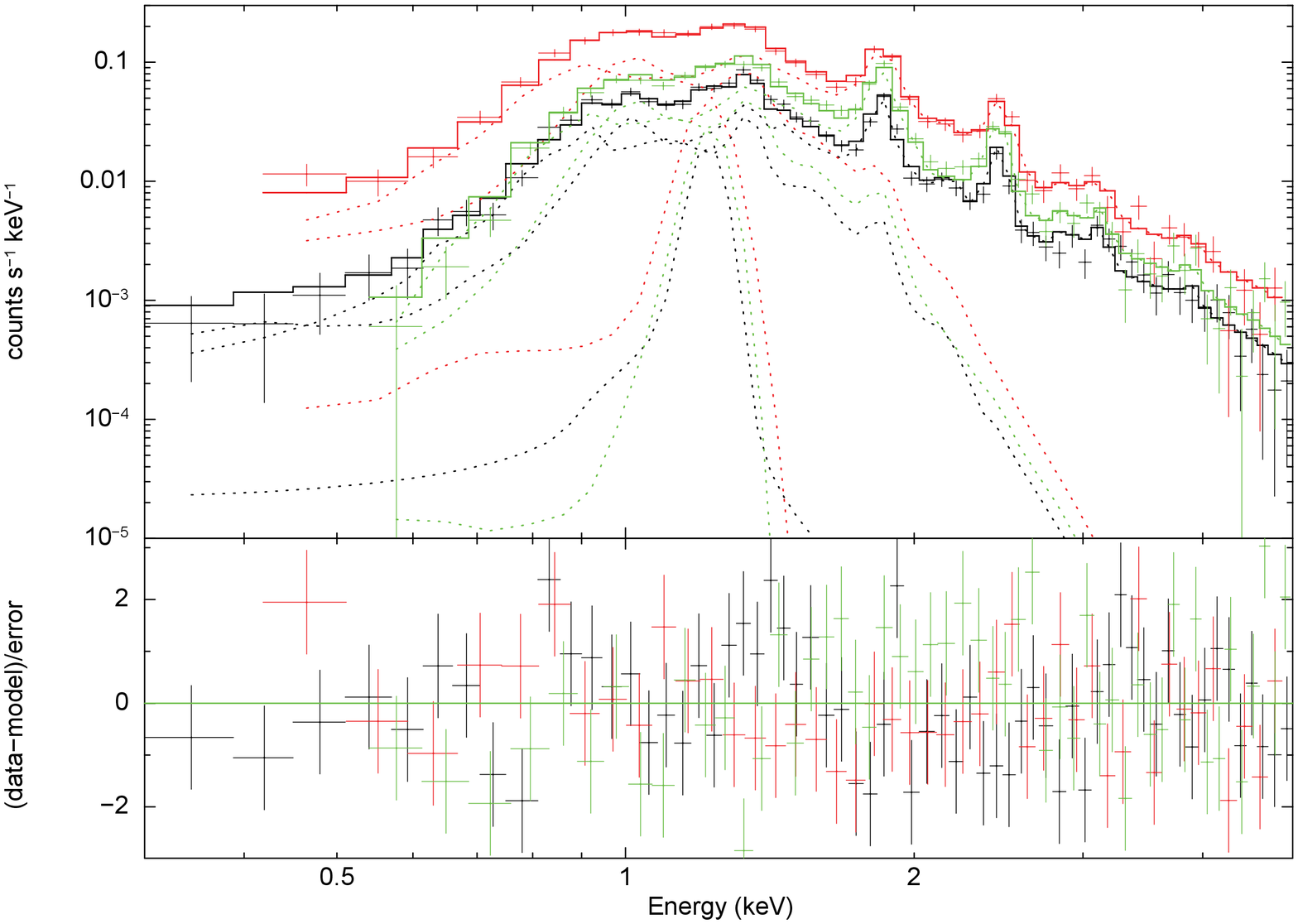}{0.45\textwidth}{reg6}
          }
\gridline{\fig{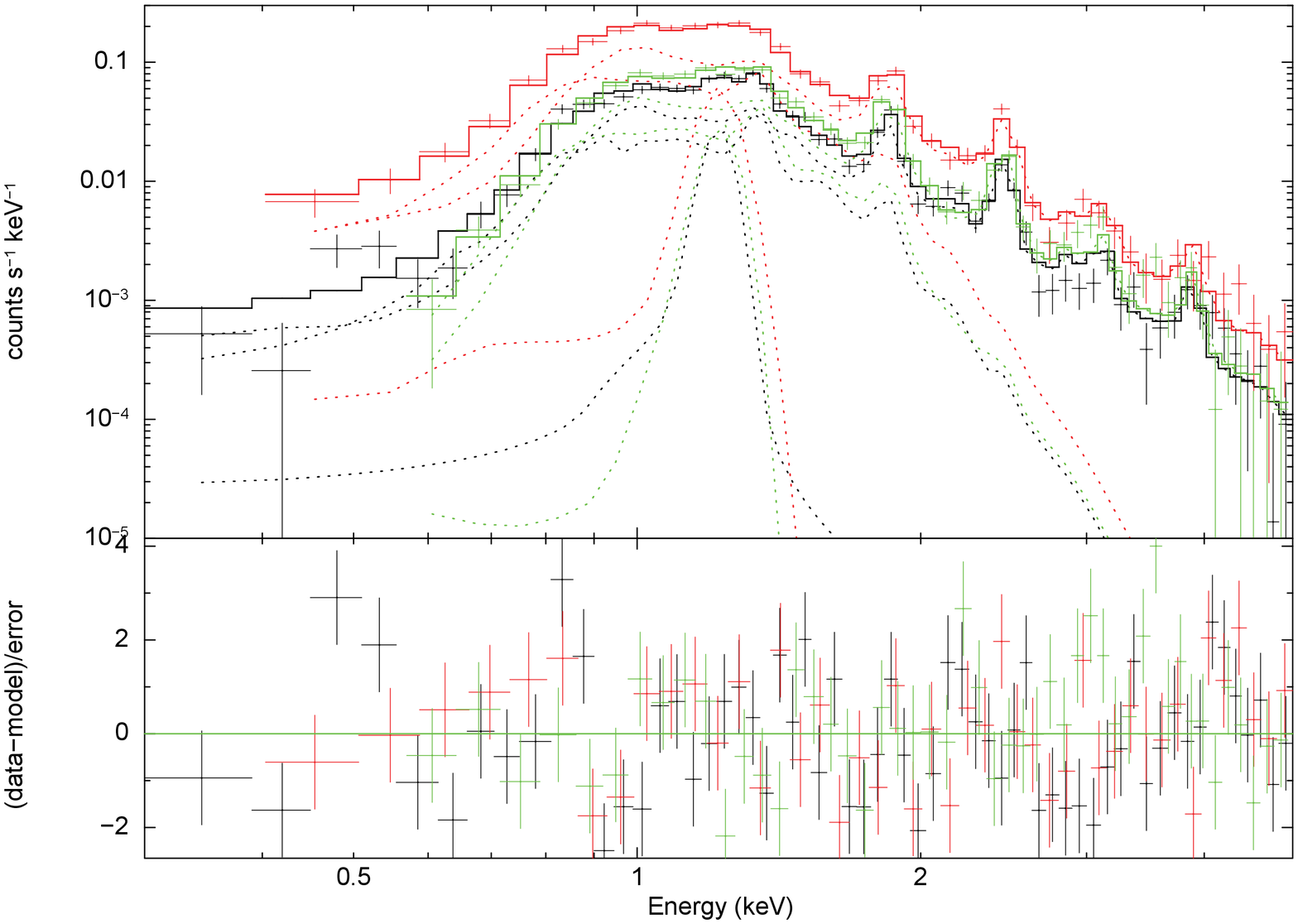}{0.45\textwidth}{reg7}
          \fig{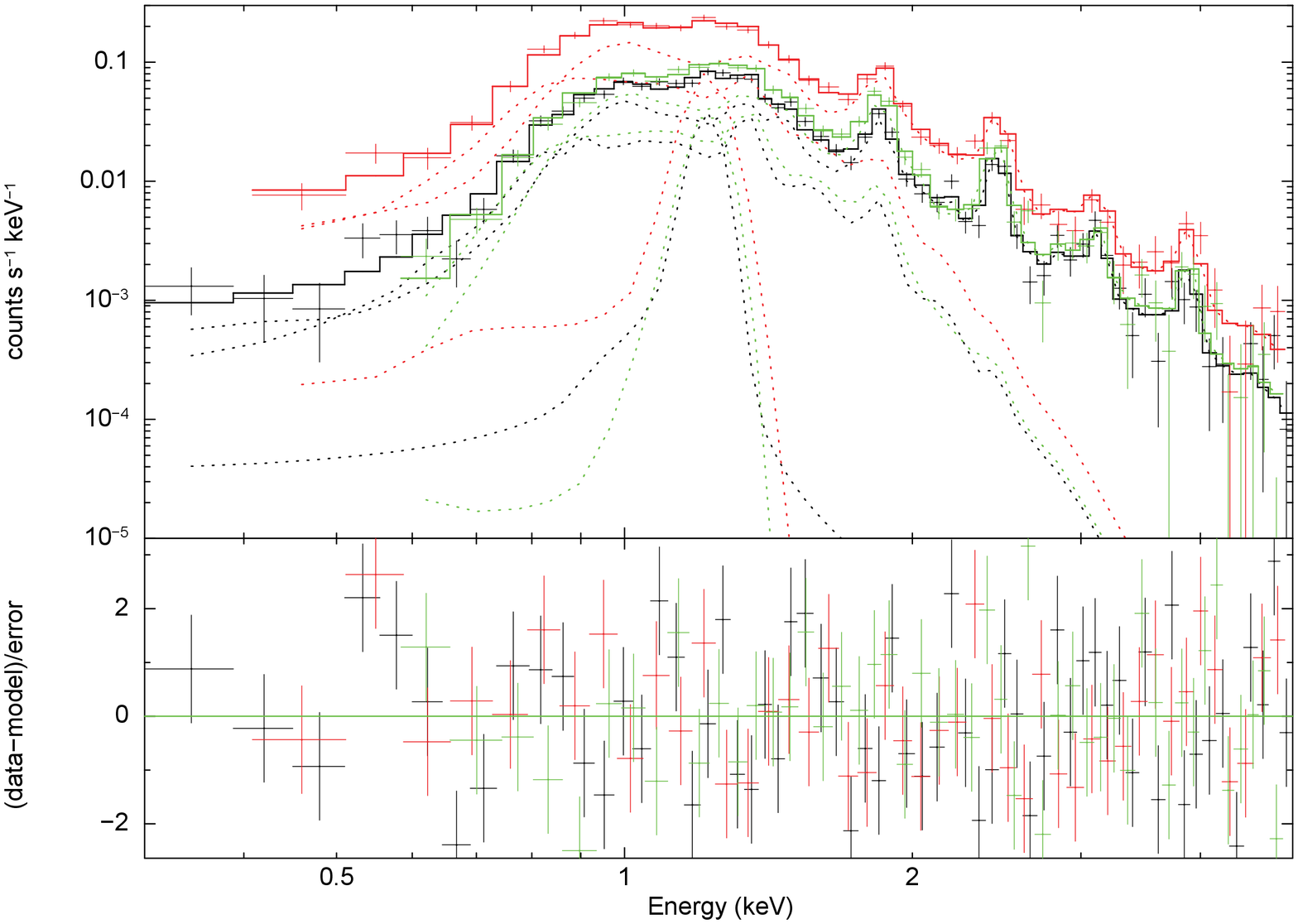}{0.45\textwidth}{reg8}
          }
\figurenum{10}
\caption{{\it Continued}}
\end{figure*}

\begin{figure*}[ht!]
\centering
\gridline{\fig{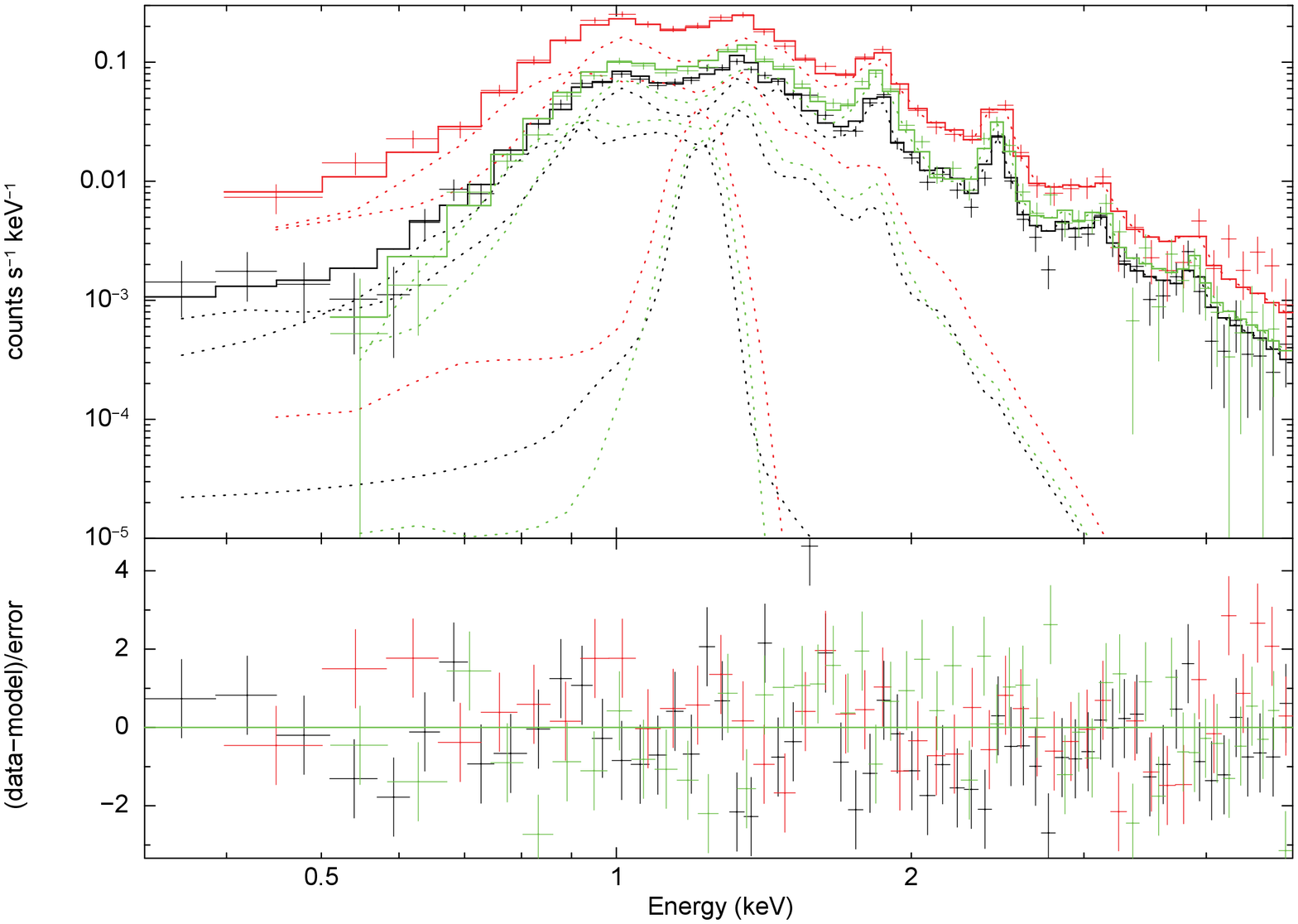}{0.45\textwidth}{reg9}
          \fig{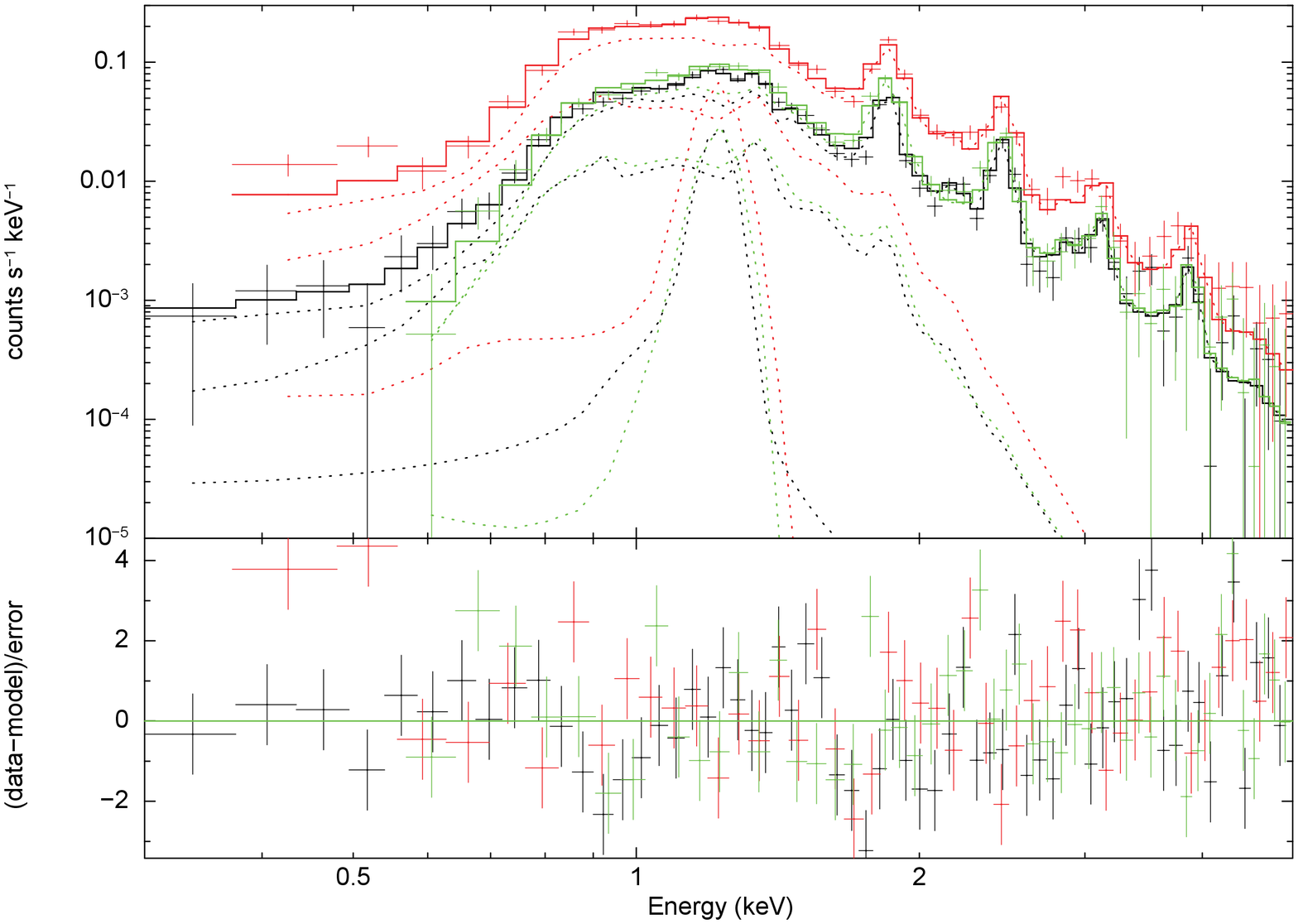}{0.45\textwidth}{reg10}
          }
\gridline{\fig{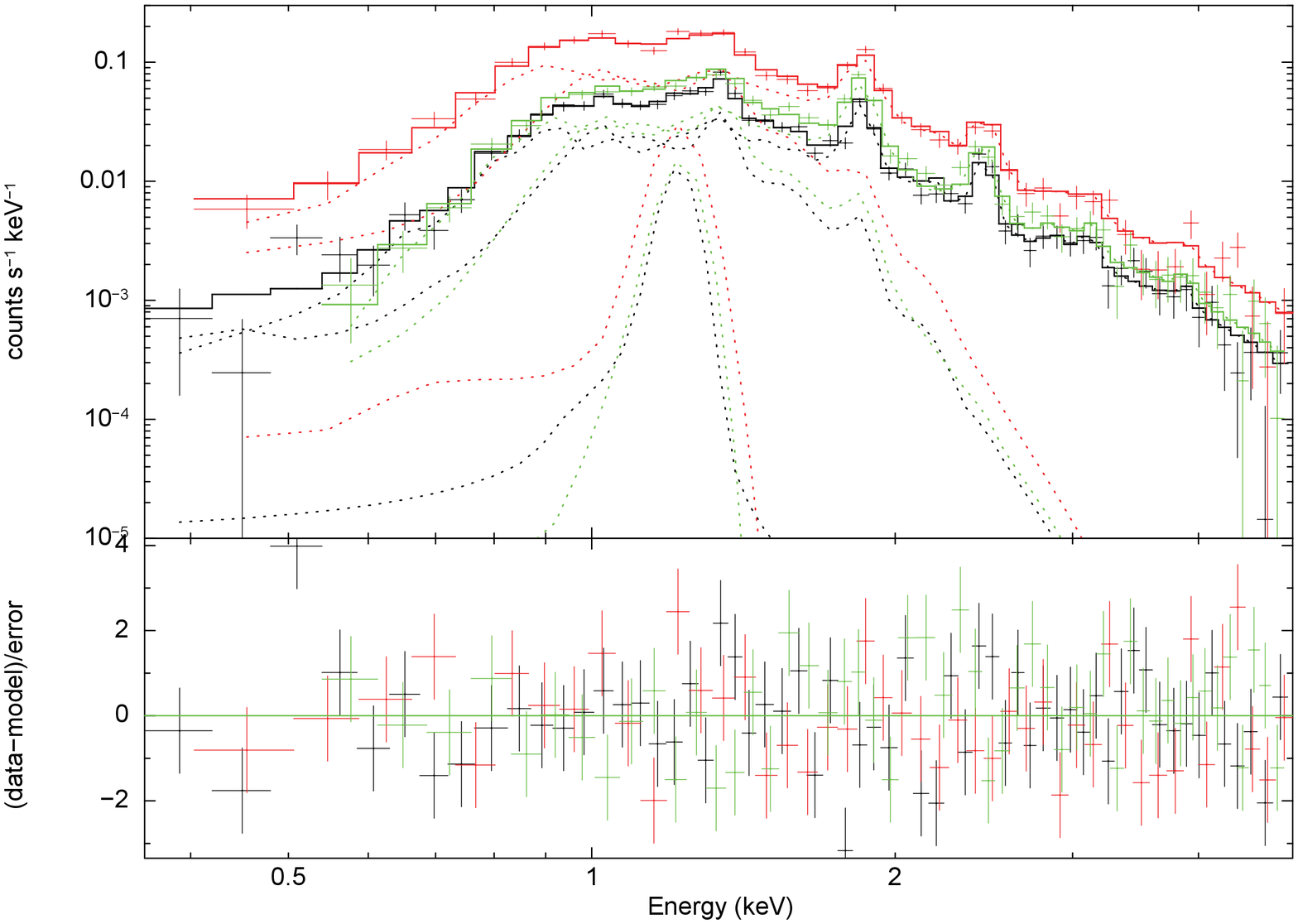}{0.45\textwidth}{reg11}
          \fig{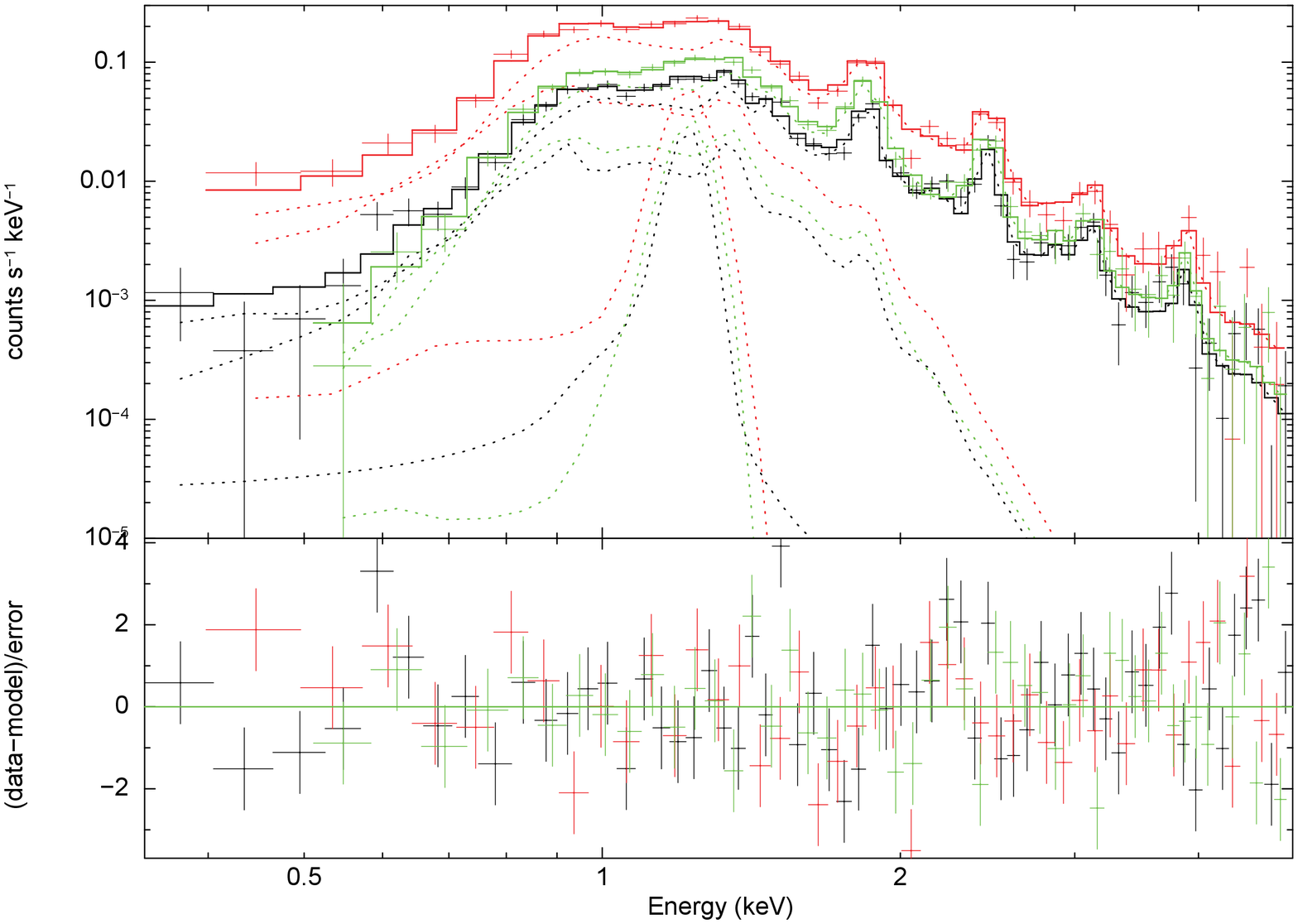}{0.45\textwidth}{reg12}
          }
\gridline{\fig{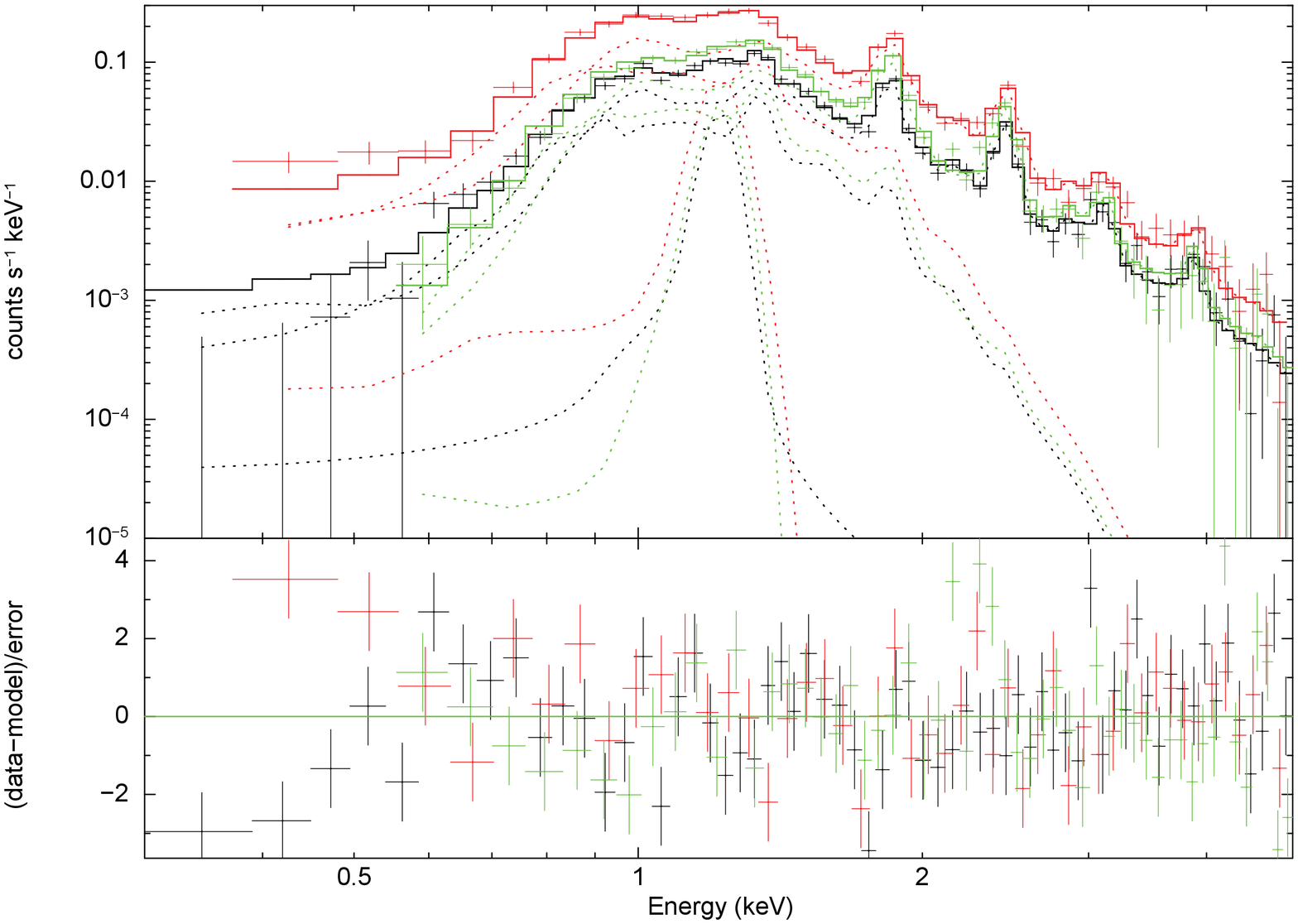}{0.45\textwidth}{reg13}
          }
\figurenum{10}
\caption{{\it Continued}}
\end{figure*}

\listofchanges
\end{document}